\documentclass{elsart}

\usepackage{graphicx} 
\usepackage{epsfig}   

\newcommand{\um}{\mbox{$\mu$m}}
\newcommand{\uA}{\mbox{$\mu$A}}
\newcommand{\degC}{\mbox{$^\circ$C}}
\newcommand{\CaF}{\mbox{CaF$_2$}}
\newcommand{\VARICH}{\mbox{VA\_RICH}}  
\newcommand{\NIM}{Nucl.\ Instr.\ Meth.\ }

\begin{document}
%
\begin{flushright}
SUHEP 20-2005\\
June, 2005
\end{flushright}

\begin{frontmatter}

\title{THE CLEO RICH DETECTOR}

\author[syr]{M.~Artuso},
\author[syr]{R.~Ayad},
\author[syr]{K.~Bukin},
\author[syr]{A.~Efimov},
\author[syr]{C.~Boulahouache},
\author[syr]{E.~Dambasuren},
\author[syr]{S.~Kopp},
\author[syr]{Ji~Li},
\author[syr]{G.~Majumder},
\author[syr]{N.~Menaa},
\author[syr]{R.~Mountain},
\author[syr]{S.~Schuh},
\author[syr]{T.~Skwarnicki},
\author[syr]{S.~Stone\corauthref{cor}},
\corauth[cor]{Corresponding author.}\ead{stone@physics.syr.edu}
\author[syr]{G.~Viehhauser},
\author[syr]{J.C.~Wang}
\address[syr]{Department of Physics, Syracuse University,  Syracuse, NY 13244-1130, USA }

\author[SMU]{T.E.~Coan},
\author[SMU]{V.~Fadeyev},
\author[SMU]{Y.~Maravin},
\author[SMU]{I.~Volobouev},
\author[SMU]{J.~Ye}
\address[SMU]{Department of Physics, Southern Methodist University, Dallas, TX 75275-0175, USA}

\author[Minn]{S.~Anderson},
\author[Minn]{Y.~Kubota},
\author[Minn]{A.~Smith}
\address[Minn]{School of Physics and Astronomy, University of Minnesota, Minneapolis, MN 55455-0112, USA}


\begin{abstract}
We describe the design, construction and performance of a Ring
Imaging Cherenkov Detector (RICH) constructed to identify charged
particles in the CLEO experiment. Cherenkov radiation occurs in LiF
crystals, both planar and ones with a novel ``sawtooth''-shaped exit
surface. Photons in the wavelength interval 135--165 nm are detected
using multi-wire chambers filled with a mixture of methane gas and
triethylamine vapor. Excellent $\pi/K$ separation is demonstrated.
\end{abstract}

\end{frontmatter}

\section{ INTRODUCTION }
The CLEO II detector was revolutionary in that it was the first to
couple a large magnetic tracking volume with a precision crystal
electromagnetic calorimeter capable of measuring photons down to the
tens of MeV level \cite{cleoii}. CLEO II produced many
ground-breaking physics results, but was hampered by its limited
charged-hadron identification capabilities that were provided by a
combination of dE/dx and time-of-flight measurements.

The CLEO III detector was designed to study decays of $b$ and $c$
quarks, $\tau$ leptons and $\Upsilon$ mesons produced in $e^+e^-$
collisions near 10 GeV center-of-mass energy. CLEO III is an
upgraded version of CLEO II, as the magnet, the calorimeter and the
muon system were kept. It contained a new four-layer silicon strip
vertex detector, a new wire drift chamber and a particle
identification system based on Cherenkov ring imaging described
herein. Information about CLEO is available
elsewhere~\cite{Artu98,Kopp96}.

Design choices for particle identification were limited by radial
space and the necessity of minimizing material in front of the CsI
crystal calorimeter. The CsI imposed a hard radial outer limit, and
the desire for maintaining excellent charged particle tracking
imposed a radial lower limit, since at high momentum the error in
momentum is inversely proportional to the square of the track
length. The particle identification system was allocated only 20~cm
of radial space, and this limited the technology choices. To retain
the superior performance of the calorimeter, a material thickness of
only 12\% of a radiation length was allowed.

The CLEO III installation including the RICH detector occurred in
the summer of 1999  and was used to study physics in the $e^+e^-$
center-of-mass energy region around 9-12 GeV. The two-ring machines
KEKB and PEP-II produced much more luminosity than CESR, and after
some excellent initial results on $B$ decays and Upsilon
spectroscopy, it became clear that there was much physics to be
explored in charm decays and studies of charmonium. The
transformation from CLEO III to CLEO-c was made in 2003, when the
CESR accelerator began operating in the 3-5 GeV center-of-mass
energy region. The inner double-sided silicon detector was replaced
with a wire drift chamber that has significantly less material and
is much better suited to the CLEO-c physics program; in any case,
the silicon was showing effects of premature radiation damage. The
magnetic field was lowered from 1.5~T to 1~T, mainly for accelerator
related considerations.

The plan of this paper is as follows:  The basic RICH detector
design will be delineated, followed by descriptions of the
individual components, namely the crystals, multi-wire chambers,
superstructure, readout electronics, and support subsystems. Issues
in the long-term operation of the RICH are then discussed. Finally,
some details of the data analysis technique and physics performance
are presented.


\section{ DETECTOR OVERVIEW }
\label{sec:overview}
\subsection{Design Choices}

The severe radial spatial constraint forces the design to have a
thin, few-cm detector for Cherenkov photons and a thin radiator.
Otherwise the photons have too little distance to travel and it
becomes very difficult to precisely measure the photon angles. In
fact, the only thin, large-area photon detectors possible in our
situation were wire chamber based, either using a reflective CsI
photocathode or a gas mixture of methane (CH$_4$) and triethylamine
(TEA) vapor to convert ultraviolet photons.\footnote{A
    TMAE-based photon detector would have been unacceptably thick
    in order to obtain the same detection efficiency,
    due to its long photon absorption length.} Use of
CsI would have allowed us to use a liquid freon radiator with quartz
windows in the system and to work in the wavelength region from
about 160--200~nm. However, at the time of decision, the use of CsI
was far from proven and, in any case, would have imposed severe
constraints on the construction process, which would have been both
technically difficult and expensive. Thus we chose a multi-wire
chamber filled with a mixture of CH$_4$ and TEA that uses Cherenkov
photons in the vacuum ultraviolet (VUV) region, 135--165~nm,
generated in a 1~cm thick LiF crystal.

\begin{figure}[tb]
  \vspace*{-0.001in}
  \begin{center}
    \includegraphics*[height=3in]{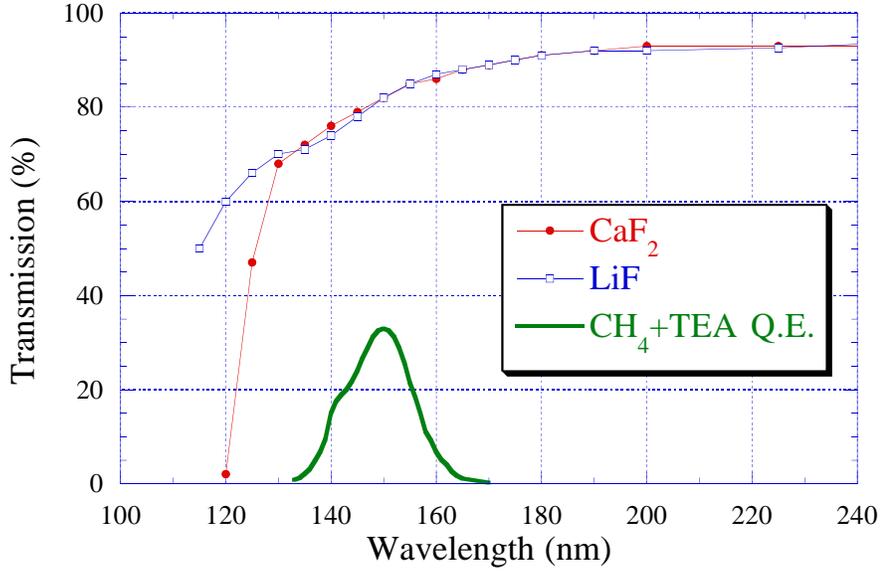}
  \end{center}
  \vspace*{-0.100in}
  \caption{\label{plane_trans}
    The transmission of the LiF and \CaF\ optics for the CLEO RICH,
    along with the measured quantum efficiency of CH$_4$ plus TEA gas~\cite{j+t}.
   }
  \vspace*{-0.001in}
\end{figure}

The quantum efficiency of the CH$_4$ and TEA mixture peaks at
$\sim$150~nm~\cite{j+t}, as shown in Fig.~\ref{plane_trans}, below
the transmission cutoff for glasses and fused silica quartz. This
short wavelength requires the use of alkali halide crystals as both
the Cherenkov-emitting medium and the entrance window to the
photosensitive volume \cite{Arno92}. We chose LiF as the Cherenkov
radiator because it has the lowest dispersion in the wavelength band
of the CH$_4$-TEA quantum efficiency~\cite{t+j}. Transparent gases
must be used between the radiator and the photon detector that are
almost entirely free of O$_2$ and H$_2$O, both of which have
1--10~Mbarn cross section for photon absorption below
175~nm~\cite{js89}.

\begin{figure}[htb]
  \vspace*{-0.001in}
  \begin{center}
    \includegraphics*[height=4in]{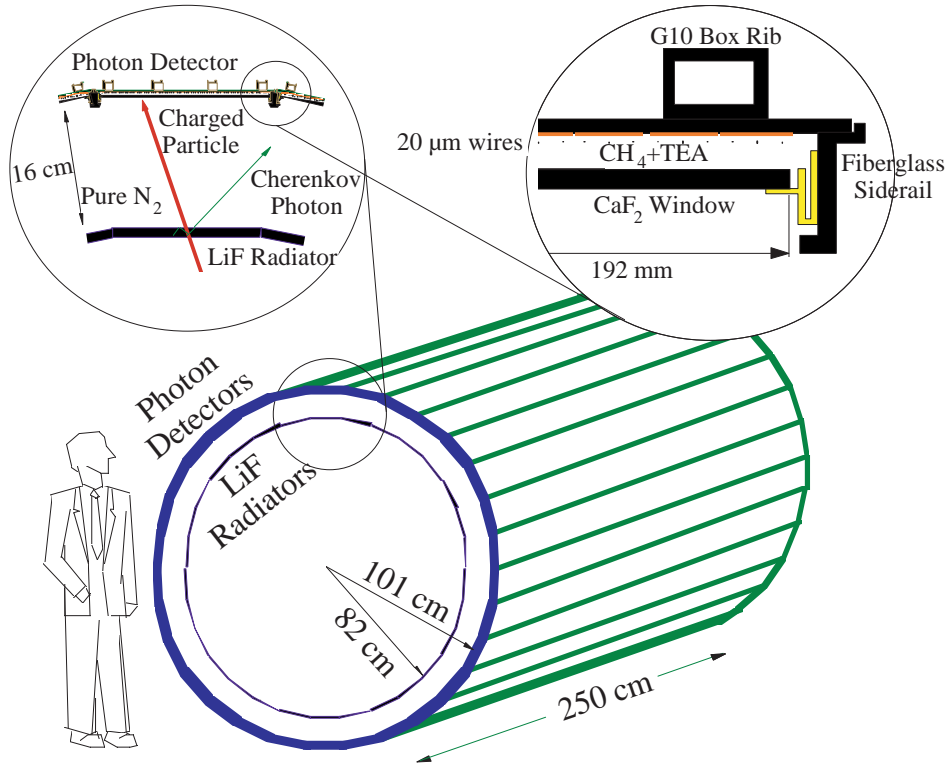}
  \end{center}
  \vspace*{-0.100in}
  \caption{\label{fig:rich_descrip} Outline of the CLEO RICH detector design. }
  \vspace*{-0.001in}
\end{figure}

\begin{figure}[htb]
  \vspace*{-0.001in}
  \begin{center}
    \includegraphics*[height=3in]{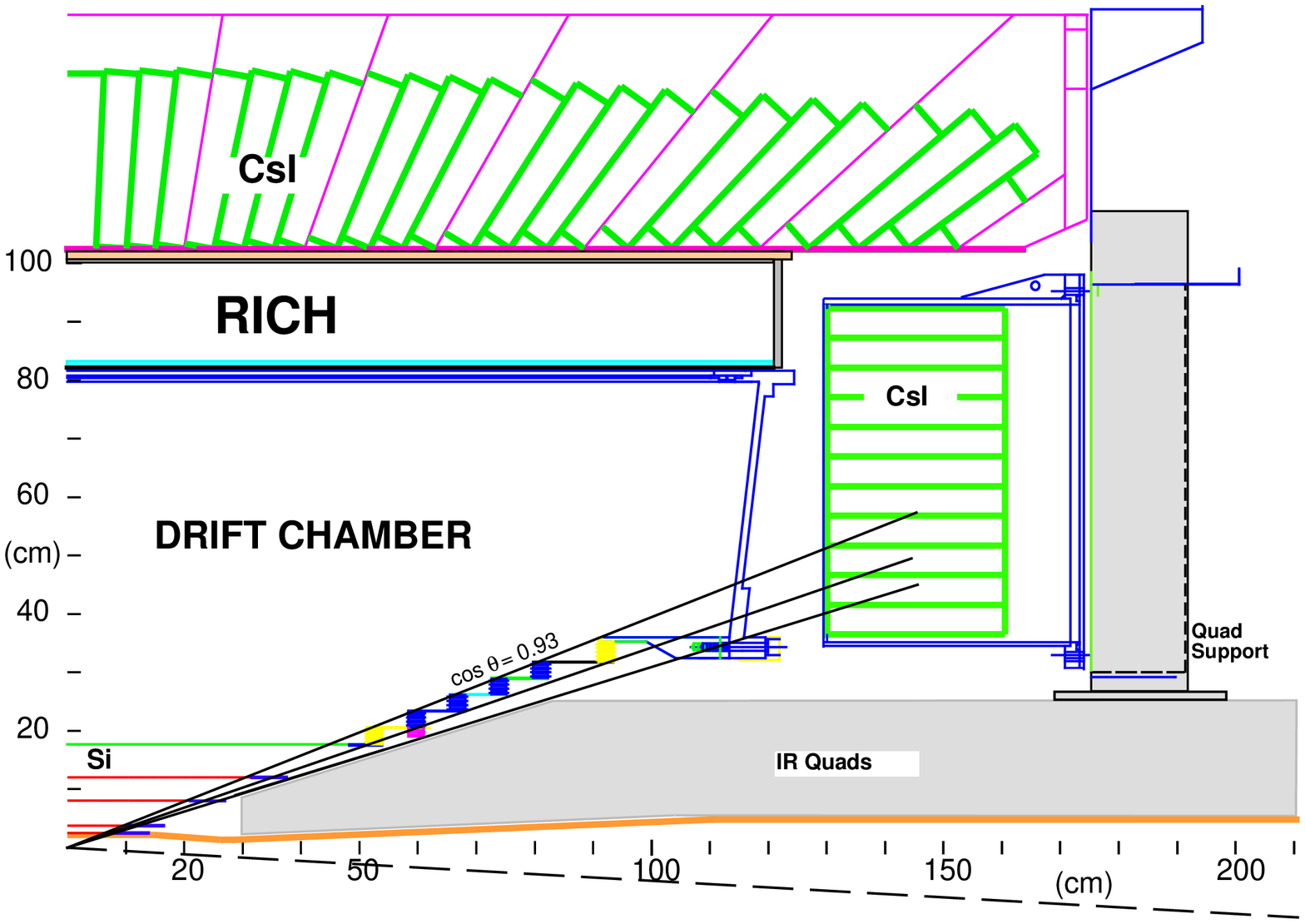} 
  \end{center}
  \vspace*{-0.100in}
  \caption{\label{fig:richcleo} The CLEO RICH, shown in CLEO-III configuration. }
  \vspace*{-0.001in}
\end{figure}

Details of the design of the CLEO RICH have been discussed
elsewhere~\cite{testbeam,Artu01}. Here we briefly review the main
elements. Cherenkov photons are produced in a LiF radiator. The
photons then traverse a free space, an ``expansion volume," where
the cone of Cherenkov light expands in size. The photons enter a
detector consisting of spatially-segmented multi-wire chambers (MWC)
filled with CH$_4$ gas mixed with TEA vapor, in which they are
converted to electrons and multiplied. Finally, signals are picked
up with sensitive low-noise electronics. No optical focusing
elements are used; this is called ``proximity-focusing''~\cite{t+j}.
The scheme is shown in the upper left of
Fig.~\ref{fig:rich_descrip}, while the placement in CLEO is shown in
Fig.~\ref{fig:richcleo}.

There are 30 individual photon detectors around the outer cylinder.
They subtend the same azimuthal angle as the radiators, which are
segmented into 30 rows and 14 rings on the inner cylinder. The
expansion volume between the radiators and detectors is filled with
pure N$_2$ gas. In fact, any transparent, pure gas could be used.
Our choice was based on the availability of very pure gas from the
super-conducting coil boil off.

The most distinctive features of the design of the CLEO RICH are:
(1) ultra-thin photon detectors, with high spatial segmentation and
almost full area coverage; (2) a large volume of high-purity,
VUV-transparent crystals including a novel ``sawtooth'' geometry;
and (3) exceptionally low-noise analog electronics readout.

\subsection{Radiators}

\begin{figure}[htb]
  \vspace*{-0.001in}
  \begin{center}
    \includegraphics*[height=2in]{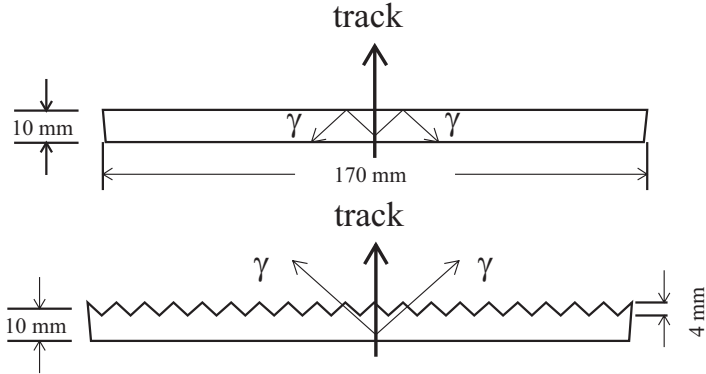}
  \end{center}
  \vspace*{-0.100in}
  \caption{\label{fig:radiators} Sketch of a plane radiator (top) and a sawtooth radiator (bottom).
    Cherenkov light paths radiated from a charged track normal to each radiator are shown. }
  \vspace*{-0.001in}
\end{figure}

LiF was chosen over CaF$_2$ or MgF$_2$, both of which are
transparent in the VUV wavelength region, due to its lower
dispersion leading to smaller chromatic error. Originally all the
radiators were planned to be 1~cm thick planar pieces.  However,
since the refractive index of LiF at 150~nm is 1.5, all the
Cherenkov light generated from tracks normal to the LiF surface
would be totally internally reflected, as shown in
Fig.~\ref{fig:radiators} (top).  We could have used these planar
radiators, but we would have had to tilt them at about a
15$^{\circ}$ angle. Besides the obvious mechanical support problem,
some tracks would produce Cherenkov photons that would cross through
the entire thickness of another radiator tile causing a loss of
efficiency and reconstruction problems. Instead, we developed novel
radiators with a serrated top surface, called ``sawtooth"
radiators~\cite{efimov}, as shown in Fig.~\ref{fig:radiators}
(bottom). Measured physical properties of the radiators have been
described previously \cite{testbeam}.

The overall radiator shape approximates a cylinder of radius 82~cm.
Individual radiator crystals are placed in 14~coaxial rings of
30~crystals each, centered around the beam line and symmetrically
positioned about the interaction point. The 30 crystals segments are
parallel to the wire chambers. Inter-crystal gaps are typically
50--100~$\mu$m. The crystals are attached to the exterior surface of
a 1.5~mm thick carbon fiber shell with a low outgassing epoxy.

The inner four rings are made of sawtooth radiators. Lengthy
production time as well as cost limited our use of these novel
objects.

\subsection{Photon Detectors}

The photon detectors have segmented cathode pads 7.5~mm (length)
$\times$ 8.0~mm (width) etched onto G10 printed-circuit boards
(PCBs). The pad array was formed from four individual boards, each
with 24 $\times$ 80 pads, the latter split into two 24 $\times$ 40
pad sections with a 7 mm gap. Each board was individually flattened
in an oven and then they were glued together longitudinally on a
granite table where reinforcing G10 ribs were also glued on. The
ribs have a box-like structure. There are 4 longitudinal ribs that
traverse the entire length. Smaller cross ribs are placed every 12
cm for extra stiffening. The total length of the pad array is 2.46
m.

Wire planes were strung with 20 $\mu$m diameter gold plated tungsten
with a 3\% admixture of rhenium; the wire pitch was 2.66 mm, for a
total of 72 wires per chamber. The wires were placed on and
subsequently glued to precision ceramic spacers every 30 cm. The
spacers extend 1 mm above the cathodes, and therefore are 3.5 mm
from the CaF$_2$ windows. We achieved a tolerance of 50 $\mu$m on
the wire to cathode distance. The spacers had slots in the center
for the glue bead.

Eight 30 cm $\times$ 19 cm CaF$_2$ windows were glued together in
precision jigs lengthwise to form a 2.4 m long window. Positive high
voltage is applied to the anode wires, while negative high voltage
is put on 100 $\mu$m wide silver traces deposited on the CaF$_2$.
The spacing between the traces is 2.5 mm, and they are connected
together by a trace running across both edges. The pad plane is
close to ground. The gain of the chamber depends on the amount of
high voltage on wires and windows. We define ``wire-gain," in the
normal manner, as the multiplication factor on a single electron. We
run the system at typical wire-gains around 30,000. ``Pad-gain" is
our usual measure of multiplication and is calculated as being 75\%
of the wire-gain.

To maintain the ability of disconnecting any faulty part of a
chamber, the high voltage is distributed independently to three
groups of 24 wires and the windows are each powered separately.

\subsection{Electronics}

\begin{figure}[htb]
  \vspace*{-0.001in}
  \begin{center}
    \includegraphics*[height=2.5in]{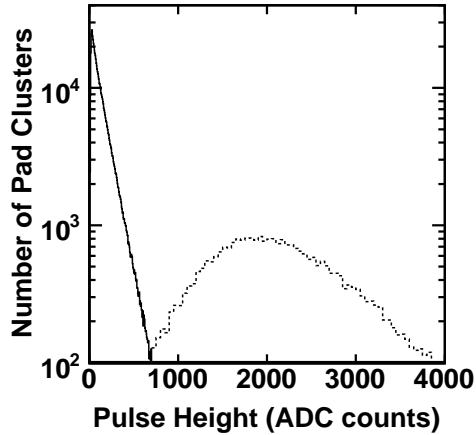}
  \end{center}
  \vspace*{-0.100in}
  \caption{\label{fig:ph-dist} Pulse height distributions
    from pad clusters containing single photons (solid histogram) and
    charged tracks (dashed histogram). The photon
    data is consistent with an exponential distribution. One ADC count corresponds to
    $\sim$200 electrons. The charged track distribution is
    affected by electronic saturation. }
  \vspace*{-0.001in}
\end{figure}

The position of Cherenkov photons is measured by sensing the induced
avalanche charge on the cathode pad array. Since the pulse height
distribution from single photons is expected to be exponential at
low to moderate gas gains~\cite{expon}, the use of low noise
electronics is required to ensure high detection efficiency. Pad
clusters in the detector can be formed from single Cherenkov
photons, overlaps of more than one Cherenkov photon, or charged
tracks. In Fig.~\ref{fig:ph-dist} we show the pulse height
distribution for single photons and charged tracks. (Note, one ADC
count corresponds to
    $\sim$200 electrons.) These can be
separated by just examining the pulse height. The charged tracks
give very large pulse heights because they are traversing
$\sim$4.5~mm of the CH$_4$-TEA mixture. We can distinguish somewhat
between the charge due to single photons and two photons because of
the pulse height shapes on adjacent pads.

To have as low noise electronics as possible, a dedicated VLSI chip,
called \VARICH, based on a very successful chip developed for solid
state applications, has been designed and produced for our
application at IDE AS, Norway~\cite{ideas}. We have fully
characterized 3,600 64-channel chips, mounted on hybrid circuit
boards. For moderate values of the input capacitance $C_{in}$, the
equivalent noise charge measured $ENC$ is found to be about
    \begin{equation}
        ENC = 130 \: e^- + (9 \: e^-\!/\mbox{pF}) \cdot C_{in}~~.
    \end{equation}
The dynamic range of the chip is between 450,000 and 900,000
electrons, depending upon whether we choose a bias point for the
output buffer suitable for signals of positive or negative polarity
or we shift this bias point to have the maximum dynamic range for
signals of a single polarity.

In our readout scheme we group 10 chips in a single readout cell
communicating with data boards located in VME crates just outside
the detector cylinder. Chips in the same readout cell share the same
cable, which routes control signals and bias voltages from the data
boards and output signals to the data boards. Two \VARICH\ chips are
mounted using wire bonds on one hybrid circuit that is attached via
two miniature connectors to the back of the cathode board of the
photon detector.

The output of the VA\_RICH is transmitted to the data boards as an
analog differential current, transformed into a voltage by
transimpedance amplifiers and digitized by a 12-bit differential
ADC. These receivers are  part of  very complex data boards which
perform several important analog and digital functions. Each board
contains 15 digitization circuits and three analog power supply
sections providing the voltages and currents to  bias the chips, and
calibration circuitry. The digital component of these boards
contains a sparsification  circuit, an event buffer, memory to store
the pedestal values, and the interface to the VME crate CPU.

Coherent noise is present. We eliminate this by measuring the pulse
heights on all the channels and performing an average of the
non-struck channels before the data sparsification step.\footnote{
    This algorithm is executed by a DSP located on the data boards
    before the data are sparsified.}
The pedestal width (r.m.s.) is reduced from 3.6 to 2.5 ADC counts
after this coherent noise subtraction. The intrinsic noise of the
system then is $\sim$500 electrons r.m.s.


\section{ CRYSTAL FABRICATION and TESTING }



The design of the CLEO RICH demands a large number of high-quality,
high-purity VUV-transparent crystals, used as the entrance window to
the multi-wire photon detectors and as the solid Cherenkov radiator
medium. In fact, nearly 420 kg of crystal was required, in the form
of 240 full-sized windows and 420 radiator tiles.
The full windows were 191.0 $\times$ 307.6 $\times$ 2.0 mm$^3$ in
size, and the planar radiators were 174.3 $\times$ 169.8 $\times$
10.0 mm$^3$.

There were several significant issues in the production of these
crystal pieces. First, the bulk absorption of VUV photons needed to
be small, so high purity raw material was required. Second, the
surface transmission needed to be appreciable, so high-quality
polishing of the surfaces was required. These two elements are
necessary to yield good transmission in the VUV region. Third, the
finished plates needed to have large dimensions, so the active area
of the detector remained high. Last, they needed to have good
mechanical stability---free of cleaves, nascent cleaves, and
geometric shape deformations---so the internal strain in the
crystals needed to be minimal. This was particularly critical for
the very thin 2~mm windows, which would cause the photon detector to
fail if a crack occurred in situ.
In order to meet these stringent requirements a R\&D program in the
areas of fluoride crystal growth and processing were undertaken.

\subsection{Crystal Production}


\subsubsection{ Crystal Growth Process }
Fluoride crystals were grown at the Optovac facilities \cite{opto}
using a modified Bridgman-Stockbarger method \cite{Stockbarger}. In
this technique, pure raw powder is packed in a graphite crucible
which is heated in a vacuum furnace to a temperature far above its
melting point. After a period of impurity removal, the crucible is
reduced to a temperature just above the melting point, and then
slowly lowered mechanically through a sharp thermal gradient into a
second volume where the temperature is below the melting point. This
effects a surface of solidification in the melt where the crystal
lattice is actually grown. The resulting solid ingot is then held at
a reduced temperature (roughly half that of the melting
point~\cite{Twyman}), at which it is annealed to reduce the bulk
strain induced by the growth process as it is slowly brought to room
temperature.

The growth process is the most important single step in crystal
production, since it not only impacts the manufacturability
directly, but the ultimate optical and mechanical properties of the
finished crystal as well.
Stockbarger recognized from the beginning the critical roles of both
the high purity of the raw material and the sharpness of the thermal
gradient between the two temperature regions. This latter impacts
the design and operation of the furnace, as does the need for
uniformity of temperature in the melt. Good control of the
crystallization zone allows for low intrinsic strain growth, and for
the impurities to segregate to the top of the ingot. Both are
necessary for the large-area RICH crystals.

The actual temperature levels and durations in the thermal cycle
were determined by repeated experimentation, and varied according to
material used and ingot size. This optimization constituted an
important line of development, in particular for the large 14.5 and
16-inch diameter ingots grown for \CaF\ windows.\footnote{Large
ingot growth also has important implications for other applications,
    such as 157~nm lithography for VLSI fabrication.}  The \CaF\ ingots
were grown using a piece of single crystal as a seed, in order to
reduce the number of domain boundaries between regions of different
crystal orientation. This was never entirely successful, and all
ingots grown were polycrystals.

The raw material used was either very pure grade LiF powder
synthesized by Merck,\footnote{
  LiF Optipur {(R)} powder, Merck SA,  Darmstadt D-64293, Germany.}
or crushed natural fluorite (\CaF), of sufficient purity for thin
windows. In order to remove residual impurities from the fluorite,
2\% PbO$_2$ was added as a getter that subsequently segregated from
the melt.
Any residual impurities reduce transmission significantly at 150~nm,
since they will either create absorption (color) centers or
scattering centers for light in the VUV region.

\begin{figure}[tb]
  \vspace*{-0.001in}
  \begin{center}
    \includegraphics*[height=3.3in]{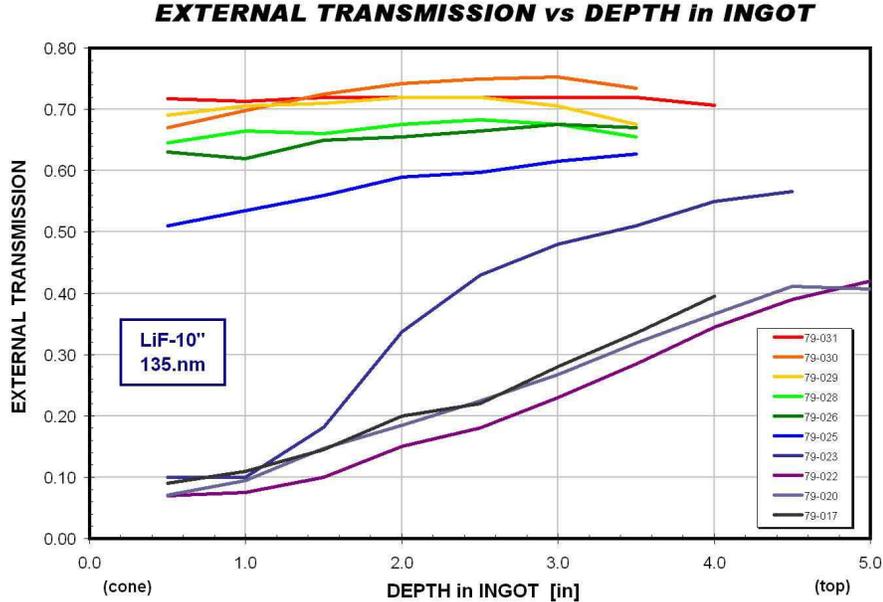}
  \end{center}
  \vspace*{-0.100in}
  \caption{\label{fig:xtal-tb-trans}
        Test bar transmission as a function of vertical position
        in the LiF ingot, at 135 nm, from 10 ingots grown in the same furnace.
        The low transmission curves show earlier bad growths
        having high levels of impurities (including Fe), while
        the high transmission curves come from later ingots,
        made after the growth process was optimized. The left end
        of the plot refers to the end of the ingot that is cone
        shaped, the bottom, while the right side shows the top.
        }
  \vspace*{-0.001in}
\end{figure}

In order to monitor the ingot quality at growth, test bars were
taken as a vertical slice from the periphery of each ingot. The
transmission of these witness pieces from a representative early bad
ingots containing relatively high impurity levels and good ingots
produced after the growth process was optimized are shown in
Fig.~\ref{fig:xtal-tb-trans}. This allowed the proofing of the
ingots, as well as a diagnostic of growth problems.

\subsubsection{ Crystal Machining }

The machining process consisted of taking a single ingot and
producing many ``blanks'' from it---each of which is a formed piece
of crystal ready for polishing in order to have good optical
transmission.
To produce the blanks, the ingot is first mounted on a plaster base
and sliced to approximate thickness on a band-saw using a diamond
blade and a gravity-feed table. It is important to use a
``well-cutting blade,"\footnote{
    A ``well-cutting'' blade is not only sharp (e.g., using a diamond abrasive),
    but is properly dressed and lubricated so as to
    prevent loading up with fine crystal silt as it cuts.
    Thus the blade will cut the ingot instead of rubbing and cracking it.
    This same criteria is applied to the other cutting tools and grinding wheels used. } to thermalize the ingot and lubricant to room
temperature, and to establish the proper feed rate of the blade
through the ingot.
Next, the slice is trimmed to approximate shape, and all edges are
bevelled by hand.
Then, the two faces of the slice are ground to within $\sim$150~\um\
of final thickness, using a Blanchard-type surface grinder with
fine-grit superabrasive pellets.
Finally, the four edges of the piece are ground to dimension using a
SX CNC machine with a very fine diamond superabrasive wheel. The LiF
radiators have two edges which are given a 6$^\circ$ angle.

For the case of windows and planar radiators, the blank is now
finished and ready for polishing. For the case of sawtooth
radiators, additional grinding steps (as well as a different
polishing technique)
are required, as described below. 

The machining process is developed by determining the proper speeds
and feed rates for each operation. Otherwise the crystal may be
easily destroyed.
Nevertheless, each of these machining steps does introduce some
actual damage to the crystal. This consists of microscopic cleaves
and irregularities at the surface and just below it, called
``subsurface damage.'' The goal of each step in the machining
process is to cut away enough material so as to get under the
subsurface damage created by the previous step, while itself
creating at worst only a finer level of damage, which is to be
removed by the subsequent operation. In this way, a piece of
sufficient transmission is ultimately produced.
 %

\subsubsection{ Crystal Polishing }

\begin{figure}[tb]
  \vspace*{-0.001in}
  \begin{center}
    \includegraphics*[height=3in]{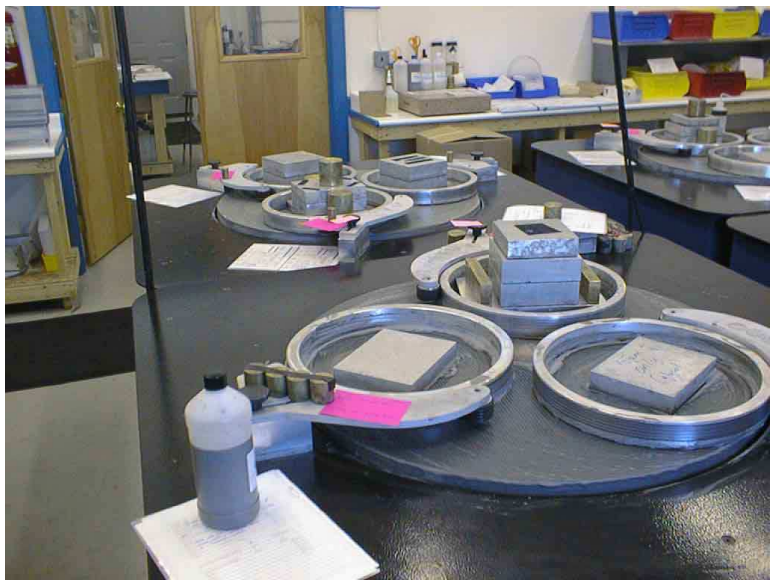}
  \end{center}
  \vspace*{-0.100in}
  \caption{\label{fig:xtal-polishtable}
        Photograph of a polishing table, showing the lapping surface
        and rotating crystal carriers.  Weights are placed on the crystals.
        }
  \vspace*{-0.001in}
\end{figure}

The polishing process consisted of placing a blank on a lapping
table with a polyurethane polishing pad and a specific polishing
media, such as fine-grit diamond powder suspended in glycol or a
commercially-available polishing slurry.
Fig.~\ref{fig:xtal-polishtable} shows a working polishing table.
The blank is weighted but free to spin within a rotating holder on
the rotating polishing pad, the net effect of which is randomized
orbits of the fine abrasive with respect to the blank.  This yields
a uniform action as it cuts away stock (mechanically or chemically)
to get under the subsurface damage.
This procedure is repeated with a finer grit abrasive, in order to
finish the surface.

This is a fine and somewhat delicately balanced procedure, taking
many hours per surface. The effectiveness is governed principally by
the total amount of stock removed, which is related to the amount of
weight applied to the piece as it glides on the polishing pad. The
rate of stock removal also depends on the material (\CaF\ or LiF)
and on the orientation of the crystal.
The effectiveness of different polishing times and weights was
extensively studied.

\subsubsection{ Production Yields }

Nearly one thousand individual crystals were produced for the CLEO
RICH.
The early stages of the crystal production saw a very low yield of
pieces. It was only after intense efforts developing all phases of
production (growth, machining and polishing) that the yields grew to
acceptable levels.

For \CaF\ production, the intrinsic strain in the thin large-area
windows needed to be minimized. Optimization of the annealing cycle,
major improvements to the machining process, and determination of
proper weighting during polishing, gave rise to increased yields.
The manufacturing yield of an ingot (i.e., the number of windows
produced divided by the number possible in an ingot) grew from below
30\% in the initial phases of the project to be routinely above
80\%, and as high as 90\%.
In the wake of these improvements, the quality of the ingot (i.e.,
the fraction of windows produced that passed CLEO requirements) also
rose to be close to 100\%.

For LiF production, proper purity of the material needed to be
maintained. Improvements parallel to those made for \CaF, as well as
critical corrections to raw material handling, contributed to slowly
increasing yields.
For the first twenty LiF ingots grown, the manufacturing yield was
under 40\% with the quality about the same. Several episodes of
material contamination then reduced the quality to effectively zero,
as indicated in Fig.~\ref{fig:xtal-tb-trans}. This was due to
insufficient segregation of impurities, or overabundance of
impurities, both of which create color centers in the bulk material
as well as enhanced domain boundaries. The latter creates ``fault
lines'' along which the crystal may separate into multiple pieces,
thereby becoming useless.
As improvements were made during the second half of the project, the
manufacturing yield grew to 70\%, albeit with large fluctuations,
due mainly to the continued presence of separating domain boundaries
in some ingots.
Also in later production, the quality of the ingots became quite
high, far surpassing CLEO requirements even at 135~nm. This is also
indicated in Fig.~\ref{fig:xtal-tb-trans}.

\subsection{Crystals for Chamber Windows} \label{sec:xtalwindows}

There were 240 full-sized chamber window crystals required. A total
of 300 were delivered, of which 272 were used in construction. To
make the number of full-sized windows used, there were 393
individual crystals, of which 151 were full-sized \CaF\ crystals,
182 were half-sized \CaF\ crystals, and 60 were half-sized LiF
window crystals. This small admixture of LiF windows was acceptable
if the machining process succeeded, and sped up the chamber
production schedule.
An additional 16\% were rejected for various reasons.

\subsubsection{ Inspection and Testing }

All window crystals were cleaned, inspected and tested individually.
Fig.~\ref{plane_trans} shows typical transmission curves for a 2 mm
thick CaF$_2$ window and for a 10 mm thick LiF planar radiator. The
main features on these curves are the drop to zero transmission at
the band edge for the crystals (120 nm for CaF$_2$ and 105 nm for
LiF). There are also several possible absorption lines at 130,
142.5, and 175 nm from water impurities. The transmission above 180
nm rises to more than 90\% where bulk impurities cause minimal loss
and only surface reflections dominate.  For speed of processing of
the crystals, most parts were scanned over their surface at 3
wavelengths only, namely 135, 150, and 165 nm, for each piece on a
grid of 30 points over its surface, using the VUV Spectrophotometer
system at Syracuse University, described in Appendix A.

Fig.~\ref{fig:xtal-win-trans} shows the distribution of
transmissions at each of these wavelengths for all window crystals.
Variations in the transmission may result from bulk impurities which
vary by ingot, from polishing non-uniformities, and from variations
in the cleaning. The variation at 135 nm are clearly larger than at
165 nm, and reflect the greater influences of water and other
impurities below 145 nm. On average, the transmission in the windows
is 71.5, 80.1, and 85.3\% at 135, 150, and 165~nm, respectively,
which matches our design criteria of 72, 80, and 85\% at these
wavelengths. All windows were tested both before and after high
voltage trace deposition.

\begin{figure}[tb]
  \vspace*{-0.001in}
  \begin{center}
    \includegraphics*[height=3in]{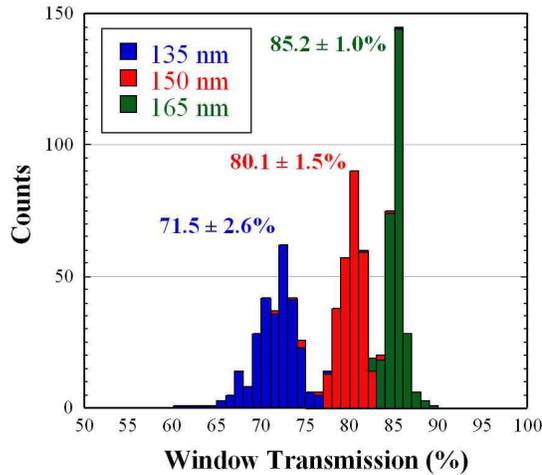}
  \end{center}
  \vspace*{-0.100in}
  \caption{\label{fig:xtal-win-trans}
        Distribution of average window transmissions at 135, 150, and 165 nm, for all full size
        and half size CaF$_2$ and LiF windows.     }
  \vspace*{-0.001in}
\end{figure}

The dominant mechanism for transmission loss was generally surface
scattering or absorption, with bulk absorption accounting for
$<$20\% of the transmission loss.  This was confirmed by
transmission measurements of identically-prepared LiF and CaF$_2$
pieces of different thicknesses to separate the two effects.

On some crystals, cleaves were found to have propagated after
production. In the case where the cleave was small and near an edge,
the pieces were hand-worked to remove the cleave and get under the
subsurface damage. If properly done, there was only a small
probability that the cleave would reappear and propagate in a
relevant amount of time. This was often successful and the best of
these reworked pieces were used as chamber windows.

\begin{figure}[tb]
  \vspace*{-0.001in}
  \begin{center}
    \includegraphics*[height=2.5in]{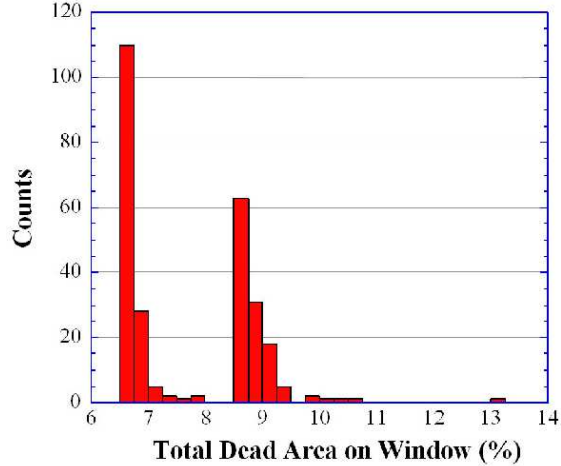}
  \end{center}
  \vspace*{-0.100in}
  \caption{\label{fig:xtal-win-lostarea}
        Distribution of optically inactive area for CaF$_2$ and LiF windows.
        The two peaks correspond to
        windows made with full plates (left) and half plates (right).
        }
  \vspace*{-0.001in}
\end{figure}

Fig.~\ref{fig:xtal-win-lostarea} shows the fraction of each window
that was optically inactive. Factors which contribute to the
inactive area include the window traces which typically contributed
a 4\% opacity, any mechanical defects such as missing corners, any
reworked areas of cleave removal as mentioned above, or surface
stains which were not easily cleaned with acetone. Additionally, due
to production schedules and the detector installation deadline, the
windows formed out of half-sized plates epoxied together in pairs
had an extra glue joint resulting in an additional 2\% loss of
active surface area. The two peaks in
Fig.~\ref{fig:xtal-win-lostarea} represent the whole and the half
windows, with the tails in each distribution coming from defects,
smudges, or coating problems.

\subsubsection{ Deposition of Traces }

The \CaF\ crystals serve a dual role as entrance windows into the
wire chambers, as well as high voltage cathodes for these chambers.
In order to apply a voltage to this plane, a 200 nm thick coating of
nickel and silver was applied to the chamber side of the window.
This coating was in the form of 100 \um\ wide strips spaced 2.54 mm
apart, and connected at each of their ends by another metallized
coating to act as a voltage distribution bus bar. During
construction, a metal conductor was glued to this bus bar using
conductive epoxy.

\begin{figure}[tb]
  \vspace*{-0.001in}
  \begin{center}
    \includegraphics*[height=3in]{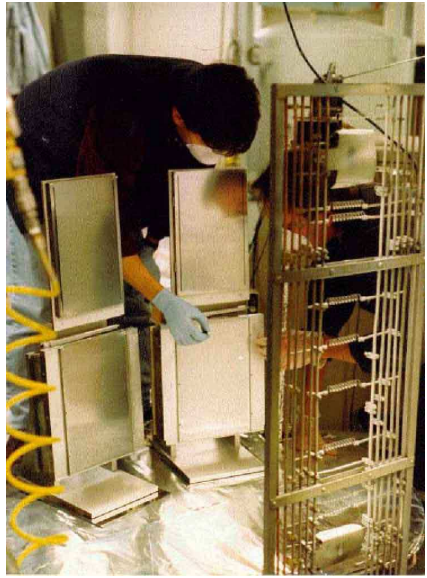}
  \end{center}
  \vspace*{-0.100in}
  \caption{\label{fig:xtal-tracedepo}
        The coating chamber for sputtering the metallized strips on the \CaF\ windows.
        In the foreground are the tungsten filaments which will vaporize the silver onto the crystals.
        The four crystals in the background lay behind steel masks with 100 \um\ slots cut into them
        to define the high voltage traces.
        }
  \vspace*{-0.001in}
\end{figure}

The metallized coating was performed by EMF;\footnote{
    Evaporated Metal Films Corp., Ithaca, NY 14850.}
the setup is shown in Fig.~\ref{fig:xtal-tracedepo}. The coating was
a standard sputtering process, in which a stream of metal ions is
created by placing loops of nickel or silver on tungsten filaments
through which a 100~Amp current passes.  The crystals and filaments
are placed in a vacuum chamber for the sputtering process.  The
nickel was found to bond better to the \CaF\ surface, so a 50~nm
layer of this was laid down first, followed by 150~nm layer of the
silver.  The strip features on the windows were defined by metal
masks placed in contact with the windows which had 100~\um\ wide
slots cut into them.  These masks were held in contact with the
crystal by placing the crystal on a machinist's magnetic chuck and
pulling on the steel mask with the field from the chuck.  Careful
preparation of the chucks were required to make them
vacuum-compatible and prevent outgassing onto the crystals. The
masks were heat-treated to make them flat enough to not rise more
than $\sim 25~\mu$m off the crystals. If the mask was not held in
intimate contact with the crystals, the sputtered silver would tend
to ``feather'' under the mask, making a broad coated area that was
then optically opaque.

\subsection{Crystals for Radiators}

There were 420 full-sized radiator crystals required, of which 300
were of planar and 120 were of sawtooth geometry. A total of 436
were delivered, from which 420 were used. To make the full radiator
array, there were 450 individual crystals, of which 270 were
full-sized LiF planar crystals, 60 were half-sized LiF planar
crystals, and 120 were full-sized LiF sawtooth crystals.
Pairs of half crystals were alternately used as the end rings to
make the radiator left-right symmetric. An additional 2\% of the
planar radiators were rejected for various reasons, whereas for
sawtooth radiators there were 23\% rejected, indicating the relative
difficulty in production.

\subsubsection{ Inspection and Testing }

All radiator crystals were cleaned, inspected and tested
individually.
For planar radiators, the transmission was measured at 135~nm,
142~nm, 150~nm, and 165~nm, for each piece on a grid of $\sim$300
points over its surface, using the VUV Spectrophotometer system at
SMU, described in Appendix A.
On average, the transmission in the planar radiators is 65.5, 77.6,
and 85.4\% at 135, 150, and 165 nm, which matches our design
criteria of 66, 77, and 85\% at these wavelengths. However, the low
value at 135~nm is misleading: it is an average which includes early
ingots having low transmission due to an excess of impurities. Later
ingots produced radiators with transmissions at 135~nm of up to
75\%.

\subsubsection {Dielectric Coating of Radiator Crystals}

Monte Carlo studies indicated that the Cherenkov angle
reconstruction efficiency is enhanced if those radiator photons
entering the RICH photon detectors which have first bounced off the
bottom surface of a radiator crystal are suppressed.
Correspondingly, we coated the bottom surfaces of all 120 sawtooth
radiator crystals with a thin layer of polystyrene. This dielectric
has an index of refraction well matched to that of LiF. Detailed
tests confirmed that this material strongly absorbs at 150 nm.

\subsubsection{ Techniques for Sawtooth Radiators }

The unique geometry of the sawtooth radiator crystals cannot be
accomplished using the fabrication methods described above. While
the lower surface and edges of the sawtooth radiators are flat and
may be polished using conventional orbital lapping techniques,
special procedures were developed to produce high-quality
VUV-polished faces in the ``vee''-shaped grooves on the exit surface
of the crystals. These new techniques could, in principle, be
adapted to other new geometries.

Sawtooth radiator blanks are cut and ground to a thickness of
12.7~mm (i.e., 0.7~mm oversized) in a manner identical to plane
radiators. The blanks are then mounted on a linear surface grinder
to define the groove shapes.  A set of ten 6-inch diameter grinding
wheels are mounted on a single spindle.  These wheels are edged with
a 90$^\circ$ ``vee'' on their edges and rough-grit superabrasive.
Grooves are ground into the upper surface of each blank to within
1.0~mm of the final depth.  Since there are 19 grooves in a sawtooth
piece, this is done in two operations. Following this step, the
group of ten wheels is replaced with a single grinding wheel with a
96$^\circ$ included angle and a fine-grit bond.  This single wheel
is then used to finish-grind the last $\sim$0.9~mm of depth in each
of the grooves, thereby minimizing groove-to-groove differences
within a single radiator crystal.  As with the planar pieces, the
fine-grit operation was found to reduce subsurface damage from the
grinding process. The wheels need to be maintained by dressing
periodically to remove LiF build-up, and to re-define the profile of
the edge which rounds after grinding. Care was taken to reduce any
spurious vibrations in the spindle during this procedure.

Polishing is done using a conventional Bridgeport milling machine
with automated travel.  Unlike the grinding, which could address
both left and right faces of each groove simultaneously, the
polishing is applied to one face of all grooves first and then
applied to the remaining face in turn. The milling head is set at a
42$^\circ$ angle from the vertical and a new head is mounted in
place of a milling bit. The polishing head was a 6-inch diameter
aluminum disk with a soft polishing pad attached on the bottom, as
is used on the usual lapping tables. The crystals are polished by
applying a small tool pressure to the groove faces with this
rotating head, and wetting the part with a polishing slurry.  The
part passes back and forth several times under the rotating head
which reaches down into the groove. Then the part is indexed to work
the next groove.  After all grooves are done, the tool pressure is
re-adjusted and a second set of passes taken. The piece is rotated
in order to polish the opposite faces on the crystal. This entire
procedure is repeated with a smaller grit abrasive, in order to
finish the piece.

\begin{figure}[tb]
  \vspace*{-0.001in}
  \begin{center}
    \includegraphics*[bb = 52 1320 1150 1820, height=2in]{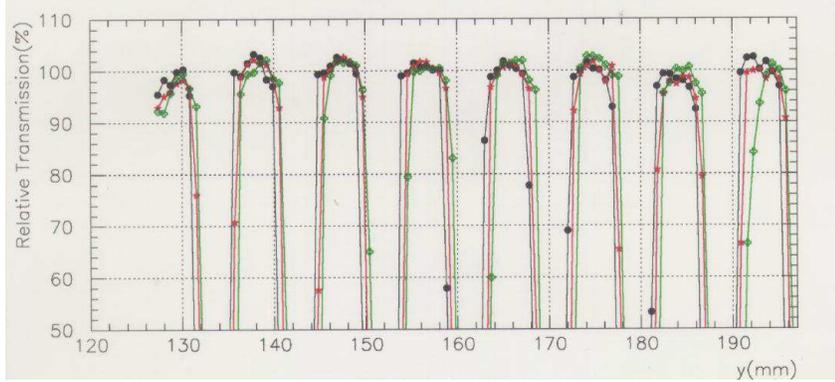}
  \end{center}
  \vspace*{-0.100in}
  \caption{\label{fig:xtal-st-scan} Relative transmission scan over
  almost one-half of a typical sawtooth crystal, showing one face
  only. The measurement scan
        is made relative to a standard polished prism,
        at three different positions along the groove.
        A slight enhancement at the middle of the face is evident,
        as are the roll-offs at the peak and valley. }
  \vspace*{-0.001in}
\end{figure}

\begin{figure}[tb]
  \vspace*{-0.001in}
  \begin{center}
    \includegraphics*[height=3.5in]{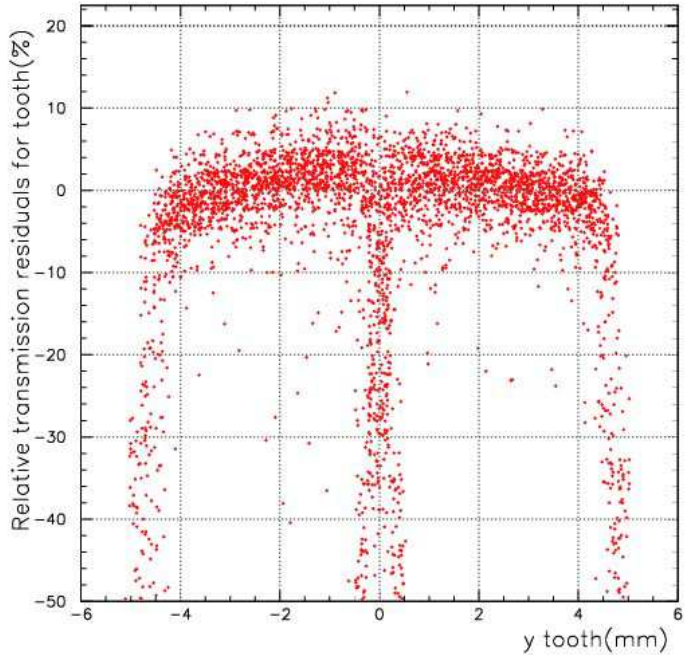}
  \end{center}
  \vspace*{-0.100in}
  \caption{\label{fig:rachid-1} Cumulative plot of relative
transmission measurements for all faces of about 40 sawtooth
crystals. The difference of the relative transmission from the
average per face is plotted.  Measurements are superimposed so as to
indicate a single tooth, with the peak at 0 mm, and the valleys at
$/pm 4.7$~mm. }
  \vspace*{-0.001in}
\end{figure}

\begin{figure}[tb]
  \vspace*{-0.001in}
  \begin{center}
    \includegraphics*[height=2.5in]{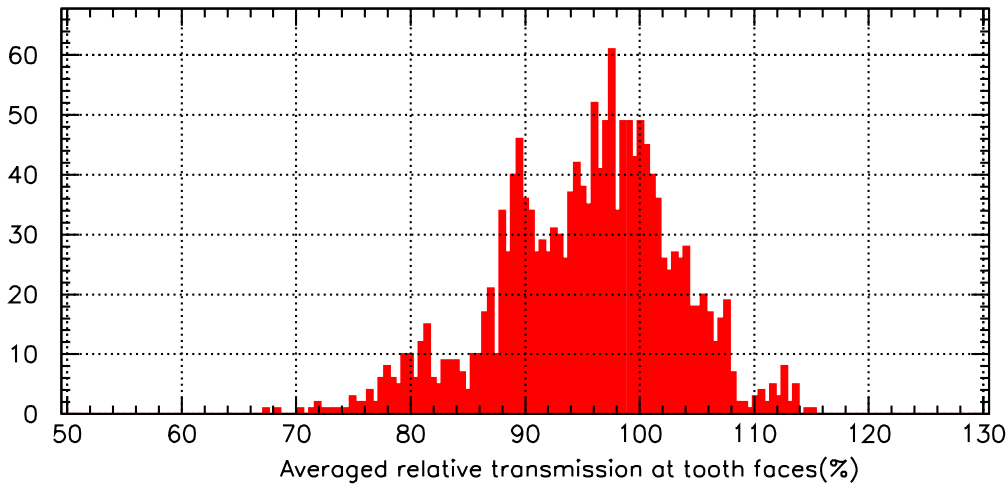}
  \end{center}
  \vspace*{-0.100in}
  \caption{\label{sawtooth-summary}Distribution of relative
transmission per face, for all faces in about 40 sawtooth crystals.
The average is calculated for each face, excluding a small region of
roll-off. }
  \vspace*{-0.001in}
\end{figure}

In order to test the optical transparency of the sawtooth radiators,
we compared them to a calibrated prism of $42^\circ$ inclination
angle which was conventionally polished and also met our
transmission specifications.\footnote{We fabricated two such prisms
and measured the transmission of the two stacked together for a VUV
beam incident at 15$^{\circ}$ angle to avoid total internal
reflection.}  To compare a sawtooth radiator to the calibrated
prism, we deflected the light incident on the flat face of the
sawtooth or prism by $15^\circ$ to avoid total internal reflection,
and the light intensity through the sawtooth is compared to that of
the prism. Using this method we scanned the left-faces of the teeth
independent of the right-faces.  Fig.~\ref{fig:xtal-st-scan} shows
the results of a transmission scan from one side of a typical
sawtooth crystal. It is possible to have a transmission that is
better than our standard prism, hence a relative transmission
greater than 100\%. As can be seen from the figure, however, it was
challenging to fully polish the sawtooth grooves all the way into
the bottom of the valley and to the top of the peak. This manifests
itself as a slight enhancement at the middle of the face, with a
roll-off at the peak and valley. This roll-off is generally limited
to 0.5--1.0 mm at the peak and valley, as indicated in the shape for
each face shown in Fig.~\ref{fig:xtal-st-scan}. A cumulative plot of
these measurements, superimposing all faces for many sawtooth
radiators, is shown in Fig.~\ref{fig:rachid-1}. The difference
between the relative transmission and the average per face is
plotted, where the average excludes a small region of roll-off at
both ends of each face.  This plot indicates the spatial uniformity
of the polishing along the faces of the teeth.
Fig.~\ref{sawtooth-summary} gives the distribution of the average
relative transmission per face.  The width of this distribution
indicates the repeatability of the sawtooth polishing process, on a
per-tooth basis.

\subsection{Radiation Damage in Fluoride Crystals}

The crystals of the CLEO RICH are not expected to be exposed to
significant levels of radiation. The RICH inner radius is 820~mm,
and its outer radius is 1020~mm.  At these radii, the CLEO-II
detector saw approximately 0.05 Rad/day.  This rate is expected to
scale very nearly with instantaneous luminosity, so that in
CLEO~III the rate would be 0.1~Rad/day, or 37~Rad/year.\footnote{%
  The scaling with luminosity is only valid at the outer radii of the CLEO detector,
  where the penetrating radiation is shielded only by the inner detector material.
  The radiation doses to which the inner detectors will be exposed will actually
  not differ much in CLEO~III from those in CLEO-II due to improved shielding near the beamline.}
In CLEO-c, the numbers are expected to be about a factor of four
higher. We have investigated what effect could be expected on the
transmission of the RICH crystals due to exposure to synchrotron
radiation.

\begin{figure}[tb]
  \vspace*{-0.001in}
  \begin{center}
    \includegraphics*[bb = 65 400 500 694, height=2.5in]{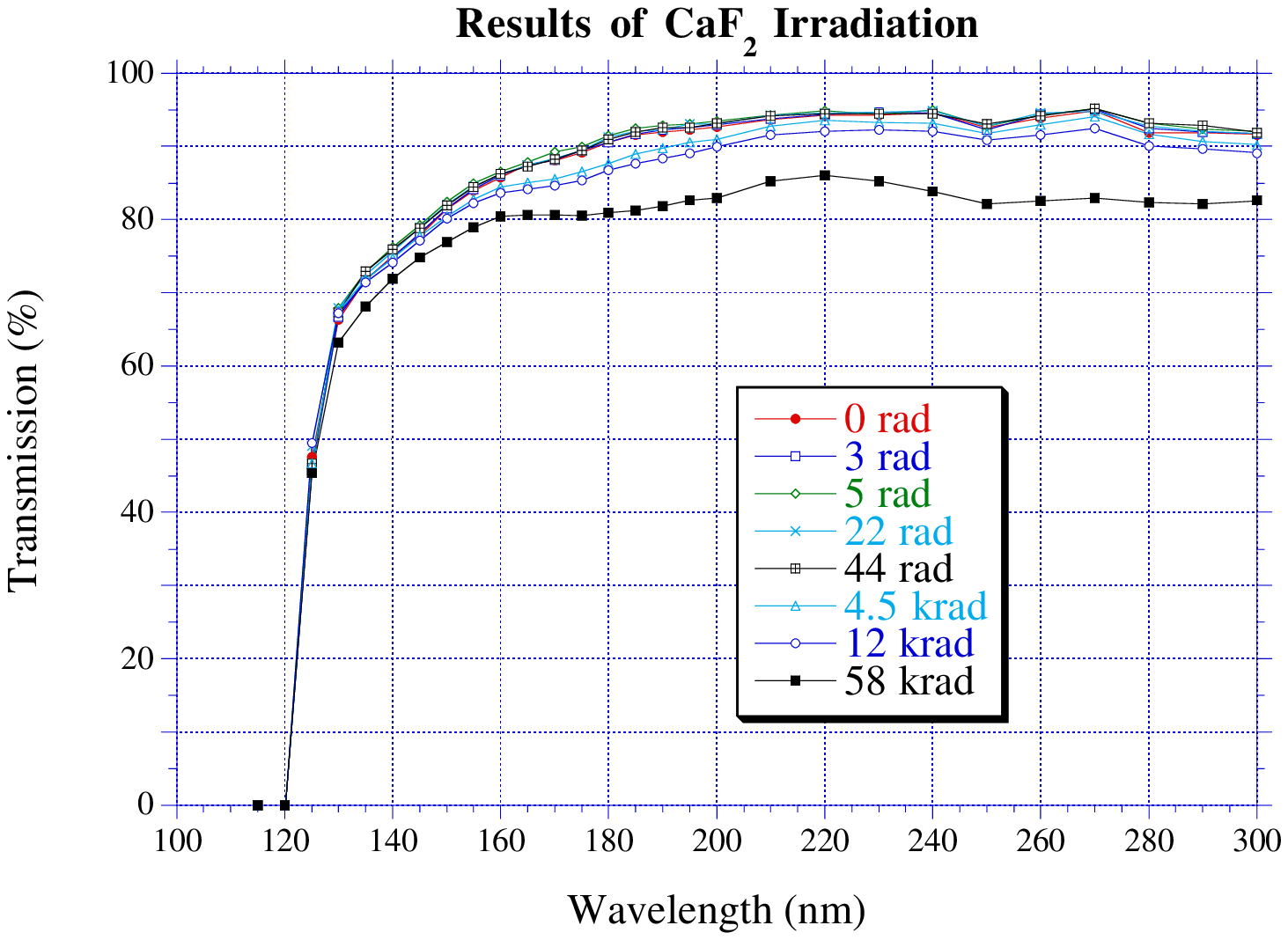}
    \includegraphics*[bb = 65 400 500 694, height=2.5in]{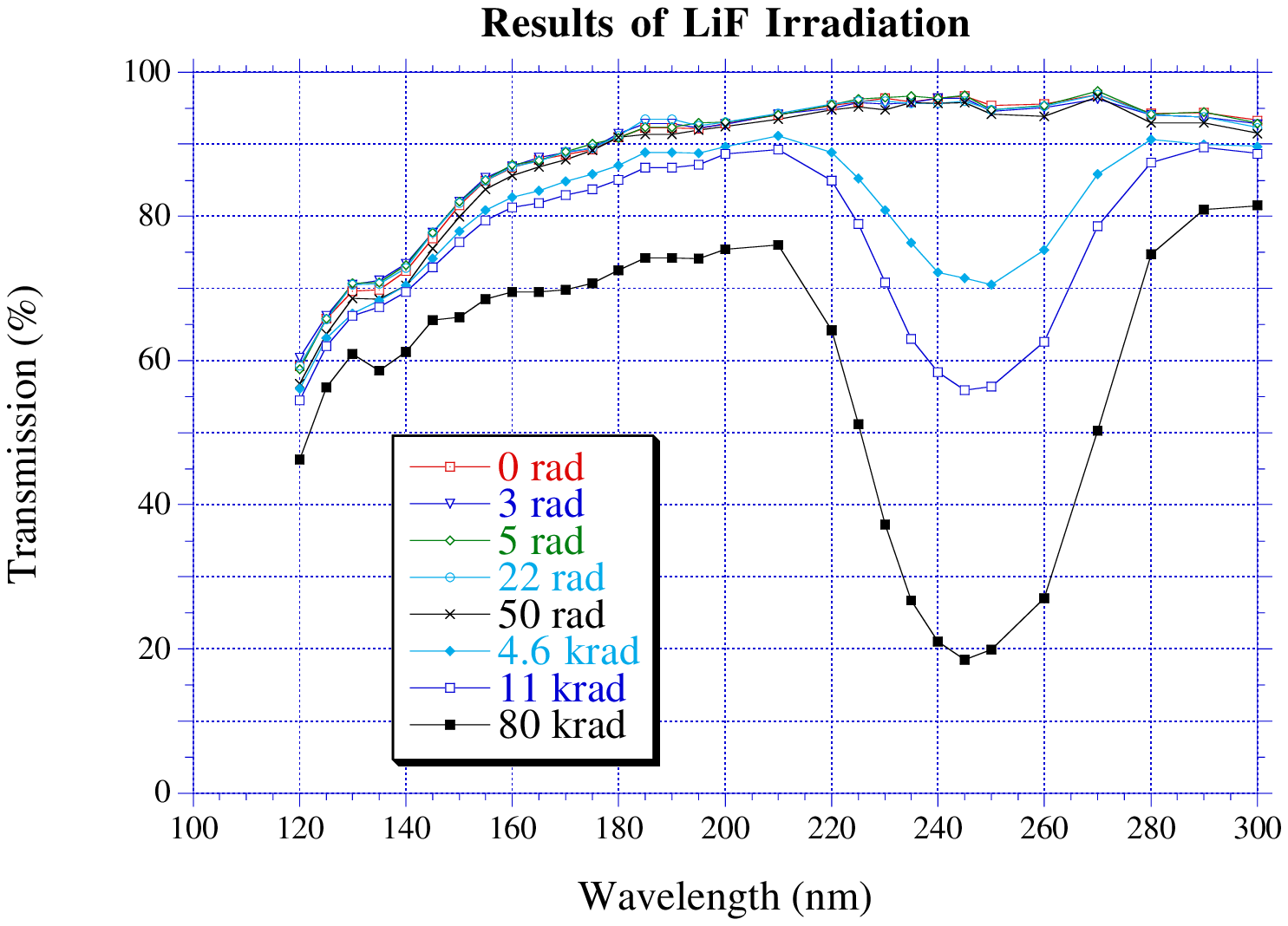}
  \end{center}
  \vspace*{-0.100in}
  \caption{\label{fig:xtal-dose}
        Transmissions of \CaF\ (top) and LiF (bottom) crystal samples
        irradiated by a $^{60}$Co source,
        for different exposures, up to 80 krad.
        }
  \vspace*{-0.001in}
\end{figure}

Twenty CaF$_2$ and 20 LiF samples of the same thicknesses, as used
in the RICH, were exposed for various times to a $^{60}$Co source at
rates of 200 to 5000 rad/hr. Transmissions of the samples were
measured before and after the exposure.  The transmissions did not
appear to depend significantly on dose rate, only on integrated
dose.  A control sample that travelled with the dosed samples but
was not exposed to the source did not experience transmission
changes greater than 1\% at any wavelength. Fig.~\ref{fig:xtal-dose}
shows the transmissions of the crystals after exposure to the
Co$^{60}$ source.  The observed absorption bands in the LiF at
$\sim$250~nm and 170~nm correspond to known color center formation
centers~\cite{Kaufman,Gorlich}, as do the 190~nm and 250~nm centers
observed in the \CaF~\cite{Williams,Gorlich2}.  At the 100--500~Rad
exposure expected in CLEO~III and CLEO-c, however, the loss in
transmission expected in our crystals would not exceed a few percent
in the 135--165~nm range.

%


\section{ MULTI-WIRE CHAMBER CONSTRUCTION and TESTING }

\begin{figure}[htb]
  \vspace*{-0.001in}
  \begin{center}
    \includegraphics*[height=3.5in]{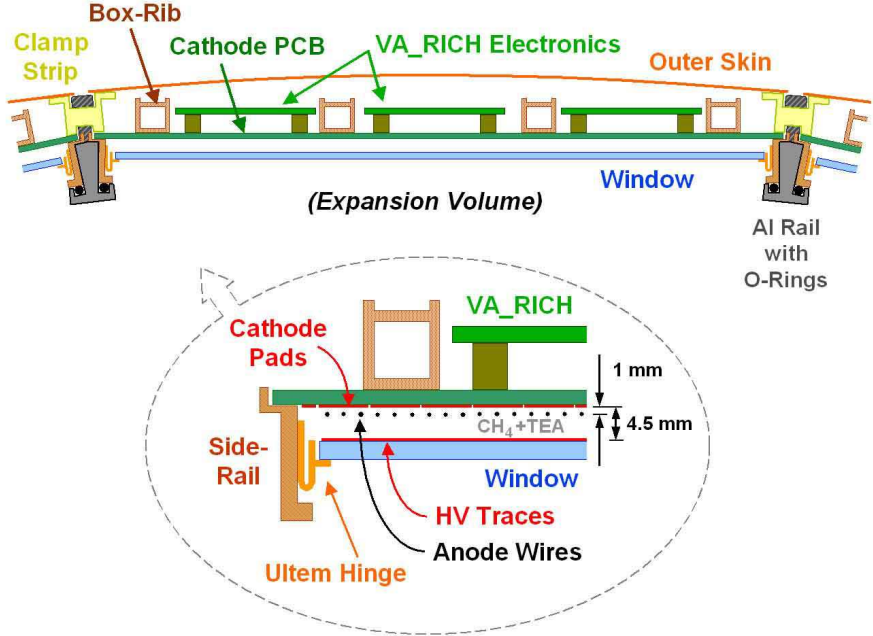}
  \end{center}
  \vspace*{-0.100in}
  \caption{\label{fig:chxsect} Cross-section of full multi-wire chamber. }
  \vspace*{-0.001in}
\end{figure}

The main issues of concern guiding the construction of the
multi-wire chamber photon detectors were: (1) field stability,
requiring electrodes to be parallel over the full area of the
detector, as well as have no local corona points, and high material
cleanliness; (2) mechanical stress relief, to avoid cracking the
thin \CaF\ windows or breaking an anode wire; (3) gas tightness, to
prevent any VUV photon-absorbing gas from leaking into the expansion
volume, or any impurities into the chamber gas;
and (4) long-term stability, since in all practicality, the
installed detector can never be accessed.
Fig.~\ref{fig:chxsect} gives a schematic view of the multi-wire
chamber.

To reduce dirt and dust, a clean room was set up, with air
filtering, and the requisite cleanliness protocols (masks, nitrile
gloves, booties, etc.). All chamber construction and testing of
component parts occurred in this clean room environment. We
purchased a set of three ten-foot long granite tables flat to better
than 0.0005 inch over the entire table, in order to establish an
accurate and stable reference.
A mechanical prototype was constructed first, which was crucial in
developing the detailed techniques and custom fixtures needed for
the construction process.

The chamber construction procedure itself was a complex operation.
The general technique was to assemble and test the two halves of
each chamber (called the ``wire plane" and the ``window-plane")
separately, then mate them together and test the completed chamber
as a single unit. This procedure allowed careful construction of the
three electrode planes, helped in problem diagnosis, and facilitated
the production schedule.

For most of the construction procedures, two structural epoxies were
used, as well as two sealant epoxies. All were chosen for their
strong adhesion properties, viscosity prior to curing, as well as
ultra-low outgassing attributes, as indicated in
Table~\ref{tab:materials}. All materials were checked for chemical
comparability using a special chamber held at elevated
temperatures~\cite{timm}.

\begin{table}[htb]
\centering \caption{ \label{tab:materials} Some special epoxies used
in RICH construction. }
\begin{center}
\begin{tabular}{llccl}
\hline
 Type & Epoxy & Consist-     & Outgassing~\cite{nasabook} & Uses \\
      &       &  ency & \mbox{\small\%TML,\%CVCM } &     \\
\hline
 Struct. & Armstrong A-12$^{(a)}$       & paste & 1.11, 0.01 & LiF, \CaF \\
 Struct. & Delta Bond {\small 152-Q/B-4}$^{(b)}$ &  stiff  & 0.49, $<$0.01 & G10, Ceramic \\ 
 Sealant    & Hysol {\small RE2039/HD3561}$^{(c)}$  &  water & n.a. & Gas seal \\
 Sealant    & Torr-Seal$^{(d)}$            &  stiff  & 0.92, 0.01 & Gas seal \\ 
\hline
\end{tabular}
\end{center}
\vspace*{0.100in}
\begin{flushleft}
{\footnotesize
$^{(a)}$ Resin Technology Group, S.\ Easton, MA 02375. \\
$^{(b)}$ Wakefield Thermal Solutions, Inc., Pelham, NH 03076.\\
$^{(c)}$ Henkel Loctite Corp., Industry, CA 91746.\\
$^{(d)}$ Varian, Inc., Palo Alto, CA 94304 . }
\end{flushleft}
\end{table}

\subsection{Wire Plane Construction and Testing}

The wire plane must be constructed in such a way as to maintain the
1 mm gap between the anode wires and the cathode-pads as uniformly
as possible over the $\sim$2.5 m chamber length along which the
wires are strung. Gain variation at the wires would result if this
spacing was not uniform, as well as variation in the pad-gain, as a
consequence of the capacitive coupling to the
cathode-pads~\cite{Arno92}. In addition, sparking could occur if the
gain becomes excessively large at any place in the chamber.
These requirement caused us to put a stringent limit of 25~\um\ on
the gap variation.

\subsubsection{Cathode Board Design }

The cathode board PCBs were designed to have cathode-pads on the
front side and connectors for the electronics readout on the back
side.
Each board was about 205 $\times$ 615 mm$^2$ in area, and nominally 1.7 mm thick. 
Four boards made up a single chamber. Each contained two 24 $\times$
40 arrays of pads. An 8 $\times$ 16 array of pads was routed to two
linear connectors in a self-contained manner, so symmetry allows
simple tiling of the plane.

\begin{figure}[htb]
  \vspace*{-0.001in}
  \begin{center}
    \includegraphics*[height=2in]{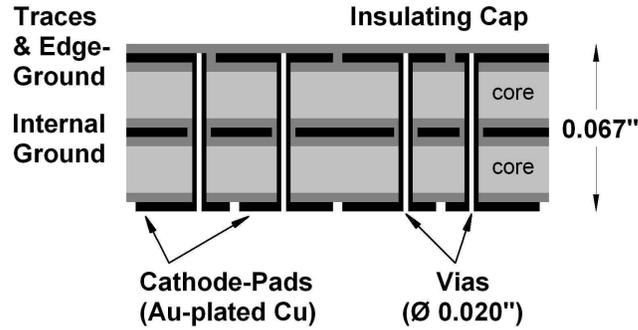}
  \end{center}
  \vspace*{-0.100in}
  \caption{\label{fig:cathodepcblayup} Diagram of the cathode board PCB,
        showing the lay-up of the pads and layers.
        Black indicates a Copper layer, light gray a G10 core layer, and dark gray sheets of G10 prepreg.
        Vertical scale is proportional; horizontal dimension is compressed.}
  \vspace*{-0.001in}
\end{figure}

As shown in Fig.~\ref{fig:cathodepcblayup}, the pads on the front
were connected to the traces on the back with small vias (0.5 mm
diameter), which must all be sealed to prevent leaks. Rather than
glue each via individually ($\sim$8000 per chamber), an additional
layer of prepreg was added to the top of the lay-up stack, which
covered all vias and had cut-outs for connectors and ground
connections. This design provided an excellent solution for sealing
the vias.

\subsubsection{Cathode Board Manufacturing and Flattening Procedure }

The large-area cathode board PCBs were manufactured,\footnote{
    Speedy Circuits, Huntington Beach CA 92649.}
under pressure in a stack in an autoclave.
After manufacture, the boards were tested for continuity, inspected
for irregularities, and measured to determine mechanical size and
deformations.
The connectors were then soldered to the board.

For a large-size thin PCB of this type, there were three major
deformations  from a geometric plane:  longitudinal bow, transverse
bow, and twist.
The magnitude of the longitudinal and transverse bows were about 1.5
mm and 1 mm on average, respectively. The twist was small on this
scale. In our design, the transverse bow was the most problematic
deformation.

A flattening procedure was developed, in order to remove this bow.
The boards were baked under weight while supported along the long
edge. The thermal cycle was 2 hr at 150\degC, followed by a slow 12
hr cool-down. The amount and placement of the $\sim$1 kg weights
depended on the initial deformations of the board. The process was
repeated as warranted. The effect of this procedure was to reduce
the transverse bow to under half of its original value, on average.
The longitudinal bow was not significantly changed.

\subsubsection{Cathode Board Assembly }

\begin{figure}[htb]
  \vspace*{-0.001in}
  \begin{center}
    \includegraphics*[height=2in]{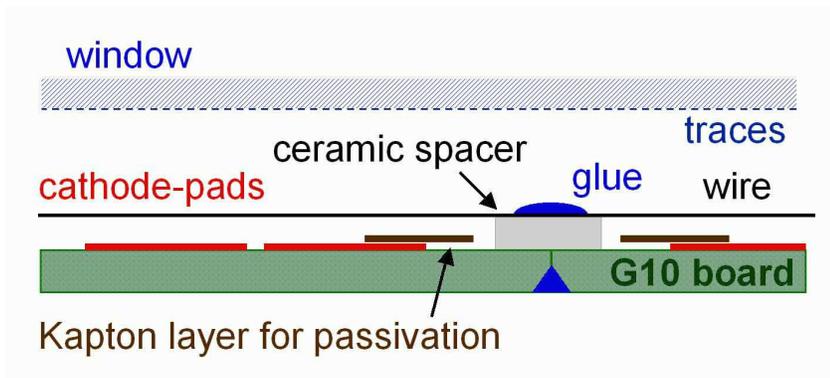}
  \end{center}
  \vspace*{-0.100in}
  \caption{\label{fig:chlsect} Diagram of longitudinal cross-section of multi-wire chamber,
    showing the ceramic spacer and
    the end-joint reinforcement of the G10 cathode board PCBs. }
  \vspace*{-0.001in}
\end{figure}

Four cathode boards were assembled into a plane by gluing them
together end to end on the granite table---weighted, edge-clamped,
and pad face down to assure geometric planarity of the finished pad
array.\footnote{
    There was a variation up to 25 \um\ in the thickness
    of the etched pads themselves, from board to board.}
The end-joint between PCBs was specially reinforced: a ``vee'' cut
was milled on the back to allow a larger contact area for epoxy and
to ensure that no epoxy came through to make the front (pad) surface
irregular. (See Fig.~\ref{fig:chlsect}.) This joint was covered on
the back by a G10 strip in a separate gluing operation for
additional strength. The appropriate ground straps between boards
were added. Measurements were made of the flatness to monitor the
gluing procedure.

Next, the four fiberglass box rib structures were screwed to
``strongback"  into inserts previously glued into the box ribs. The
strongback was made of full-length 1 $\times$ 2 inch aluminum box
channels. The box ribs were then epoxied longitudinally on the back
of the cathode PCBs. Enough glue was applied that any surface
non-uniformities in the cathode boards or ribs would be accommodated
by the glue. In addition,  G10 cross-pieces were epoxied
transversely between the ribs. (See Fig.~\ref{fig:chxsect}.) The
main purpose was to provide requisite stiffness to the thin cathode
board, when mounted in its final configuration. A second purpose was
to remove the residual deformations in the cathode board (the
longitudinal bow being removed more effectively than the transverse
bow). Excessive twisting, for example, would increase the
possibility of breaking wires.

 The strongback  remained
connected throughout the whole construction procedure until the
completed chambers were attached to the cylinder superstructure.

A precision ceramic spacer was epoxied each 30 cm along the
cathode plane, near the window connections. 
This holds the wires at a precise distance from the cathode-pads, as
well as allow for containment of the failure mode in which there was
a broken wire.
The ceramic spacer was cleaned before gluing, and care was taken in
handling it, so as not to allow any grease or dirt to provide an
eventual current path from the wires to the cathode-pads. The wires
were eventually epoxied to the ceramic strip which had a slot
running down the center to contain the glue bead. The area next to
the spacer on the cathode board was passivated by a strip of Kapton,
as indicated in Fig.~\ref{fig:chlsect}. This extended over the edge
of the nearest pad, and was done in order to remove the potential
problem of any corona points that could lead to high voltage
instability over time. Our experience with an early prototype showed
that this was necessary.

The anode PCBs, on which the wires were to be soldered, were then
epoxied into individually-milled grooves on top surface of cathode
board, in order to have precision control of the wire to cathode-pad
distance.

\subsubsection{ Wire Stringing }

The chamber wires were strung on a custom-made jig, consisting of a
PCB and a precision comb aligned and fixed at each end of a rail
structure, approximately 3 m long (longer than a chamber). The 70
central field wires were 20~\um\ diameter Au-plated W
wires\footnote{
    LUMA Type 861-60, W with 3--5\% Au and 3\% Re, LUMA-METALL AB, 391 27 Kalmar, Sweden.}
held at 60 g tension. The two outside wires would produce higher
fields so we used larger diameter 30~\um\ Au-plated W wire\footnote{
    Type F-77, 4\% Au, Philips Elmet Corp., Lewiston, ME.}
held at 90 g tension to keep the gain approximately at the same
level as the central wires. Each wire was held at the appropriate
tension by means of a frictionless pulley with a weight. The wire
was carefully positioned in the comb using transverse locator
screws. It was then soldered at each end to the PCBs. When done for
all wires, a ``temporary'' plane of wires was created. As much as
possible, a single spool was used for the field wires of a given
plane.

\begin{figure}[htb]
  \vspace*{-0.001in}
  \begin{center}
   \includegraphics*[height=2in]{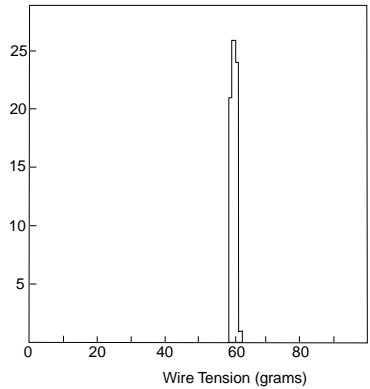}
  \end{center}
  \vspace*{-0.100in}
  \caption{\label{fig:wirtens} Typical distribution of tension in wires, as strung. }
  \vspace*{-0.001in}
\end{figure}

This temporary plane was tested for wire tension using the standard
resonance frequency method. Any wire out of tolerance ($\pm 2$ g)
was replaced. Fig.~\ref{fig:wirtens} shows the distribution of
tensions for a representative sample of wires.

\subsubsection{ Wire Transfer }

After the cathode plane was prepared, as described above, the
temporary wire plane was flipped over and lowered onto it by means
of precision scissor jacks. This was a delicate operation, since the
wires had to be aligned to the solder pads on the anode PCBs and
just touching the ceramic spacers. When in position, each wire in
turn was soldered to two consecutive pads on each end, approximately
5 mm in length, separated by 5 mm of non-pad surface.  We chose to
use solder with silver added.\footnote{
    Ersin Type Sn62, Eutectic, Tin/Lead/Silver, Multicore Solders, Westbury, NY 11590.}

The quality of the solder joint was then assessed, and inspected for
sharp points (important since the solder joints will sit in the
chamber gas volume). The wire was cut behind the back solder joints,
and checked for electrical continuity. Then anode PCBs were given a
first cursory cleaning with isopropyl alcohol. The solder joints
were covered completely in Delta Bond glue to prevent the solder
from being attacked by the TEA in the gas. This process transferred
the wires at tension from the temporary jig to the real cathode
plane.

\begin{figure}[htb]
  \vspace*{-0.001in}
  \begin{center}
    \includegraphics*[height=2in]{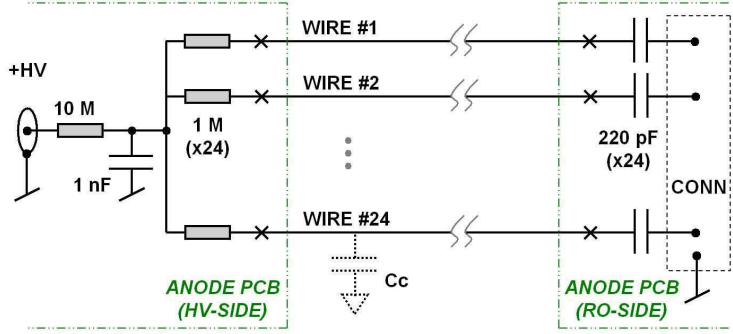}
  \end{center}
  \vspace*{-0.100in}
  \caption{\label{fig:mwccircuit} Circuit diagram of one high voltage cell.
        There are three such cells per multi-wire chamber.
        The blocking and decoupling components are shown.
        The wire-to-pad coupling C$_c\approx 0.1$~pF/pad.}
  \vspace*{-0.001in}
\end{figure}

The anode PCBs were then populated with their requisite components,
as indicated in Fig.~\ref{fig:mwccircuit}. This was a mixture of
discrete and surface-mount components. The novel feature here is
that the large HV surface-mount capacitors (size 1812) were mounted
on their {\it side\/} edge, in order to retain the tight-packing
demanded by the wire spacing and the constraints on the overall
chamber length.  A technique was developed which used a special jig
and both solder paste and solder cord, and this worked acceptably
well for all chambers. Afterwards, the PCBs were brushed with
isopropyl alcohol, and the electronic components tested.

The wires were then epoxied to the ceramic spacer. The glue bead was
well-contained in a groove atop the ceramic spacer, such that the
bead itself was smooth and there was no wicking along the wires. A
bead of glue was also made between the two solder joints on the
anode PCBs at each end, ensuring no loss of tension over time.

The last stage in the wire transfer procedure was a thorough final
cleaning of the anode PCBs. This was accomplished by immersing the
entire end of the wire-plane in a large ultrasonic bath filled with
isopropyl alcohol, and covering all solder joints.  The wire-plane
(via its strongback) was mounted on a wall at a 23$^\circ$ angle in
order to submerge the anode PCB in the bath. Each end was bathed for
30 mins at a time, and repeated three times with fresh isopropyl
alcohol. Afterward, it was rinsed with isopropyl alcohol and dry
nitrogen. This was not a completely efficient procedure, since in
many cases, there remained a whitish residue around some components
afterwards. This is a well-known effect due to solder resin, and
simply needs to be cleaned by hand.

\subsubsection{ High Voltage Testing }

After ground connections were made, the wire-plane was placed into a
polycarbonate testing box, filled with CH$_4$ gas and tested for
high voltage stability.
All wire-planes were tested to 1600 V, with $\leq$ 10~nA current
draw, in conditions of $\sim$22\degC\ and $<$15\%\ relative
humidity.

In total, there were 35 wire-planes constructed. One was destroyed
in subsequent testing, leaving 34 functional planes. They were
stored in a low humidity tent until mating with the window-planes.

The wire connections have been very reliable (due to the
conservative tension). One wire in the completed detector broke
early in the operation due to the chamber exceeding the operating
temperature, which went undetected due to insufficient slow control
monitoring at the start of the experiment.

\subsection{Window-plane Construction and Testing}

After production of the windows as described in
Section~\ref{sec:xtalwindows}, the full window-plane was
constructed.  This consisted of a window frame into which the
full-length ladder of eight window segments was epoxied.


The individual window segments, wrapped in teflon, were epoxied in
pairs. They were butt-jointed with Torr-Seal, with a well-controlled
glue bead. The pair was held together in a custom-made jig, under
gentle compression, during the curing period. The interior surface
of the butt-joint was ``painted" using Hysol. When completed, this
procedure was repeated for two pairs, and then once again, for a
full-length ladder of eight window crystals.\footnote{Half-plate
windows
    were made into full-plate windows by the same technique.}


The full window frame consisted of a long fiberglass side-rail, G10
end pieces, and an Ultem plastic hinge, all epoxied together, as
shown in Fig.~\ref{fig:chxsect}. The G10 end pieces had holes bored
through for gas flow to the completed chamber.

The Ultem hinge was specially designed to allow some flexibility
when the crystal window was glued to it, in order to take up any
mismatch in thermal expansion between the crystals and the stiff
frame. The full-length ladder of windows was lowered onto the frame,
and glued to the Ultem hinge using Armstrong A-12. At the same time,
a small metal foil was wrapped around the edge of each individual
crystal window and glued to the interior trace bus-bar by conductive
epoxy.  On the exterior, this foil was soldered to a teflon-coated
high voltage lead wire.\footnote{
 Gore Type F01A080 Wire, W.L.~Gore \& Associates, Inc., Newark, DE 19711.}

\begin{figure}[htb]
  \vspace*{-0.001in}
  \begin{center}
    \includegraphics*[height=3in]{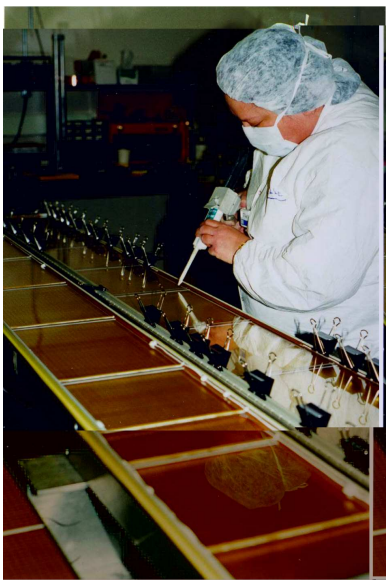}
  \end{center}
  \vspace*{-0.100in}
  \caption{\label{fig:chphoto} Construction operations being performed on two multi-wire chambers. }
  \vspace*{-0.001in}
\end{figure}

After curing, another bead of epoxy (Hysol) was put on top of the
existing one, in order to provide a secondary gas seal around the
perimeter. Additionally, the butt-joints between window crystals
were reinforced by gluing a narrow G10 strip over the joint,
externally.  This was again reinforced by a glue bead (Torr-Seal) on
either side. The Torr-Seal was also used to tack the high-voltage
wire along the frame. This operation may be seen in
Fig.~\ref{fig:chphoto}.
The end-joints of the windows overlap the position of the ceramic
spacers on the cathode plane to minimize the blockage of Cherenkov
photons.

The critical mechanical item used in this construction procedure was
a set of custom-made (multilayered) jigs, which established the
proper referencing surfaces to hold the tolerance in
window-to-cathode board spacing.
In total, there were 34 window-planes constructed. After testing
procedures, 33 usable planes were produced.

\subsection{ Full Chamber Construction and Testing}

\begin{figure}[htb]
  \vspace*{-0.001in}
  \begin{center}
    \includegraphics*[height=2in]{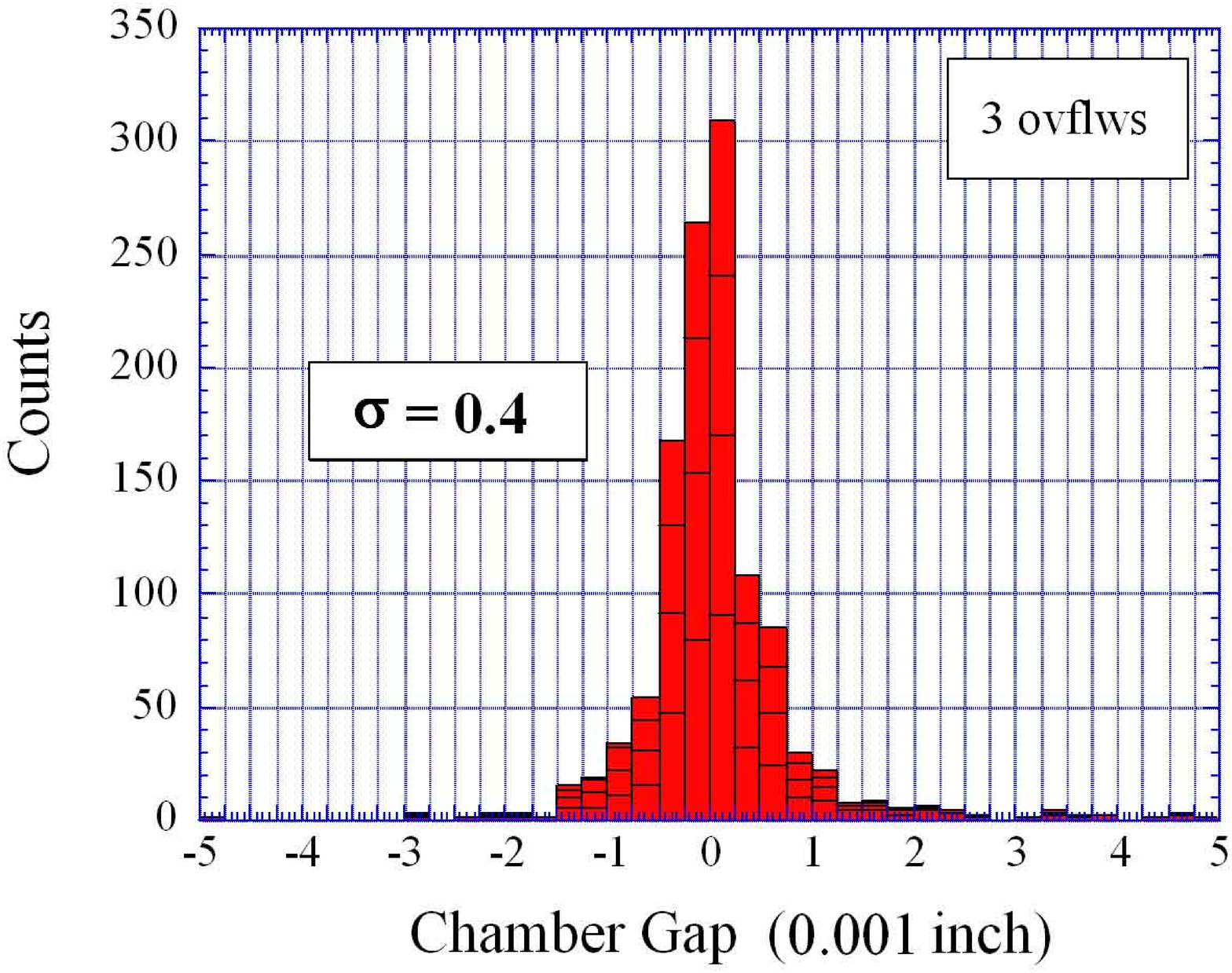}
  \end{center}
  \vspace*{-0.100in}
  \caption{\label{fig:mwcgap} Distribution of the residual about the
    nominal cathode-pad to window gap.
     }
  \vspace*{-0.001in}
\end{figure}

In the next step a wire-plane was matched with a mechanically
suitable window-plane, and they were clamped together temporarily.
 The cathode-pad to window gap was measured at many points along
the length of the chamber.  The distribution about the nominal
4.5~mm gap size is shown in Fig.~\ref{fig:mwcgap} for all chambers
using approximately 32 measurements per chamber.

The mated chamber was tested with CH$_4$-TEA gas, with nominal high
voltage on both wires and windows for a full month. Four special
test boxes were used in this testing phase, so that tests could be
performed simultaneously. The chamber was scrutinized heavily during
this period. Relative chamber gain was measured from the wires, upon
excitation by a $^{106}$Ru $\beta$-source. Current draws were
monitored.  Problems were diagnosed and fixed. Typically, if the
chamber drew more than 10 nA, it was opened up and cleaned of stray
dirt or residual construction materials acting as corona points.
These were the most prevalent problems.

If a mated chamber passed this test period, the given window and
wire-planes were epoxied together and the chamber was completed. The
full chamber was then tested again for a month, under these same
conditions.

One problem encountered later, when the chambers were mated with the
radiators, was leaking of chamber gas. This was attributed to
certain of the fiberglass side-rails, which were epoxy-starved and
microscopically split under small torsion applied during handling.
This situation was rectified, using a rather painful procedure, that
required painting a coat of Hysol epoxy over the side-rail, and
reinserting the chamber before the glue dried; this often had to be
repeated.

A total of 33 chambers passed all tests. On the average, this phase
of construction took $\sim$0.5 months per chamber. Parallel
operations allowed a maximum of eight chambers to be constructed in
a single month.

The last stage in individual chamber construction was mounting the
electronics chip-carriers and cables on the back.  The hybrids
(described in Section~\ref{sec:elec}) were screwed into small
standoffs epoxied to the connectors, thus attaching them with a
squeezing action instead of pressing action which could deflect the
cathode board and possibly break wires.

%
\section{ RADIATOR CYLINDER CONSTRUCTION }


\subsection{ Inner Carbon Fiber Cylinder }

The RICH Inner Cylinder has $12\,{\rm m^2}$ of LiF crystals attached
to a cylindrical $1.64\,$m outer diameter carbon fiber support
shell~\cite{rcubed} of $2.48\,$m length and $1.5\,$mm skin
thickness. The shell has mean density $1.42\,{\rm g/cm^3}$, mean
Young's modulus $E\simeq 70\,$GPa, and is built from wrapping
multiple layers of pre-impregnated unidirectional tape\footnote{
    Type RS-3/AS-4 unidirectional tape,
    YLA, Inc., Benicia, CA 94510.} 
around a drum-shaped steel mandrel. After serving as the form for
the carbon fiber tape during autoclaving, the mandrel provides
mechanical support for the shell during transportation, radiator
assembly, insertion into the Outer Cylinder of photon detectors, and
installation of the entire RICH detector into CLEO.

The mandrel rotates on a shaft that allows the shell to be rotated
to an arbitrary azimuthal position. A large box-beam
frame~\cite{psl} supports the shaft so that the total mass of the
radiator system, including all its auxiliary mechanical fixturing,
is $3500\,$kg.

\subsection{ Radiator Crystal Alignment and Mounting }

The radiator cross-section is essentially a 30-sided polygon
(``triacontagon'') with each side corresponding to a longitudinal
row of 14 radiators. (Each row contains 13 full-sized crystals and
two half-sized crystals at one end.) Construction was accomplished
by first assembling and aligning rows of crystals and then attaching
the rows sequentially to the support shell. All radiator assembly
procedures are performed inside a class 100,000 clean room with
typical 25\% relative humidity to reduce the risk of contaminating
crystal surfaces with particulate matter or excessive water vapor.
Test crystals placed in the clean room were periodically measured
for VUV transmission to verify that the ambient environment did no
damage to radiator crystals.

To construct a radiator row, a set of 13+2 crystals were selected on
the basis of VUV transmission and compatible mechanical dimensions.
Each crystal was attached temporarily, by a $50\,\mu$m thick Kapton
belt, to a picture frame jig that rests directly on top of it and
that has vertical posts in each corner to provide for temporary
attachment to a transport jig. Each crystal with its attached jig
was then placed on an optical stage assembly with 6 degrees of
freedom (3 orthogonal linear displacements and 3 Euler angle
rotations). The 15 stage assemblies were all attached to a single
rigid optical rail.

The open picture frame allows a pair of $2.2\,$cm diameter dowel
pins to be placed directly on the crystal top surface. These pins
support a precision level so that each surface is made horizontal to
a typical angular precision of $150\,\mu$rad. A similar technique is
used to limit vertical offsets between top surfaces of adjacent
crystals to a precision of $\sim50\,\mu$m.

A $100\,\mu$m clearance gap between adjacent crystals was reliably
set by the thickness of the Kapton belts from adjacent crystal jigs.
Pushing the same edge of each crystal flush against a set of tooling
balls running parallel to the optical rail ensures that one edge of
a crystal row was straight within $75\,\mu$m over its entire length.

\begin{figure}[htb]
  \vspace*{-0.001in}
  \begin{center}
    \includegraphics*[height=3in]{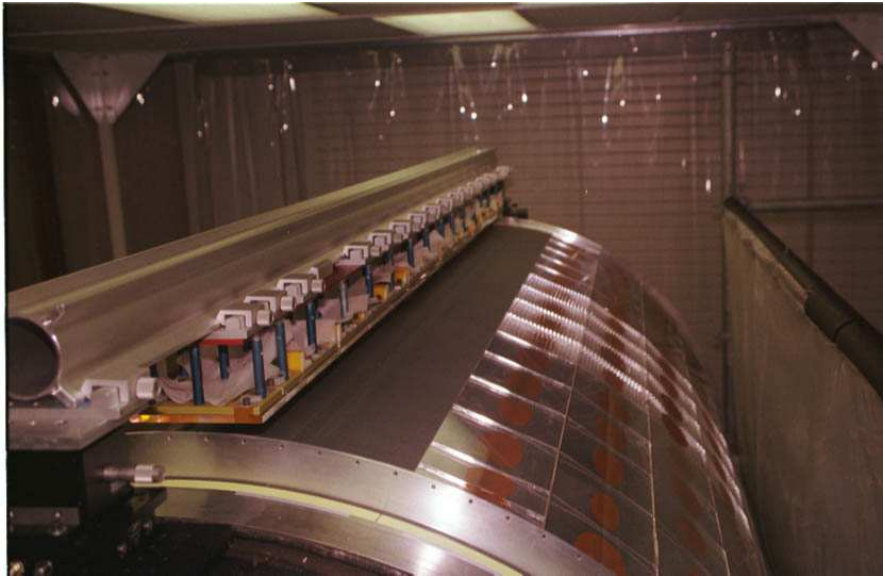}
  \end{center}
  \vspace*{-0.100in}
  \caption{\label{fig:cfc} Photograph of one row of crystal radiators
    being aligned on the inner carbon fiber cylinder.
    Previously epoxied rows are shown on the right half of the cylinder.
    The inner half of the end flange is visible in the forefront. }
  \vspace*{-0.001in}
\end{figure}

After a crystal row has been aligned, a second ``transport'' optical
rail is positioned over the crystal jigs. Quick-setting epoxy
temporarily fixes the transport rail to the vertical posts of each
crystal jig.  The transport rail, along with the attached row of
crystals, was then moved to the carbon fiber cylinder.

Linear translation optical stages positioned at opposite ends of the
support shell near its surface at the ``twelve o'clock'' position
receive the transport rail, as shown in Fig.~\ref{fig:cfc}. Reamed
holes in these stages mate with dowel pins in the transport rail so
that the axis of the new crystal row is aligned with the axis of the
support shell. These stages eventually set the epoxy gap between the
crystal bottom surfaces and the cylinder surface to $25\,\mu$m
precision. Gaps between adjacent crystal rows on the cylinder were
controlled by the azimuthal rotation of the mandrel, set by a
micrometer actuator to a precision of $\sim50\,\mu$m.

After a crystal row has been appropriately positioned about the
shell surface, Armstrong A-12
was applied to the shell just below the center of each crystal. No
special preparation of the shell surface was performed. The crystal
row was then lowered by the stages to make a nominal $150\,\mu$m gap
between shell and crystal, forcing the epoxy to make a contact patch
with nominal surface area of $100\,{\rm cm^2}$. The jigging was left
in place for a minimum of 8 hours before removal. The shell was then
azimuthally rotated $12^{\circ}$ to a new position and the entire
alignment and epoxy process repeated until all 30 crystal rows were
attached.


\subsection{Radiator Transport}

Care was taken to prevent the LiF crystals from experiencing
excessive vibration or humidity during radiator transportation from
its construction site (Dallas, TX) to the detector integration site
(Syracuse, NY). A dedicated temperature-regulated shipping truck was
used for the 2500~km trip. A local atmosphere close to 0\% relative
humidity (measured as $<$1\%) was achieved by erecting a temporary
plastic film cocoon around the radiator and flowing dry N$_2$ gas
through the cocoon from gas bottles attached to the shipping frame.

Air bags attached to the four corners of the shipping/rotation
fixture cushioned the radiator against mechanical shock and
vibration. Acceleration of the radiator and temperature inside the
truck as functions of time were measured, digitized and written to
local non-volatile memory by special instrumentation\footnote{
    Model EDR-3C, Instrumented Sensor Technology, Inc.,
    Okemos, MI 48864.} 
installed on the mandrel support frame. No damage to the radiator
was observed at the trip's conclusion.

%
\section{ SUPERSTRUCTURE CONSTRUCTION } \label{sec:superstructure}

The guiding principles for the design and construction of the full
cylinder mechanical superstructure were: (1) low mass, so as not to
degrade the excellent performance of the electromagnetic calorimeter
surrounding the RICH; (2) no internal obstructions, so as not to
shadow the photon detectors; and (3) an excellent gas seal.

The overall design of the superstructure consisted of an ``Outer
Cylinder" for the multi-wire chambers, and an ``Inner Cylinder" for
the crystal radiators.  The expansion gap separated these and was
contained by two large end flanges \cite{stadco}.

\subsection{End Flanges}

\begin{figure}[htb]
  \vspace*{-0.001in}
  \begin{center}
    \includegraphics*[height=4in]{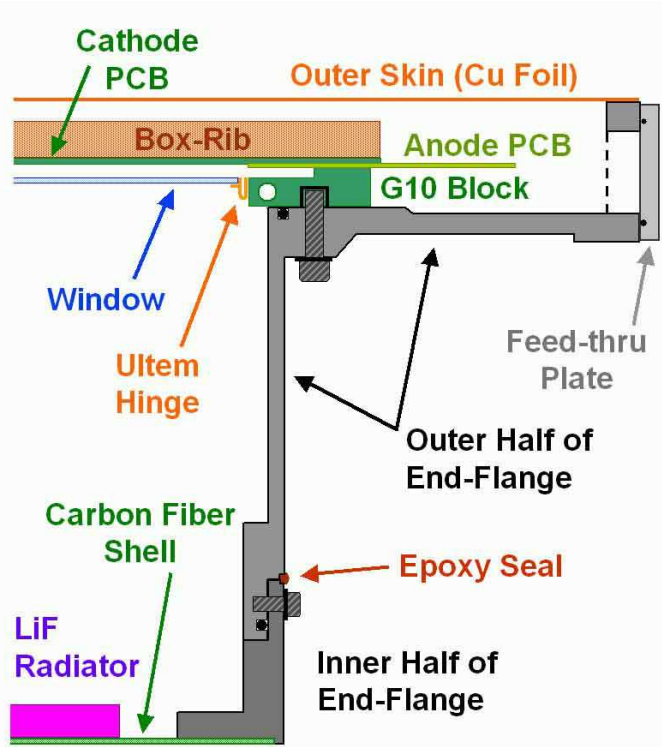}
  \end{center}
  \vspace*{-0.100in}
  \caption{\label{fig:richendxsect} Cross-sectional view of one end of the RICH detector,
        showing end flange construction.
    }
  \vspace*{-0.001in}
\end{figure}

For construction purposes, the two annular end flanges were each
made of two concentric parts, as shown in
Fig.~\ref{fig:richendxsect}. The outer half was a structural element
for the Outer Cylinder, and the inner half was fixed to the carbon
fiber shell of the Inner Cylinder. These two parts fit together with
an double gas seal,
as described below.  
The end flanges were made of structural aluminum alloy, and coated
on the interior surface by low-outgassing black paint.

\subsection{Outer Cylinder}

\begin{figure}[htb]
  \vspace*{-0.001in}
  \begin{center}
    \includegraphics*[height=3in]{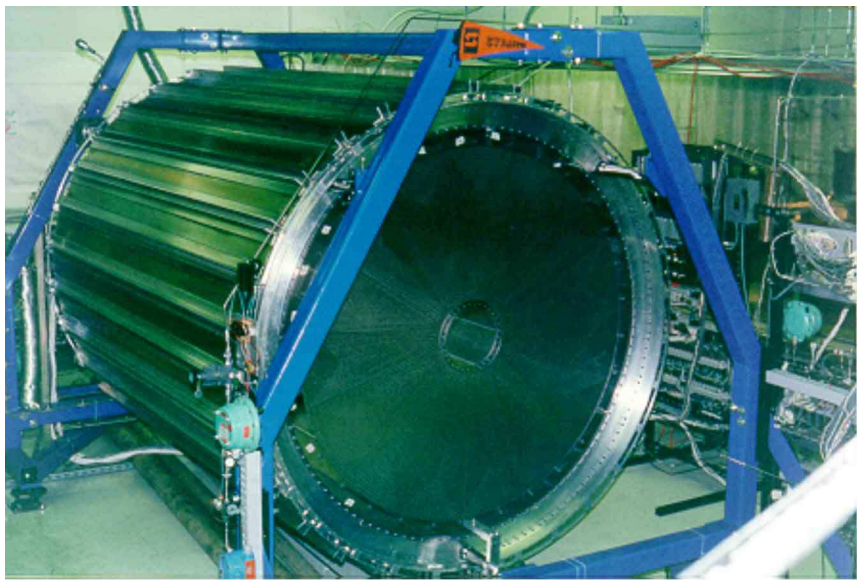}
    \end{center}
  \vspace*{-0.100in}
  \caption{\label{fig:outcylext} Photograph of Outer Cylinder during construction,
        supported on the A-Frame, with ``dummy panels" in place. }
  \vspace*{-0.001in}
\end{figure}

The construction of the Outer Cylinder used a scaffold, on which the
multi-wire chambers were mounted, as can be seen in
Fig~\ref{fig:outcylext}. It had long thin aluminum rails supported
between the outer halves of the end flanges. The rails were machined
to fit between the chambers, and contained an O-ring groove that was
the primary gas seal for the expansion gap at the Outer Cylinder. As
it was not very stiff but had to maintain the requisite compression
on the O-ring, a clamp-strip was used to squeeze the chamber on the
O-ring, with screws spaced on 1~inch centers. Hence the chambers
themselves effectively formed the outer radius of the detector. This
is another reason why the box rib reinforcements were needed.

\begin{figure}[htb]
  \vspace*{-0.001in}
  \begin{center}
    \includegraphics*[height=2in]{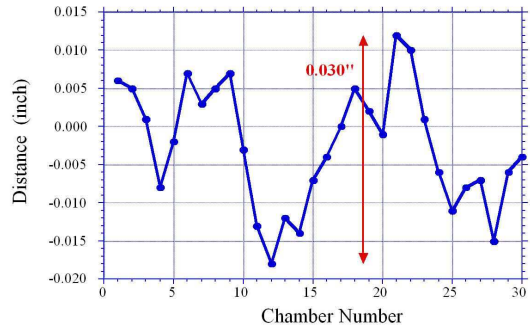}
  \end{center}
  \vspace*{-0.100in}
  \caption{\label{fig:endflange} Out-of-roundness deformation of end flange, under no-load condition. }
  \vspace*{-0.001in}
\end{figure}

During assembly, the Outer Cylinder was held on a large
custom-designed A-Frame structure with three-point kinematic mounts,
allowing free rotation of the cylinder. Initially, the inner area of
the end flange was plugged, and
the thirty chamber spots were taken up by ``dummy panels,'' stiff blanks that allowed for 
gas tightness testing of the cylinder. In this condition, the
out-of-roundness deformation was measured, as shown in
Fig.~\ref{fig:endflange}, to have a small eccentricity.

\begin{figure}[htb]
  \vspace*{-0.001in}
  \begin{center}
    \includegraphics*[height=3in]{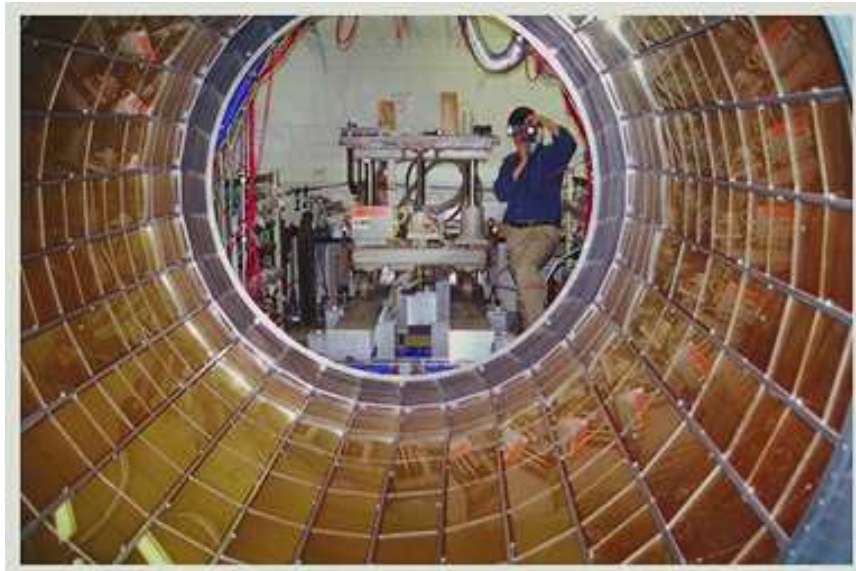}
    \end{center}
  \vspace*{-0.100in}
  \caption{\label{fig:outcylint} Photograph of inside of Outer Cylinder with photon
  detectors installed.
         }
  \vspace*{-0.001in}
\end{figure}

As chambers were completed, the dummy panels were removed and the
chambers were mounted on the cylinder. A photograph of the interior
during this procedure is given in Fig.~\ref{fig:outcylint}. It was
only at this time that the strong-backs could be removed.
Leak-checking could then be done using He gas for high sensitivity.
This was particularly important at the corners of the long single
O-ring seal around the perimeter of the chamber.  Often, the joints
and corners had to be hand-worked to seal at the level of 10$^{-4}$
ml/s.

%

\subsection{Cylinder Mating and Gas-Sealing} \label{sec:mating}

The Inner and Outer Cylinders were mated in a precision
      mechanical process.  The inner half of one end flange was first
      epoxied to the carbon fiber shell while the other inner flange
      was free. The Outer Cylinder was rotated azimuthally to align
      the fixed inner-half flange with its corresponding outer half
      and O-rings were installed in both outer flanges. The Inner
      Cylinder remained on the mandrel and slipped into the Outer
      Cylinder, riding on two large box-beams. The halves of the end
      flanges were screwed together, providing the mechanical coupling
      between the Inner and Outer Cylinders. An O-ring provided the
      primary gas seal at this junction. Following our dictum to use
      redundant gas seals on all mated surfaces, a secondary gas seal
      was made by a glue bead around the joint.  The previously free
      inner half of one end flange was epoxied to the carbon fiber
      shell as the mechanical connection was made.

Thereafter, the completed structure could be supported by the
mandrel. It was then transported for installation in CLEO by a
method similar to that used for the radiator cylinder.
Chambers and gas seals were then all tested again, in situ. After
additional iterations of hand-working and testing, the expansion
volume was sealed. Most leaks occurred when the inner and outer
cylinder were mated.


%
\section{ READOUT ELECTRONICS } \label{sec:elec}


\subsection{System Description}

The CLEO RICH electronics design is driven by two important
considerations. First of all, we need to operate the chambers at
moderate gain, to improve the stability of operation and the
lifetime of the detector system. This requirement is very important
as the system is designed to operate throughout the lifetime of CLEO
without any access for repair. In this regime the single photon
response of the MWC used as photon detectors has an exponential
distribution, as shown previously in Fig.~\ref{fig:ph-dist}. Thus,
low noise is a critical requirement to maintain good efficiency. On
the other hand, the exponential distribution spans a wide dynamic
range. Moreover we would like to be able to reconstruct the charge
deposited by a minimum ionizing particle, shown as the bump on the
right-hand side in the distribution of Fig.~\ref{fig:ph-dist}. This
implies that a wide dynamic range is also very important.

The cathode-pad segmentation is such that the charge signal induced
by the avalanche is spread around more than one pad, thus analog
charge weighting produces a better spatial resolution and also
allows for easier separation of the charge clusters produced by two
nearby photons. Thus we chose to implement an analog readout.

\begin{figure}[htb]
  \vspace*{-0.001in}
  \begin{center}
    \includegraphics*[height=1.65in]{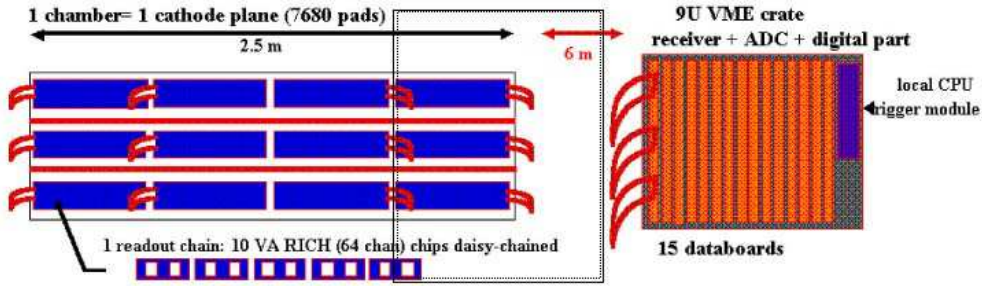}
  \end{center}
  \vspace*{-0.100in}
  \caption{\label{fig:archi} A schematic view of the readout architecture
    for a single multi-wire chamber. }
  \vspace*{-0.001in}
\end{figure}

Fig.~\ref{fig:archi} shows a schematic view of the readout
architecture for a multi-wire chamber. Each chamber is divided into
four sectors. Each sector contains three daisy-chained rows,
connected with 50-conductor shielded cable to the back-end
electronics, located in VME crates about 18 meters from the detector
cylinder.

\begin{figure}[htb]
  \vspace*{-0.001in}
  \begin{center}
    \includegraphics*[height=3in]{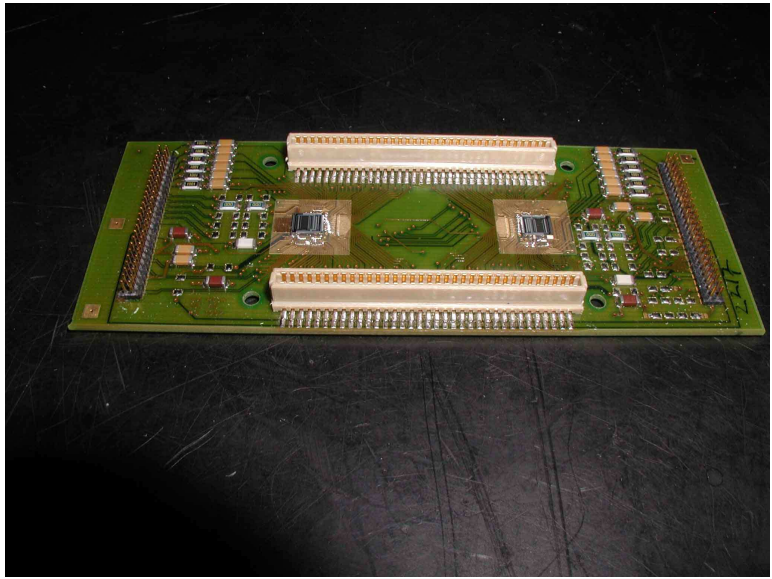}
  \end{center}
  \vspace*{-0.100in}
  \caption{\label{fig:fe-hybrids} Photograph of the front-end hybrids of the CLEO RICH.
        Two \VARICH\ ASICs are visible, which are wire-bonded to the PCB.  No protective
        caps are used. }
  \vspace*{-0.001in}
\end{figure}

The front-end hybrids, shown in Fig.~\ref{fig:fe-hybrids}, are
mounted on the back of the cathode board in a mother board-daughter
board configuration. Five hybrids are daisy chained and share the
same bias sources. Each hybrid has an independent differential
output for faster data processing.

\begin{figure}[htb]
  \vspace*{-0.001in}
  \begin{center}
    \includegraphics*[height=2in]{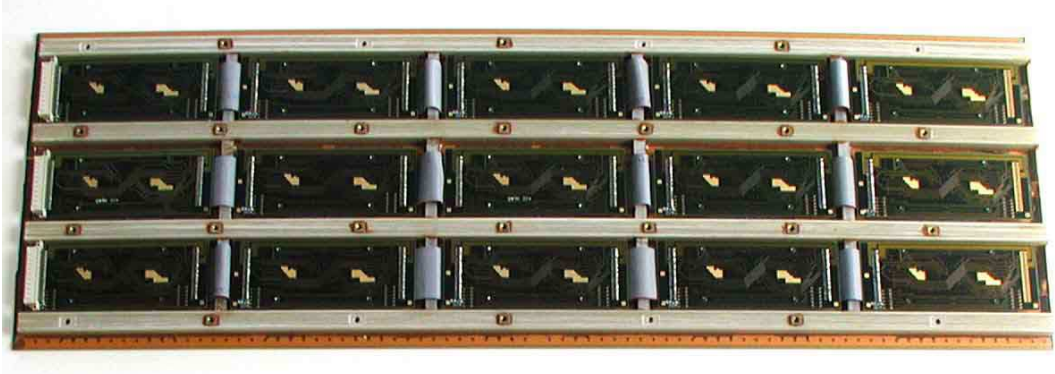}
  \end{center}
  \vspace*{-0.100in}
  \caption{\label{board} Photograph of 1/4 of a cathode plane. }
  \vspace*{-0.001in}
\end{figure}

A photograph of a quarter-section of the cathode board with the
hybrids attached is shown in Fig.~\ref{board}.

\subsection{The Front-End ASIC}

The heart of the front end electronics is the \VARICH\ ASIC. It was
designed for our application by the engineering team at IDE AS,
Norway~\cite{ideas}, and is an adaptation of the basic design of the
very successful VA family, originally developed to process the
signals of silicon microstrip detectors, tailored to our noise and
dynamic range requirement.

\begin{figure}[htb]
  \vspace*{-0.001in}
  \begin{center}
    \includegraphics*[height=3in]{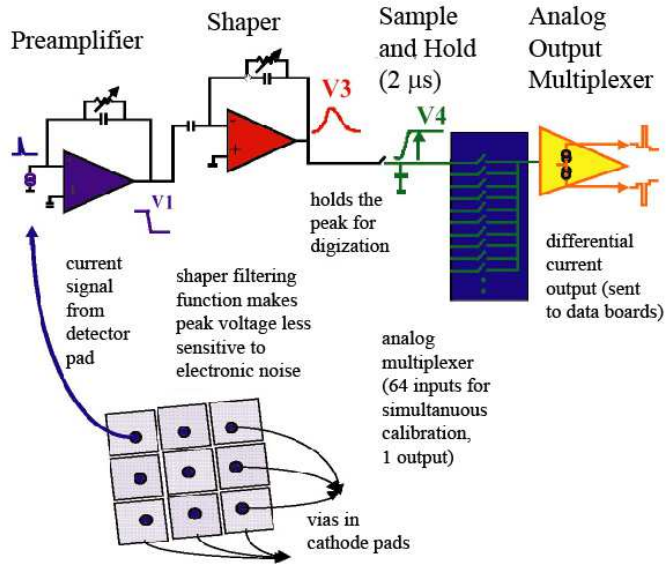}
  \end{center}
  \vspace*{-0.100in}
  \caption{\label{fig:varich} A conceptual diagram of the \VARICH\ ASIC. }
  \vspace*{-0.001in}
\end{figure}

Fig.~\ref{fig:varich} shows a conceptual diagram of this device. It
features 64 individual channels including a semi-Gaussian
preamplifier and shaper circuit. The peaking time can be adjusted by
changing the biases of the shaper circuit around a typical value of
2 $\mu$s. We tuned the peaking time to match the time when a Level 1
decision is achieved. This section is followed by a sample and hold
circuit, which is designed to hold the peak level out of the shaper
until the output multiplexer is ready to transfer this level as a
differential current. The individual inputs are connected through a
64:1 multiplexer to a calibration circuit, which allows injection of
a test charge into each individual channel. The ASIC is implemented
in 1.2~\um\ AMS CMOS technology.

\begin{figure}[htb]
  \vspace*{-0.001in}
  \begin{center}
    \includegraphics*[height=2in]{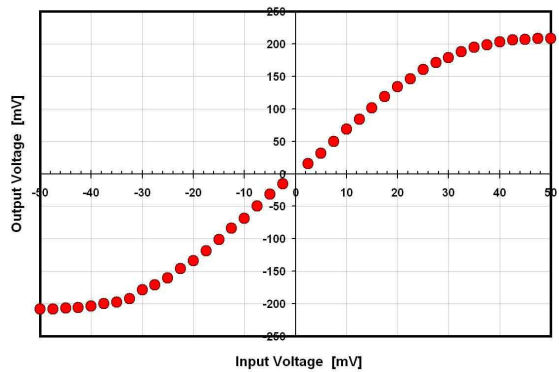}
  \end{center}
  \vspace*{-0.100in}
  \caption{\label{fig:an-out} Typical \VARICH\ output response curve. }
  \vspace*{-0.001in}
\end{figure}

The equivalent noise charge dependence upon the input capacitance
was measured on prototype single chip hybrid carriers, using a set
of calibrated capacitors. The measured performance matches the
predictions from the ASIC simulation~\cite{einar}. In our
application, we expect a total equivalent noise charge of 300~e$^-$,
without the cathode boards connected. Fig.~\ref{fig:an-out} shows
the analog output voltage on 500 $\Omega$ resistors. Saturation
occurs at an input charge level of 80~fC. In the linear region, the
preamplifier and shaper gain is 40~mV/fC, for a load of
500~$\Omega$.

\subsection{Bench-Test Characterization}

The total number of hybrids produced was 2200, over a period of two
years. The hybrids produced were tested for functionality at IDEAS
and then shipped to Syracuse for more complete characterization. Our
tests involved a noise measurement by taking pedestal data, followed
by a detailed mapping of the individual channel gain with
calibration pulses of different amplitudes. Only 1800 were installed
in the detector.

The noise measurement required great care in the grounding and
shielding of the hybrids and careful routing of the analog power
supplies. Our goal was to achieve a noise of the order of 400--600
e$^-$ with a simple and relatively quick set-up of the measurement.

After this initial characterization, we performed a burn-in test of
the hybrids, maintaining them biased in their nominal working point
at elevated temperature (70\degC) and performing electronic
calibration cycles at regular intervals. In order to perform these
tests at a rate compatible with our installation schedule, we
produced a dedicated set-up where 32 hybrids could be biased and
monitored in parallel. After one week in the burn-in set-up, the
hybrids were tested for noise and gain with a quicker calibration
procedure. The hybrids that were rejected upon this procedure were
only a few per thousand.

%

\subsection{CLEO RICH Data-Boards}

\begin{figure}[htb]
  \vspace*{-0.001in}
  \begin{center}
    \includegraphics*[height=3in]{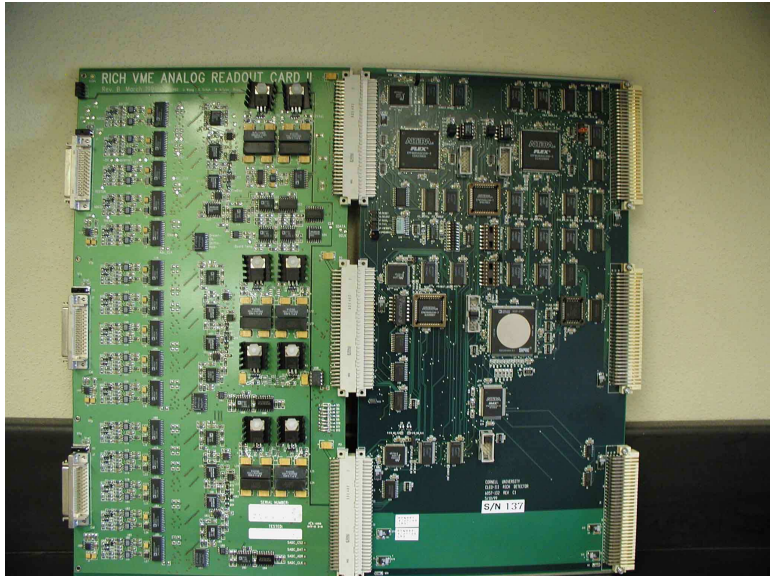}
  \end{center}
  \vspace*{-0.100in}
  \caption{\label{fig:databoards} Photograph of the CLEO RICH Data Boards:
        Analog board on the left, Digital board on the right. }
  \vspace*{-0.001in}
\end{figure}

Fig.~\ref{fig:databoards} shows a picture of the CLEO RICH data
boards. They are 9U mixed analog and digital environment boards that
perform several very complex functions. Note that the board is
physically composed of two different sections. The first is an
analog section, providing the biases needed for the functionality of
the \VARICH\ ASIC, the transimpedance receivers, 12-bit ADCs
digitizing the serial analog information, and slow control
monitoring ADCs. The second is a pure digital section,
based on a common CLEO data acquisition framework, and containing
two components specific to our application: a sequencer, based on
the ALTERA MAX FPGA, that contains the firmware necessary to operate
the \VARICH\ ASIC, and an Analog Devices ADSP-21061 DSP, used to
perform the common mode suppression and data sparsification
described below.

The analog section features 15 receiver channels, organized into
three cells sharing the same ribbon cable interconnection to the
front end hybrids. Each cell encompasses dedicated $\pm$2V analog
power and several DACs that allow the adjustment of the bias
currents and voltages that influence the working point of the analog
front end, as well as a slow control section that monitors the
values of these voltages and currents, and the temperature on
different locations of the detector cylinder and on the data boards
themselves. In addition, local pulse generators are used for the
electronics calibration. All the cells share a common regulated
$\pm$2V digital section.

When an event occurs, all the 15 channels in a board receive the
synchronous serial information from the corresponding hybrid. In
order to simplify the digital section
the 5 ADC channels in a cell are multiplexed into a common FIFO
buffer, with an offset of 23.4 ns, determined by the local 42 MHz
clock.

To maintain the low thresholds that are required to optimize the
efficiency, the hardware common mode noise suppression algorithm is
crucial to suppress the adverse effects of coherent fluctuations of
all the channels in an ASIC.


\section{ SUPPORT SYSTEMS } \label{sec:support}

\subsection{Gas System }


The gas system supplies several distinct volumes. The system must:
supply CH$_4$-TEA to 30 separate chambers, supply ultra-pure N$_2$
to the expansion volume, supply ultra-pure N$_2$ to a sealed single
volume surrounding all the chambers, called the ``electronics
volume,'' since this is the region where the front-end hybrid boards
are present. In addition we need to test the CH$_4$-TEA mixture for
the ability to detect photons and test the output N$_2$ for purity.

It is of primary importance that the gas system must not destroy any
of the thin CaF$_2$ windows. We use computerized pressure and flow
sensors with programmable logic controllers (PLC).

\subsubsection{System Design}

The performance and mechanical integrity of the CLEO~III RICH are
critically dependent on the performance of its gas systems.  In
order to achieve its design resolution, the RICH must efficiently
detect 14-21 photons emitted by charged tracks traversing the
radiator. The efficiency of the RICH depends on the expansion volume
transparency.  The photosensitive detector gas and transparent
nitrogen of the expansion volume are separated by fragile windows on
the inner faces of the detector modules. These windows could be
destroyed by a slight pressure difference between these volumes (of
order 15''~H$_{2}$O). Such an overpressure would be catastrophic to
the RICH.

The gas systems were carefully designed to protect against such
damage.  A highly automated design was chosen in order to minimize
the possibility of operator error and to provide fast response to
dangerous conditions.  Most valves are controlled by the automated
control system and most sensors are read out and monitored
electronically.

The two major subsystems are the expansion volume gas system (EVGS),
which purges the expansion volume with a large flow of ultra-pure
nitrogen, and the detector gas system (DGS), which supplies the
thirty detector modules with photo-sensitive methane/TEA gas.
Pressures in the detector and expansion volumes are referenced to
atmospheric pressure.  Both the EVGS and DGS employ triply-redundant
overpressure and underpressure protection.  The first tier of
protection is based on readings from pressure transmitters connected
to the expansion volume and each of the thirty detector volumes, and
has trip points  at +0.75''~H$_{2}$O and -0.5''~H$_{2}$O.  The
second is based on readings from pressure switches which have trip
points set at +1.5''~H$_{2}$O and -1.0''~H$_{2}$O.  The final level
of protection comes from  mechanical relief valves with set points
at +2.5''~H$_{2}$O and -1.5''~H$_{2}$O.   Valves which perform the
most critical operations, such as those which shut off the gas
inputs,  are equipped with position indicating sensors and, in the
most critical, redundant valves are implemented.

\subsubsection{Control System}

All critical operations of the RICH gas systems are controlled by a
programmable logic controller (PLC).   The PLC is an industrial
control system designed to be reliable and modular.  The control
system utilizes 112 analog input, 8 analog output, 96 discrete
output, and 128 discrete input channels.  The gas system is
controlled through a small number of buttons on the main control
panel.

The basic operation of the EVGS and DGS is similar.  Each has four
states of operation: RUN, IDLE, STOP, and ALARM.  RUN mode is the
normal operating state of the system.  In this state gas flows
through the detector and the pressure is actively regulated by the
PLC at 0.5''~H$_{2}$O to an accuracy of $\pm 0.02$''~H$_{2}$O.  IDLE
mode is similar to RUN mode except that the pressure is not actively
regulated.  IDLE mode is used in the system start up sequence.
ALARM mode is the safe state to which the system defaults in case of
an emergency.  In this state, the input gas flow is shut off and the
exhaust side of the detector is opened to track atmospheric
pressure.  After the system is stopped by an alarm it is important
that it be restored to RUN or IDLE mode as soon as possible to
prevent degradation of the CaF$_{2}$ and LiF crystal transparency
when exposed to ambient humidity.  After the offending condition is
fixed, alarms may be reset using a button on the control panel.
When the alarm is cleared, the system changes to STOP mode, which
has the same physical configuration as ALARM mode. The system start
up sequence is fully automated and is initiated by pressing a single
button.  The start up sequence consists of a purge of the input
lines, followed by a ramp of the input flow, and finally a ramp of
the pressure.  All transitions in the system are gradual in order to
protect the RICH from possible pressure spikes.

All critical components of the gas system are powered through an
uninterruptable power supply which is backed up by a diesel-powered
generator.

Interfaces to the operator and to the CLEO~III detector control
system
 are provided through the gas system monitor (GMON), consisting of
LabView$^{\rm TM}$~\cite{labview} programs running on a PC.  The
GMON programs collect sensor readings, status, and alarm information
from the PLC which it displays in several ``active schematics''
corresponding to each of the major components of the EVGS and DGS.
The GMON relays this information to the CLEO~III slow control
system.

The GMON also handles the display of alarms.  ``Fatal'' alarms are
generated by the PLC and result in the system defaulting to ALARM
mode in order to protect the system from damage.  All other
``non-fatal'' alarms are generated by the GMON code, which monitors
the status of all system parameters.  The system is heavily
instrumented with sensors in order to allow problems to be easily
identified and repaired.

The GMON programs are also used to control some non-critical
operations of the gas system, such as the gas quality monitoring,
and to set some system parameters. The gas system does not depend on
the PC or GMON programs in order to run, however.

\subsubsection{Expansion Volume Gas System (EVGS)}

It is critical that the purity of the expansion volume be maintained
in order for the RICH to achieve its design resolution. Most
impurities will absorb the UV photons before they reach the
detectors and must be kept below a few ppm concentration.  This
purity is obtained by flowing nitrogen at a rate of approximately
1500 $\ell$/h in a single-pass configuration.  Boil-off from liquid
nitrogen is used as the source for the expansion volume gas.  Most
of this system is constructed from electropolished stainless steel
and other materials with low vapor pressure in order to minimize
impurities.

Nitrogen gas from the dewar is processed by an automated purifier
subsystem.  This purifier injects a regulated flow of hydrogen
corresponding to 40 ppm, which reacts with the oxygen inside a
catalyst cartridge to form water.  A slight excess of hydrogen is
acceptable since it is known to be transparent over the wavelength
range of interest. Water and other large molecules are removed by a
large molecular sieve trap.  The molecular sieve trap must be
regenerated occasionally by baking at  300$^{\circ}$~C for a day.
The purifier subsystem contains two parallel purifiers to allow for
continuous operation while one of the sieves is being baked.  The
PLC controls the flow ramping and switching between purifiers as
well as the temperature control and gas purge for the baking sieve.

The operating flow of nitrogen to the expansion volume is set with a
manual metering valve near the input to the expansion volume.  Gas
exiting the expansion volume passes through ten 1/2'' stainless
steel lines to a 1'' manifold outside of the CLEO~III solenoid
return yoke, where pressure is regulated by a PLC-controlled
metering valve based on feedback from a pressure sensor.

\subsubsection{Detector Gas System (DGS)}

The detector gas system is responsible for providing thirty separate
photon detector volumes with a mixture of methane and TEA.  Because
TEA is corrosive to most materials, 316 stainless steel components
were used wherever possible.  Any other materials used in the system
were first subjected to stringent chemical compatibility tests.

TEA vapor is introduced into the stream of methane gas by bubbling
it through liquid TEA at 15$^{\circ}$~C.  The TEA bubbler system is
automated using programs on the PLC for reliable and continuous
long-term operation.  It utilizes a temperature-controlled bubbler
chamber and elevated TEA reservoir with enough TEA for approximately
two months of operation.  The level of TEA in the bubbler chamber is
regulated by the PLC which fills from the reservoir as needed.  The
reservoir may be depressurized and filled without interrupting the
operation of the system.  Sensors monitor the level and temperature
of the TEA in the bubbler and the level of the TEA in the reservoir
and notify the user when refilling is required.

In order to provide adequate pressure control, it is necessary to
supply gas to the chambers in a parallel configuration rather than
in series. The flow is split into thirty separate streams to provide
even flow to all 30 detectors.  The flow in each branch is set
manually and monitored electronically.  The flow is ramped up and
down automatically during start up and purge operations. A total of
ninety 1/4'' stainless steel lines pass through the CLEO~III
solenoid return yoke steel to connect to the thirty photon detector
modules: one input, one output, and one pressure sensing line.

Exhaust from the thirty modules is collected into a single exhaust
manifold.  Gas from this manifold passes through a PLC-controlled
pressure regulating valve which maintains the average pressure of
the thirty modules at 0.5''~H$_{2}$O.

\subsubsection{Gas Quality Monitoring}

 A nitrogen transparency monitoring system measures the UV transmission of
the expansion volume gas as a function of photon wavelength.
Monochromatic light is produced by passing light from a deuterium
lamp through a computer-controlled monochromator. The monochromatic
photon beam passes through a gas sample tube and is detected by a
photomultiplier.  The sample tube can be switched to sample the
expansion volume exhaust gas or an ultra-pure argon reference gas.
A comparison of these two gives the transparency.  The GMON controls
the transparency monitor and other gas monitoring devices.
Automated transparency scans are taken once per day.  Graphical
panels allow the user to configure the scans and view the results.

Oxygen is the most likely and problematic contaminant, since it
absorbs strongly at 150~nm.  The oxygen concentration of the output
gas from the expansion volume is also monitored using a precision
oxygen sensor.

The CH$_4$-TEA mixture is monitored by an electron capture detector.
This device consists of a cylindrical proportional drift chamber
which utilizes a beta source to produce ionization. Electrons
produced near its outer diameter drift through a $\sim$5~cm long
path to an anode wire at the center of the chamber where they are
collected.  A 0-5~VDC signal proportional to the time-integrated
drift current is monitored by the control system. Impurities or
changes in the gas composition can be detected as a change in the
output of this device.

\subsubsection{Electronics Volume Purge System}

The gas system also provides a purge flow of nitrogen gas around the
electronics volume of the RICH in order to minimize the
concentration of impurities adjacent to the gas volumes of the RICH.
This helps to reduce the infiltration of oxygen into the expansion
volume.

\subsubsection{Performance}

The gas system has been in operation since the early commissioning
of the RICH and has been running in its final form since the Summer
of 1999.  The most critical aspect of its performance is that it has
protected the extremely fragile CaF$_2$ windows from damage.  It has
delivered nearly continuous service for several years, with only a
few short down-times for repairs. Identification and diagnosis of
problems has been straightforward due to the wealth of information
provided by the system.

\begin{figure}[htb]
  \vspace*{-0.001in}
  \begin{center}
    \includegraphics*[height=2in]{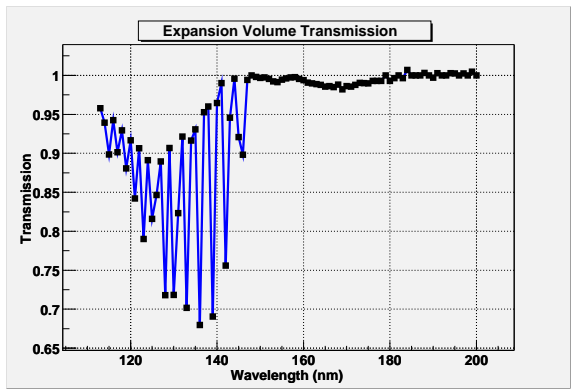}
  \end{center}
  \vspace*{-0.100in}
  \caption{\label{fig:n2-transparency} A typical spectrum of
    the expansion volume gas transparency.  The RICH response is centered at
    150~nm and extends from approximately 135~nm to 165~nm.  The structure
    below 150~nm is due to the vibrational-rotational absorption bands of nitrogen.
    The transmission at 150~nm is greater than 99.5\%. (Note the suppressed zero of the vertical scale.) }
  \vspace*{-0.001in}
\end{figure}

The expansion volume transparency exceeds the specifications for the
RICH design.  The EVGS typically delivers transparencies of
$>$99.5\% at 150~nm, as shown in Fig.~\ref{fig:n2-transparency}.
This wavelength is near the peak of the RICH sensitivity and is
particularly vulnerable to oxygen contamination.  The recovery time
of the expansion volume gas system is quite short due to the high
purge rate and careful choice of materials. After exposure to air,
the expansion volume can typically be restored to acceptable
transparency within a half day.

The RICH photon detectors show no evidence of aging due to reactions
of TEA with materials in the gas system and RICH.  The DGS has
provided reliable and simple operation.  In particular, maintaining
and filling the liquid TEA bubbler has proved to be quite simple and
trouble-free.

The highly automated design of the RICH gas system has proved to be
very effective at protecting the RICH and simple to operate. It has
exceeded expectations of performance and reliability.

%
\subsection{High Voltage System}


The High Voltage System for the CLEO RICH was required to supply 90
channels of anode field wire voltage (about $+$1500~V) and 240
channels of cathode window trace voltage (about $-$1200~V), with a
current monitoring and trip circuit on each channel.
The system was based on the LeCroy ViSyN System~\cite{hv-lrs}, and
consists of a Model 1458 mainframe containing slots for 16 Model
1469 modules, which have either positive or negative polarity. In
our implementation, we use 5 positive and 11 negative modules. The
module outputs are ganged together such that there are eight HV
channels per so-called ``bulk'' connector, a specially-designed LHV
connector. There are three bulk connectors per module. The system
allows voltage setting at the bulk connector level, so each channel
in the bulk connector has the same voltage, but may have different
current trip levels. The channels fan out to a patch-panel so the
proper high voltage can be routed to each chamber channel.
The value of the voltage setting on each of the 330 detector
channels was determined by an optimization procedure, as discussed
in Section~\ref{sec:gaineq}.

Control of the high voltage system is accomplished by a stand-alone
server program, on a dedicated linux station, that is integrated
with the CLEO DAQ System in order to synchronize the state of the
system with accelerator operation.
 %

Ramping of voltages are made at rates of 40~V/s (increasing) and
50~V/s (decreasing) for both polarities. Care is taken to raise all
voltages in a given chamber simultaneously, in order to minimize
electro-mechanical deflections.
All channels have current trip levels set to 10 \uA, and are
monitored at a rate of $\sim$2~kHz. Individual channel status,
currents and voltages are read back and monitored as well. Any
currents or voltages out of tolerance will cause an alarm, requiring
immediate operator intervention.
During detector operation, communication between the server and the
mainframe is maintained via TCP/IP, and the status of all voltages
and currents is displayed in a representational color-coded GUI.


%
\subsection{Cooling System}


The total power output of the VA\_RICH chips is approximately 360~W.
In order to provide heat removal necessary for the mechanical
stability of the chambers, four nylon cooling tubes run over the
back of each chamber, and are each modestly thermally coupled to the
\VARICH\ hybrids by two-component RTV.\footnote{
    RTV 577 / RTV 9811, GE Silicones, Wilton, CT 06897.}
Supplying these tubes with the hydrocarbon coolant PF-200IG is a
specially-designed CLEO cooling farm~\cite{cooling}. This coolant is
used rather than water because of concern that a leak could reach
the water sensitive CsI crystal, which would melt; an additional
advantage is that PF-200IG is non-conductive. The RICH cooling
circuit is one of several driven by the cooling farm through
active-manifold platforms, with a coolant reservoir temperature of
18\degC\ and flow rate of $\sim$1~$\ell$/min. Each RICH chamber line
is equipped with an embedded tip-sensitive $100\,\Omega$ platinum
resistance temperature sensor (RTD). For these RTDs, an array of
``hockey-puck'' style 4--20~mA transmitters route signals to a
custom-built multiplexer circuit used to read out these transmitters
every 10~s. The temperature signals are fed into a small logic
controller for real-time control and monitoring. The typical range
of exit temperatures from the 30 chambers when in operation is
21--23\degC, each held stable to $\pm$0.5\degC.

The status of the cooling system is continuously monitored by a
LabVIEW program which provides web-based diagnostic tools for
real-time viewing of the performance parameters on site and
monitoring worldwide.

%
\subsection{Slow Control System}


A bit-serial data bus was implemented on the P3 connector on the
RICH AVME data board using differential signaling technique. This
bus is referred to as the SBUS. The maximum transfer rate of the
SBUS is approximately 1 Mbit per second. Each RICH VME crate is
equipped with one SBUS crate controller. Each SBUS crate controller
can be addressed uniquely. All eight RICH VME crates are daisy
chained to one SBUS controller module located in a separate SBUS
crate at the pit level.

The SBUS system is responsible for monitoring the operational status
of the VA\_RICH chips as well as that of the AVME data board.
Important quantities such as the reference voltages, the reference
currents, and the temperature on the chip carrier chain inside the
RICH detector are being monitored periodically through the SBUS
system. Analog signals from the RICH detector are converted into
digital signals with a 10-bit ADC (TI TLC1542/3CDW) on the AVME data
board for the SBUS read out. The SBUS can also be used to download
some of the EEPROMs on the DVME data board.

A server program running on an on-line computer communicates with
the SBUS controller module via CORBA calls. A monitor request is
sent out every 15 minutes. This sampling interval is programmable
and was chosen to ensure both adequate monitoring and maximum system
wide stability. The data are shipped back to the server, archived
locally in plain text files and then processed to provide
information on the front-end electronics status in an easy-to-read
HTML format on a Web server.


\section{ OPERATING EXPERIENCE } 

The CLEO RICH detector has been in operation since September 1999.
All but $\sim$2\% of the detector is functioning for data-taking. We
lost 1.7\% due to the breaking of one wire after about one year of
operation due, most likely, to the un-monitored heating of chamber
due to problems with slow control software. We have also lost 2\% of
the electronics chips and suffered one broken output cable, so the
total number of lost channels is $\sim$5\%.

\subsection{Normal Operation}


In order to ensure proper functioning of the RICH over the long
period of its operation, a series of calibration and monitoring
activities are performed periodically as a part of its normal
operating procedure.

To ensure the highest detection efficiency of the VA\_RICH chips,
the electronics pedestals need to be measured routinely. This is
done by performing a special calibration procedure called
``SmallCal.'' During a SmallCal, the VA\_RICH chips are set in
calibration mode and the electronics noise from all 230,400
electronic channels are read out. The pedestals are then calculated
on-line and loaded to the VME crates to be used for sparsification
during successive data-taking runs. This SmallCal procedure is
carried out regularly every eight hours.

Another type of calibration, the WirePulse run, is performed once
per week. The purpose of the WirePulse run is to measure the
response of the RICH electronics to a given input signal. During a
WirePulse run, the anode wires are pulsed with a sequence of several
predefined waveforms of differing amplitudes. The output from all
electronic channels are analyzed off-line (see below).

As detailed above in Sec.~\ref{sec:support}, the high voltage server
program monitors the RICH HV status, by reading voltage and current
values, and watching for trips.
Also as detailed above, the RICH gas system and the cooling system
adequately control critical operating parameters, but also provide
for web-based monitors which are checked twice per hour during
normal operation. For the gas system, the expansion volume
transparency, the TEA bubbler temperature and the expansion volume
oxygen content are closely monitored. For the cooling system, the
chamber temperatures are watched.
The operating parameters of the readout electronics is checked using
the web-based utilities of the Slow Control system.

In addition, RICH performance is supervised offline, with a set of
quality-monitoring plots produced from a prescaled sample of
incoming events, from the so-called CLEO pass1 analysis. These are
examined after each run, and include distributions of Cherenkov
angles and photon yields from fast tracking, as well as the more
pedestrian distributions of raw hits and cluster pulse heights (as
described in Sec.~\ref{sec:ana}).

This information has been sufficient to ensure that the CLEO RICH
detector produced data with a high degree of stability over many
years.

\subsection{Chamber Gain Equalization} \label{sec:gaineq}

The high voltage operating point for all chambers was determined by
a gain equalization procedure, which sought to make equal the
pad-gain for single photons averaged over each window module
individually. The gains were set to be below 25,000 in order to
avoid discharges.

Gain changes were measured as a function of window and wire
voltages. The gains were determined for each window-sized module by
fitting an exponential curve to the pulse-height spectrum. We
parameterized the gain as a function of the wire voltage, for each
window voltage.

\begin{figure}[htb]
  \vspace*{-0.001in}
  \begin{center}
    \includegraphics*[height=2.5in]{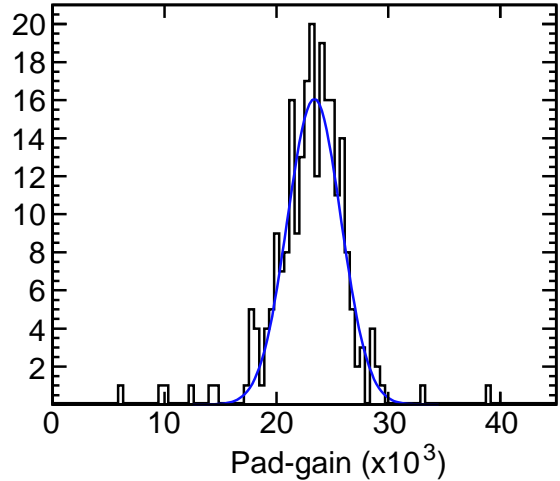}
  \end{center}
  \vspace*{-0.100in}
  \caption{\label{fig:gain-dist}
  Distribution of measured pad-gains,
        segmented for each window module separately. }
  \vspace*{-0.001in}
\end{figure}

Gains of all of the chambers were varied by changing the voltages in
an iterative manner in order to make them equal. Our goal was to
have a pad-gain of $\sim$23,000. Fig.~\ref{fig:gain-dist} shows the
distribution of gains as measured after the gains had been
equalized. We find a mean of 23,400 with a fitted r.m.s. spread of
10\%. Fig.~\ref{fig:gain-dist} shows the distribution of gains as
measured after the gains had been equalized. We find a mean of
23,400 with a fitted r.m.s. spread of 10\%. This has remained stable
during detector operation.

\subsection{Electronics Performance in CLEO III and CLEO-c}


\begin{figure}[htb]
  \vspace*{-0.001in}
  \begin{center}
    \includegraphics*[height=2.5in]{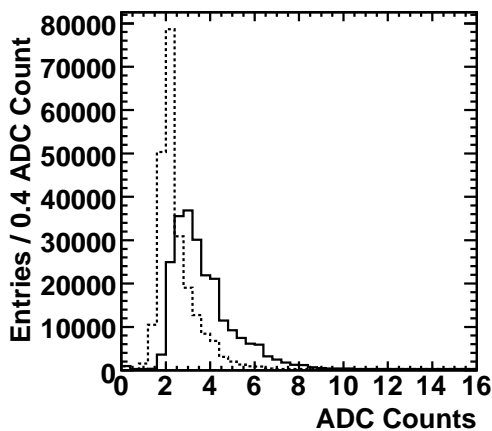}
  \end{center}
  \vspace*{-0.100in}
  \caption{\label{fig:noise-distr} Typical noise distributions
  for all channels in units of ADC channels (1 ADC channel equals
  $\sim$200 e$^-$). The solid line shows the total noise while the
  dashed line shows the incoherent component after the coherent
  noise subtraction.
         }
  \vspace*{-0.001in}
\end{figure}

During data taking we routinely perform two sets of measurements to
check the electronics performance. We measure the pedestal
periodically, to verify that no baseline shifts occurred. These data
also allow us to monitor the value of the total and incoherent noise
of each channel. Typical noise distributions are shown in
Fig.~\ref{fig:noise-distr}. The noise reduction produced by the
coherent noise suppression can be clearly seen. The noise levels are
quite low, the peak of the incoherent noise is at 425 electrons.

\begin{figure}[htb]
  \vspace*{-0.001in}
  \begin{center}
    \includegraphics*[height=2.5in]{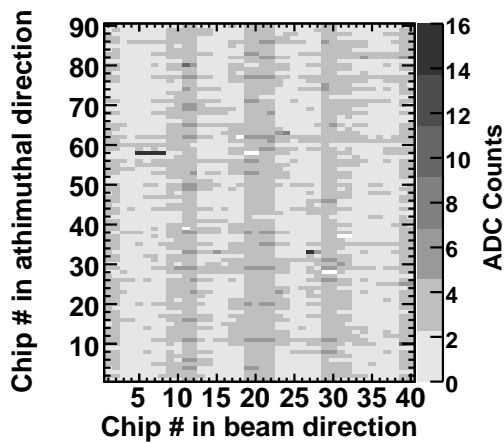}
  \end{center}
  \vspace*{-0.100in}
  \caption{\label{fig:noise-profile} Spatial distribution of the typical intrinsic noise
        on each \VARICH\ chip, given in units of ADC counts.
        (The RICH cylinder is unfolded onto a plane.)
        There are slightly elevated noise bands at each end of a chain.  }
  \vspace*{-0.001in}
\end{figure}

\begin{figure}[htb]
  \vspace*{-0.001in}
  \begin{center}
    \includegraphics*[height=2.5in]{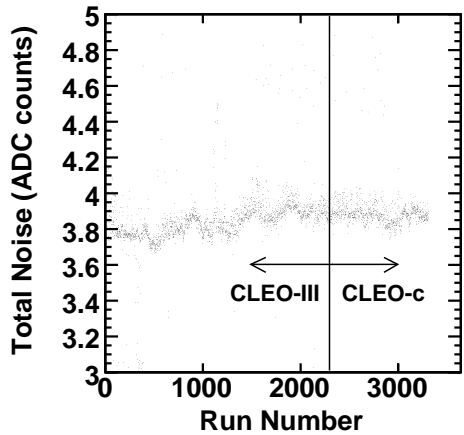}
  \end{center}
  \vspace*{-0.100in}
  \caption{\label{fig:noise-trend} Time evolution of the mean of the total noise. The
  time covered is about 5 years. } 
  \vspace*{-0.001in}
\end{figure}

Fig.~\ref{fig:noise-profile} shows a scatter plot of the intrinsic
noise distribution across the detector. The horizontal axis
corresponds to the length of the chambers and the vertical axis
corresponds to the chamber number. It can be seen that the profile
is rather uniform. The darker areas correspond to break points
between different chamber sectors, and are associated with
additional noise sources at the boundary between two adjacent
sectors. Cross-talk or additional digital noise in the chip carrier
at the end of a chain may cause this higher noise level.

The overall noise performance has been extremely stable throughout
the years. Fig.~\ref{fig:noise-trend} shows the time evolution of
the mean value of the total noise and its very stable value.

\begin{figure}[htb]
  \vspace*{-0.001in}
  \begin{center}
    \includegraphics*[height=2.5in]{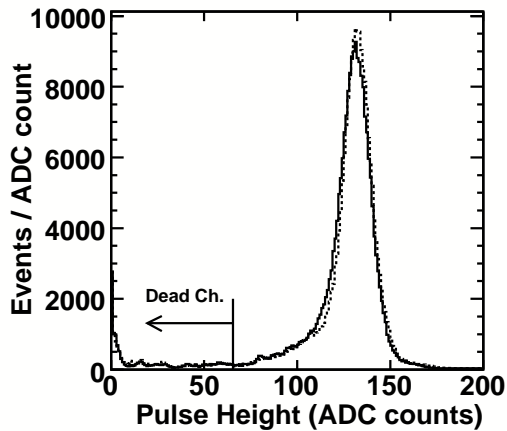}
  \end{center}
  \vspace*{-0.100in}
  \caption{\label{fig:wire-pulse} Distribution of signals on all pads
    from injecting pulses on the wires of $\pm5$ mV. The solid
    line shows positive signals and the dashed line negative
    signals.
    Data is from the CLEO-c era. }
  \vspace*{-0.001in}
\end{figure}

An additional quantity that needs to be monitored is the gain of the
front end electronics. In principle, we could undertake the same
sort of electronics calibration that was performed to verify that
the hybrids were compliant with all our specifications. However, the
amount of data to be transferred is beyond the capabilities of the
data acquisition system. Thus, we use a method that is much quicker
and simpler, although not as accurate. We pulse the MWC wires and
the capacitive coupling between wire and pad produces the current
pulse amplified by the front end electronics. The advantage of this
calibration procedure is that the current originates at the same
location as the real signal. Thus it tests the integrity of the
whole processing chain, including the wire bond between input
channel and hybrid trace and all the connectors along the signal
path. On the other hand, the pulse shape is not perfect because of
the improper termination, thus the gain measurement is not very
accurate, as illustrated in Fig.~\ref{fig:wire-pulse}, that shows
the signal distribution on all the pads for input pulses of +5 mV
and $-5$ mV respectively. The wire pulse distribution is also used
to determine the number of dead channels. We have about 5\%\ dead
channels, and the number has been relatively stable throughout the
duration of the experiment. These losses include one damaged
multi-wire ribbon cable and ASIC damage as well as lost wirebonds.

The most notable source of ASIC loss has been occasional latch-up of
these devices, due to the activation of parasitic PNPN paths in the
device. The symptoms include large current draw and a different
pedestal profile along the ASIC, much ``flatter'' than in a
functioning ASIC. This phenomenon occurred more frequently in the
early stages of the experiment, due to occasional faulty start-up
procedures, that caused some device destruction. After we refined
the start-up procedure, to prevent regenerative loops from
occurring, the number of flat ASICs has remained stable, with some
signs of recent recovery.

In general, we can say that the operation of this system has been
very stable and reliable and no tuning of the bias voltages and
currents has been necessary, since the initial adjustment performed
upon installation.



\DeclareGraphicsRule{*}{eps}{*}{} \graphicspath{{figs/}}


\section{ DATA ANALYSIS AND PHYSICS PERFORMANCE } \label{sec:ana}


\subsection{Introduction}
The CLEO III detector was used for studies at the
$\Upsilon(1S)-\Upsilon(5S)$ resonances from August of 2000 to March
of 2003. The CLEO detector was then modified by replacing the
silicon strip vertex detector with a low mass wire chamber. The
magnetic field was also lowered from 1.5 T to 1.0 T, to help
increase the machine luminosity. Data was then taken from October of
2003 until April of 2005. The results in this section refer to the
first period as CLEO III and the second period as CLEO-c. More
CLEO-c data will be forthcoming.


Coherent noise suppression and data sparsification are performed
on-line to eliminate the Gaussian part of the electric noise. A
small non-Gaussian component of the coherent electric noise is
eliminated off-line, by using an algorithm too complicated for use
in the data board DSP. The incoherent part of non-Gaussian noise was
eliminated by off-line pulse height thresholds adjusted to keep
occupancy of each channel below 1\%. Finally, we eliminate clusters
of cathode pad hits that are extended along the anode wires, but are
only 1--2 pads wide in the other direction.

\begin{figure} [tb]
    \vspace{-0.01cm}
    \begin{center}
        \epsfysize
        1.38in\epsffile{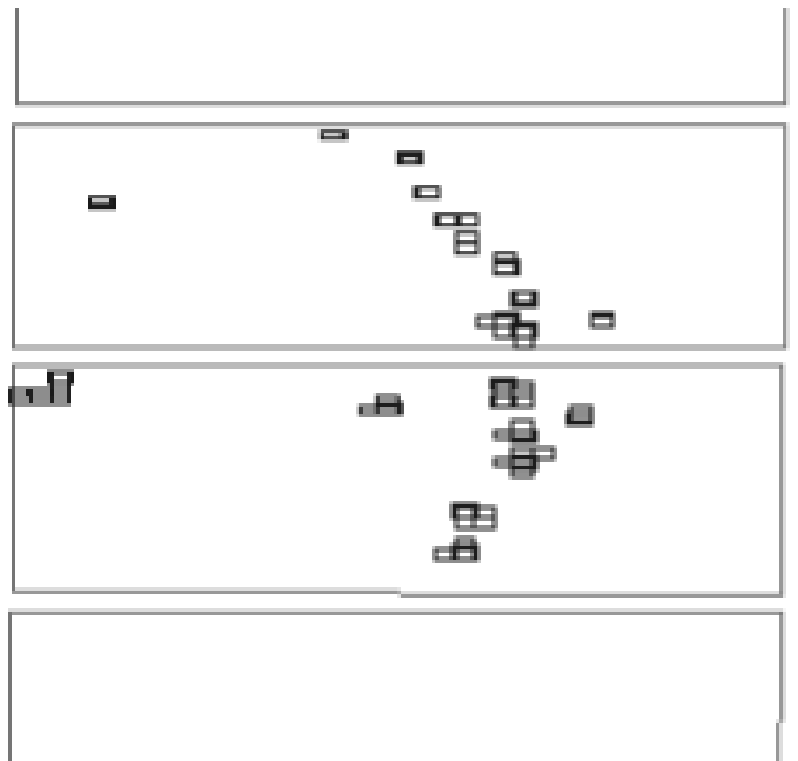}\hspace{0.7in}
        \epsfysize 1.43in\epsffile{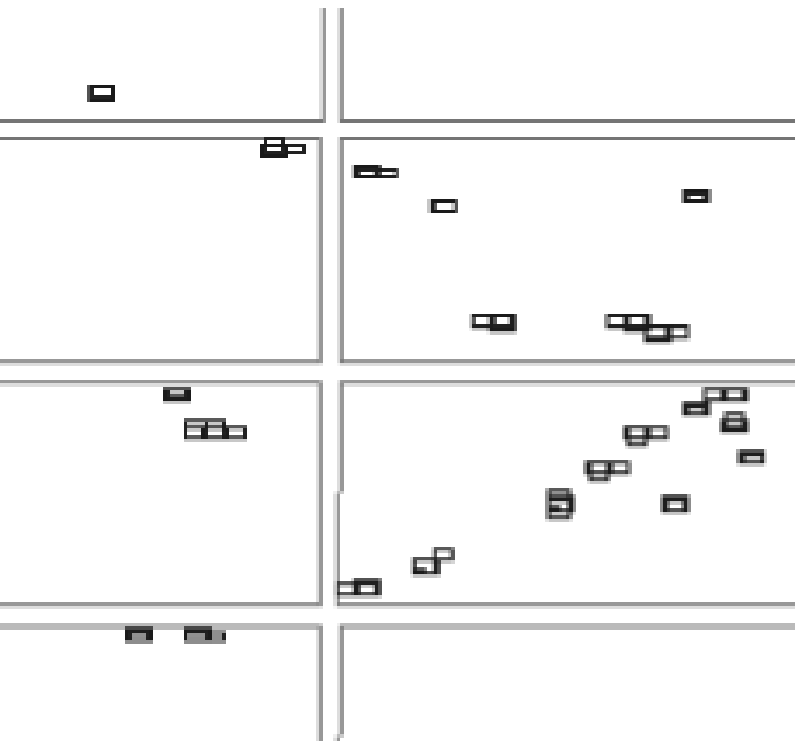}
    \end{center}
    \vspace{-0.01cm}
    \caption{\label{fig:images}
        Raw hit patterns produced by the particle passing through the plane
        (left) and sawtooth (right) radiators. Each rectangle indicates a pad
        channel above threshold.}
    \vspace{-0.01cm}
\end{figure}

We show in Fig.~\ref{fig:images} the hit pattern in the detector for
a Bhabha scattering event ($e^+e^-\to e^+e^-$) for track entering
the plane (left image) and sawtooth (right image) radiators. The
shapes of the Cherenkov ``rings" are different in the two cases,
resulting from refraction when leaving the LiF radiators. The hits
in the centers of the images are produced by the electron passing
the RICH multi-wire chambers.

\subsection{Clustering of Hits}

The entire detector contains 230,400 cathode pads, which are
segmented into 240 modules of $24\times40$ pads separated by the
mounting rails and anode wire spacers. We cluster pad hits in each
module separately. Pad hits touching each other either by a side or
a corner form a ``connected region." Each charged track
reconstructed in the CLEO tracking system~\cite{Peterson} is
projected into the RICH and matched to the closest connected region.
If the matching distance between the track projection and the center
of the connected region is reasonably small and the total pulse
height of the connected region sufficiently high, we associate this
group of hits with the track. Local pulse height maxima in the
remaining connected regions, so-called ``bumps," are taken as seeds
for Cherenkov photons. We allow the pulse height maxima to touch
each other by corners if the pulse height in the two neighboring
pads is small relative to both bump heights. Hits adjacent to the
bumps on the sides are assigned to them in order of decreasing bump
pulse height.

To estimate the position of the photon conversion point we use the
center-of-gravity method corrected for the bias towards the central
pad. For many Cherenkov photons we are able to detect induced charge
in only one pad. Since the pad dimensions are about
$8\times8$~mm$^2$, the position resolution in this case is
$8\:\mbox{mm}/\sqrt{12}=2.3\:\mbox{mm}$. For charged track
intersections, which induce significant charge in many pads, the
position resolution is $0.76$~mm. The position resolution for
Cherenkov photons which generate multiple pad hits is somewhere in
between these two values. In any case, the photon position error is
not a significant contribution to the Cherenkov angle resolution
(see below).

\subsection{Corrections to the Track Direction}

The resolution of the CLEO tracking system is very good in the
bending view (the magnetic field is solenoidal in
CLEO)~\cite{Peterson}. The track position and inclination angle
along the beam axis is measured less precisely. The r.m.s. of the
observed RICH hit residual is 1.7 mm. Since the RICH hit position
resolution is 0.76 mm as measured by the residual in the
perpendicular direction, the RICH can clearly help in pinning down
the track trajectory. This, in turn, improves Cherenkov resolution,
especially for the flat radiators for which we observe only half of
the Cherenkov image and thus are quite sensitive to the tracking
error. The improvement is as much as 50\% in some parts of the
detector.

\subsection{Reconstruction of the Cherenkov Angle}

Given the measured position of the Cherenkov photon conversion point
in the RICH, the charged track direction and its intersection point
with the LiF radiator, we calculate a Cherenkov angle for each
photon-track combination ($\theta_\gamma$). We use the formalism
outlined by Ypsilantis and S\'{e}guinot \cite{t+j}, except that we
adopt a numerical method to find the solution to the equation for
the photon direction, instead of simplifying it to a fourth-order
polynomial. The latter would allow an analytical solution, but at
the expense of introducing an additional source of error.
Furthermore, using our numerical method, we calculate derivatives of
the Cherenkov angle with respect to the measured quantities which
allows us to propagate the detector errors and the chromatic
dispersion to obtain an expected Cherenkov photon resolution for
each photon independently ($\sigma_\theta$). This is useful since
the Cherenkov angle resolution varies significantly even within one
Cherenkov image. We use these estimated errors when calculating
particle ID likelihoods and use them to weight each photon when
measuring the average Cherenkov angle for a track.

\subsection{Performance on Bhabha Events}

\begin{figure}[tb]
  \vspace{-.28cm}
  \begin{center}\hspace{-.3 cm}
    \epsfxsize2.48in \epsffile{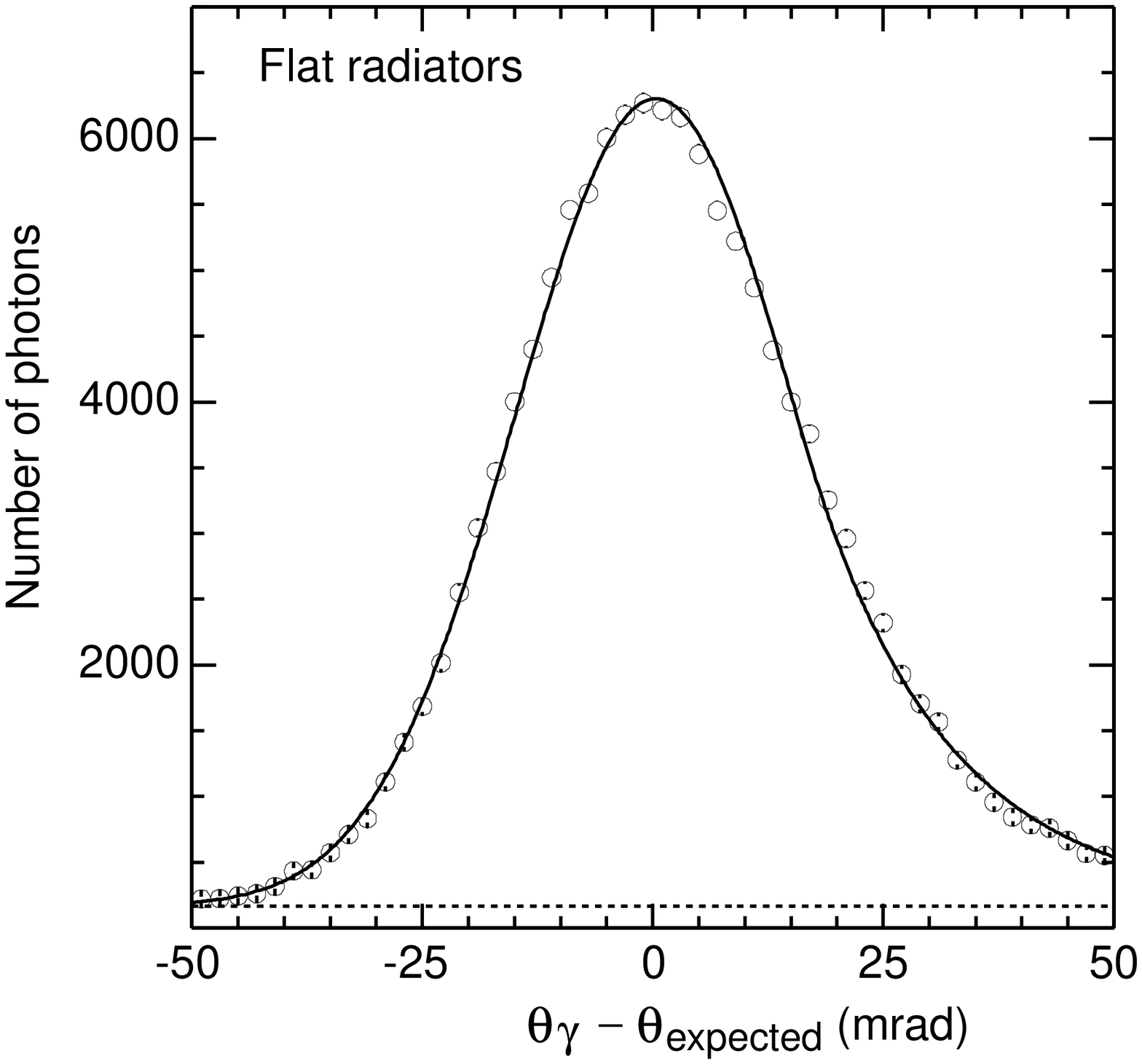}
  \end{center}
  \vspace{0.2cm}
  \begin{center}
    \epsfxsize2.48in \epsffile{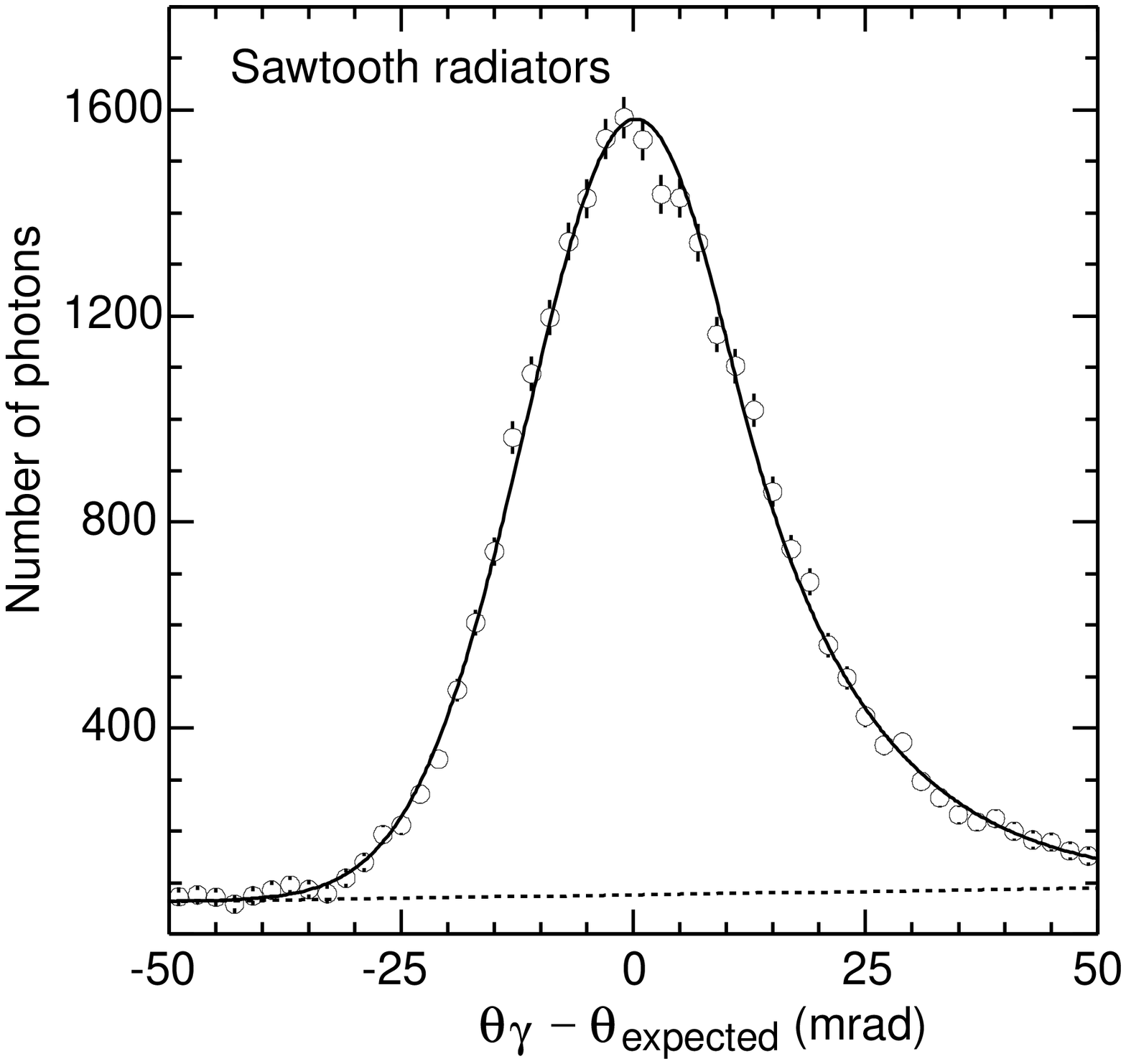}
  \end{center}
  \vspace{.001cm}
  \caption{\label{fig:single_photon}
    The measured minus expected
    Cherenkov angle for each photon detected in Bhabha events,
    (top) for plane radiators and (bottom) for sawtooth radiators.
    The curves are fits to a special line shape function (see text),
    while the lines are fits to a background polynomial.}
\vspace{-0.007cm}
\end{figure}

We first view the physics performance on the simplest type of
events, Bhabha events, and then subsequently in hadronic events. The
distribution of Cherenkov angles measured for each photon in Bhabha
events is shown in Fig.~\ref{fig:single_photon}.

We note that Bhabha events have very low multiplicity compared with
our normal hadronic events. They have two charged tracks present
while the hadronic events have an average charged multiplicity of
approximately 10 in CLEO III. In addition, the hadronic events have
on the average 10 photons, mainly from $\pi^o$ decays. All of these
particles can interact in the calorimeter and the splash-back can
hit the RICH photon detector.

The Cherenkov angle spectrum for single photons has an asymmetric
tail and modest background. It is fit with a line-shape similar to
that used when extracting photon signals from electromagnetic
calorimeters~\cite{CBL}. The functional form is
\begin{equation}
P(\theta|\theta_{exp},\sigma_{\theta},\alpha,n)=
\end{equation}
\vspace{-2mm}
\begin{eqnarray*}
&A\cdot{\rm exp}\left[-{1\over 2}\left({{\theta_{exp}-\theta}\over
\sigma_{\theta}}
\right)^2\right]~{{\rm for}~\theta<\theta_{exp}-\alpha\cdot\sigma_{\theta}}\\
&A\cdot{{\left({n\over \alpha}\right)^n e^{-{1\over 2}\alpha^2}
\over \left({{\theta_{exp}-\theta}\over \sigma_{\theta}}+{n\over
\alpha}-\alpha\right)^n}}
~~~~~~~~~{{\rm for}~\theta>\theta_{exp}-\alpha\cdot\sigma_{\theta}},\\
&A^{-1}\equiv \sigma_{\theta} \left[{n\over \alpha}{1\over
{n-1}}e^{-{1\over 2}\alpha^2} +\sqrt{\pi\over 2}\left(1+{\rm
erf}\left({\alpha\over\sqrt{2}}\right) \right)\right].
\end{eqnarray*}
Here $\theta$ is the measured angle, $\theta_{exp}$ is the ``true''
(or most likely) angle and $\sigma_{\theta}$ is the angular
resolution. To use this formula, the parameter $n$ is fixed to value
of about 5.

\begin{figure}[tb]
  \vspace{-1.1cm}
  \begin{center}
    \epsfxsize 2.6in \epsffile{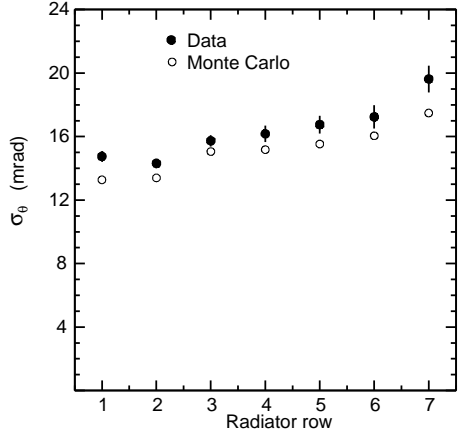}
  \end{center}
  \vspace{-0.007cm}
  \caption{\label{fig:bhares_ang}
    The values of the Cherenkov angular resolution
    for single photons for data compared with Monte Carlo simulation
    as a function of radiator ring number.
    The four-rings of sawtooth radiators are in rings numbered 1 and 2,
    at the center of the detector.}
  \vspace{-0.007cm}
\end{figure}

The data in Fig.~\ref{fig:single_photon} are fit using this signal
shape plus a polynomial background function. We compare the results
of these fits for the resolution parameter $\sigma_{\theta}$ as a
function of radiator ring\footnote{Effectively,
    this shows the dependence of the resolution on polar angle.}
for data and Monte Carlo simulation in Fig.~\ref{fig:bhares_ang}.
Here we use the symmetry of the detector about the center to map two
full physical radiator rings into a single ring number, with ring 1
being closest to the middle. The single photon resolution averaged
over the detector solid angles are 14.7 mr for the flat radiator and
12.2 mr for the sawtooth.

\begin{figure} [tb]
\vspace{-.9cm} \centerline{
    \epsfxsize 2.5in \epsffile{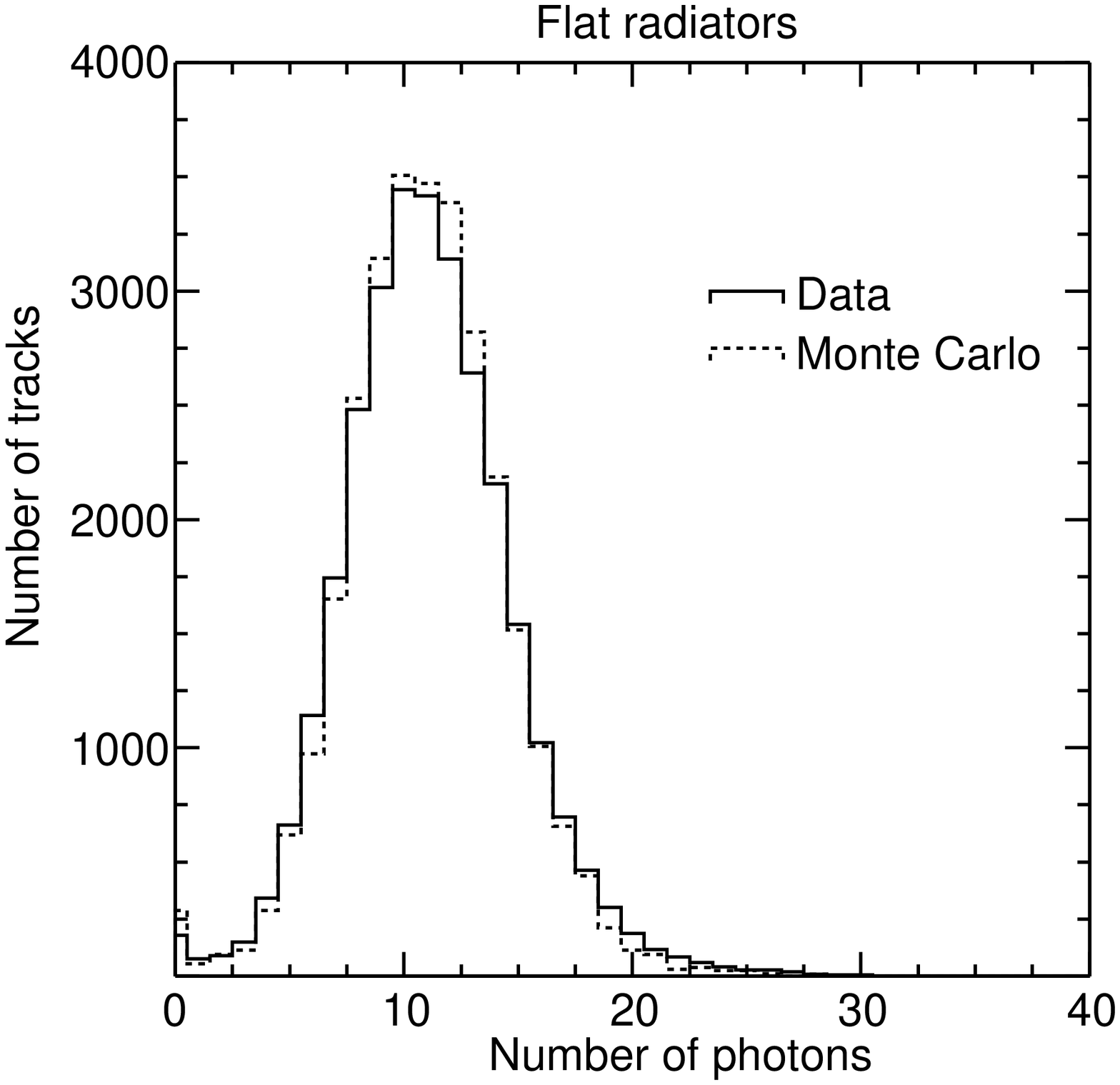}
    } \vspace{-0.7cm}
\centerline{
    \epsfxsize 2.5in \epsffile{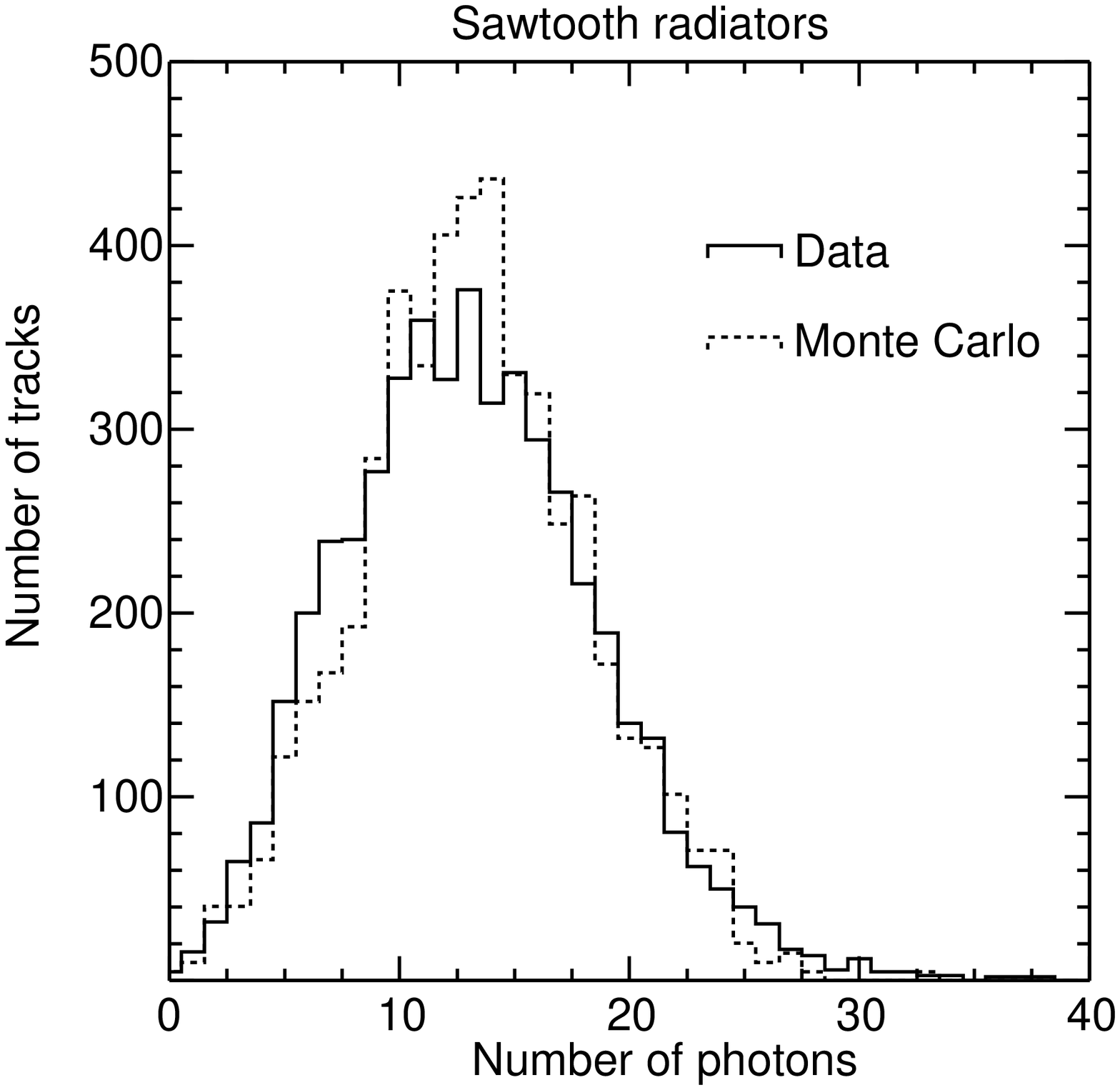}
    }
\vspace{-.3cm} \caption{\label{fig:ngamma} The number of photons
detected on Bhabha tracks (top) for plane radiators and (bottom) for
sawtooth radiators. The dashed lines are predictions of the Monte
Carlo simulation.} \vspace{0.2cm}
\end{figure}
\begin{figure} [tb]
\vspace{.6cm}
\centerline{\hspace{1in}
    \epsfxsize 1.3in \epsffile{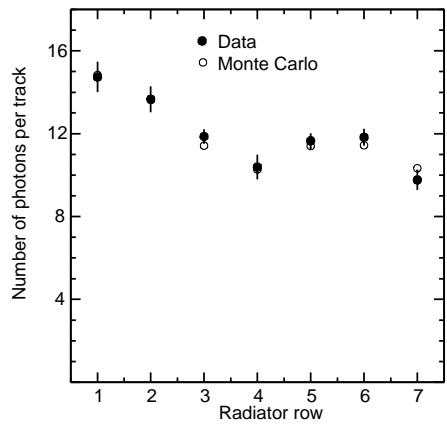}
    }
\vspace{.2cm} \caption{\label{fig:bha_photon} The number of photons
as a function of radiator row for Bhabha events. (Sawtooth radiators
are in rings 1 and 2.)} \vspace{-0.007cm}
\end{figure}

The number of photons per track within $\pm 3\sigma$ of the expected
Cherenkov angle for each photon is shown in Fig.~\ref{fig:ngamma}
and shown as a function of radiator row in
Fig.~\ref{fig:bha_photon}. Averaged over the detector, and
subtracting the background, we have a mean number of 10.6 photons
with the flat radiators and 11.9 using the sawtooth radiators.

\begin{figure} [tb]
\vspace{1cm} \centerline{\hspace{1in} \epsfxsize .922in
\epsffile{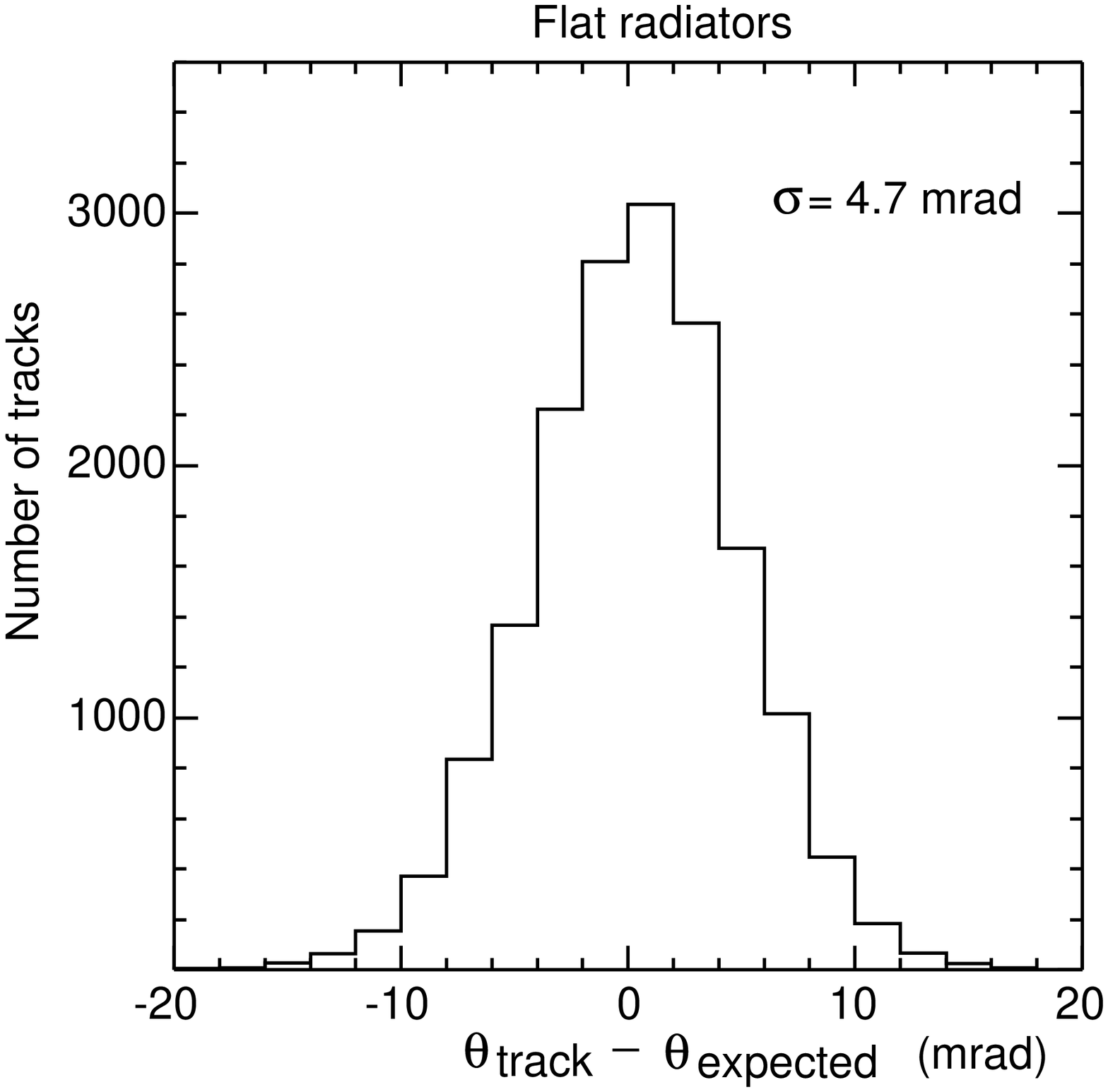} } \vspace{-0.4cm} \centerline{
\epsfxsize 2.65in\epsffile{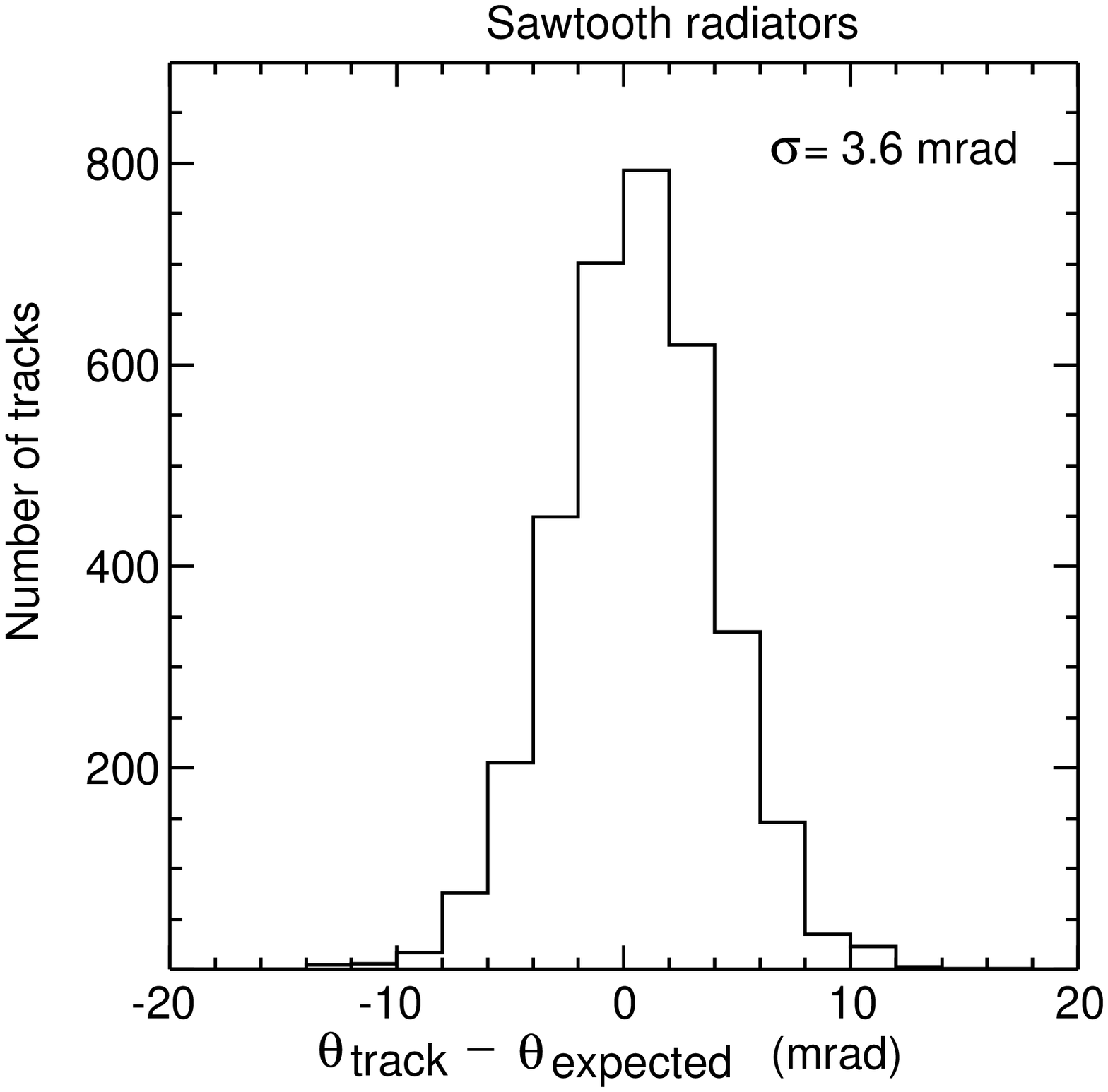} } \vspace{-.2cm}
\caption{\label{fig:trk_res}
    Measured Cherenkov angle for each track in Bhabha events,
   (top) for plane radiators and (bottom) for sawtooth radiators.}
\vspace{-0.007cm}
\end{figure}

The resolution per track is obtained by taking a slice within $\pm
3\sigma$ of the expected Cherenkov angle for each photon and forming
an average weighted by $1/\sigma^2_{\theta}$. These track angles are
shown in Fig.~\ref{fig:trk_res}.

\begin{figure} [tb]
\centerline{\hspace{.5in}
    \epsfxsize 1.3in \epsffile{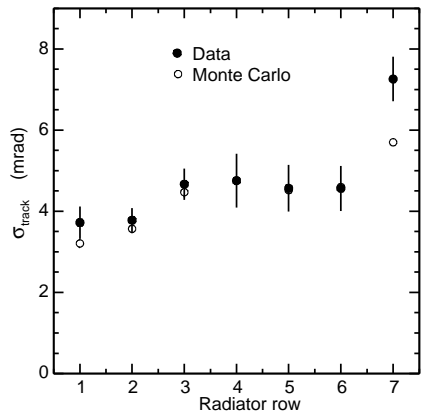}
    }
\vspace{.2cm} \caption{\label{fig:bha_track}
    Cherenkov angle resolutions per track
    as a function of radiator row for Bhabha events. (Sawtooth
radiators are in rings 1 and 2.)
     }
\vspace{-0.007cm}
\end{figure}

The r.m.s. spreads of these distributions are identified as the
track resolutions. We obtain 4.7 mr for the flat radiators and 3.6
mr for the sawtooth. The resolutions as a function of radiator row
are shown in Fig.~\ref{fig:bha_track}.

\begin{figure}[tb]
  \vspace*{-0.008cm}
  \begin{center}
     \hspace{0.4in}
    \epsfxsize 2.5in \epsffile{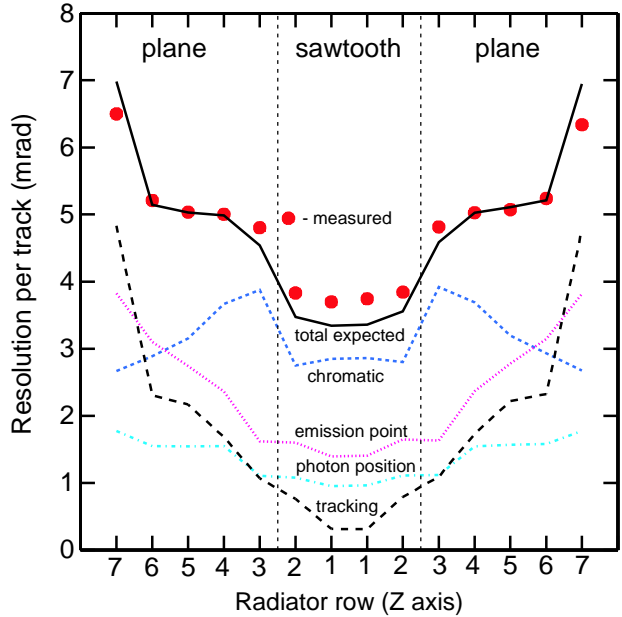}
  \end{center}
  \vspace*{0.2cm}
  \caption{\label{fig:res}
    Different components contributing to the Cherenkov angle
    resolution per track shown as a function of radiator row for Bhabha
    events. The points are data and the solid line is the sum of
    the predicted resolution from each of the individual components
    indicated on the figure.}
  \vspace*{0.1cm}
\end{figure}

The Cherenkov angular resolution is comprised of several different
components. These include: error on the location of the photon
emission point, the chromatic dispersion, the position error in the
reconstruction of the detected photons, and finally the error on
determining the charged track's direction and position. These
components are compared with the data in Fig.~\ref{fig:res}.

\subsection{Performance on Hadronic Events in CLEO III}

\begin{figure}[tb]
  \vspace*{-0.008cm}
  \begin{center}

    \epsfxsize 2.5in \epsffile{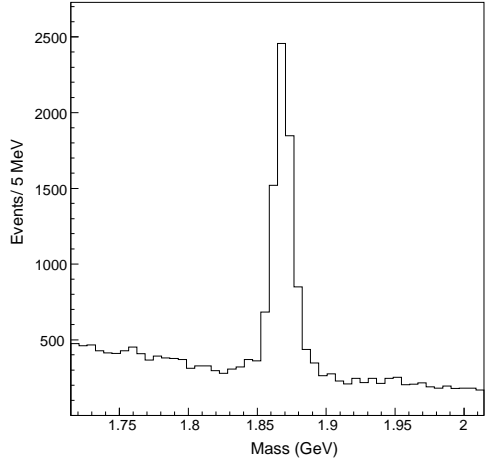}
  \end{center}
 \vspace*{0.2cm}
  \caption{\label{D0mass}
    The $K^{\mp}\pi^{\pm}$ mass after selecting events with
   $K^-\pi^+\pi^+$ - $K^-\pi^+$ mass differences close to the known $D^*-D$ mass difference.}
\end{figure}

To resolve overlaps between Cherenkov images for different tracks we
find the most likely mass hypotheses. Photons that match the most
hypothesis within $\pm3\sigma$ are then removed from consideration
for the other tracks. To study the RICH performance in hadronic
events in CLEO III\footnote{Until now CLEO-c has been running below
threshold for the production of $D^*$'s.} we use $D^{*+}\to\pi^+
D^0$, $D^0\to K^-\pi^+$ events. The charge of the slow pion in the
$D^{*+}$ decay is opposite to the kaon charge in subsequent $D^0$
decay. Therefore, the kaon and pion in the $D^0$ decay can be
identified without use of the RICH detector. The effect of the small
combinatorial background is eliminated by fitting the $D^0$ mass
peak in the $K^-\pi^+$ mass distribution to obtain the number of
signal events for each momentum bin. The $K^-\pi^+$ invariant mass
distribution selected by requiring that the $K^-\pi^+\pi^+$ -
$K^-\pi^+$ mass difference be within 2.5 r.m.s. widths of the known
mass difference is shown in Fig.~\ref{D0mass}. Here both the kaon
and the pion are required to have momenta $>0.6$ GeV/c.

\begin{figure} [tb]
\vspace{-.008cm} \centerline{\hspace{.5in}
    \epsfxsize1.5in\epsffile{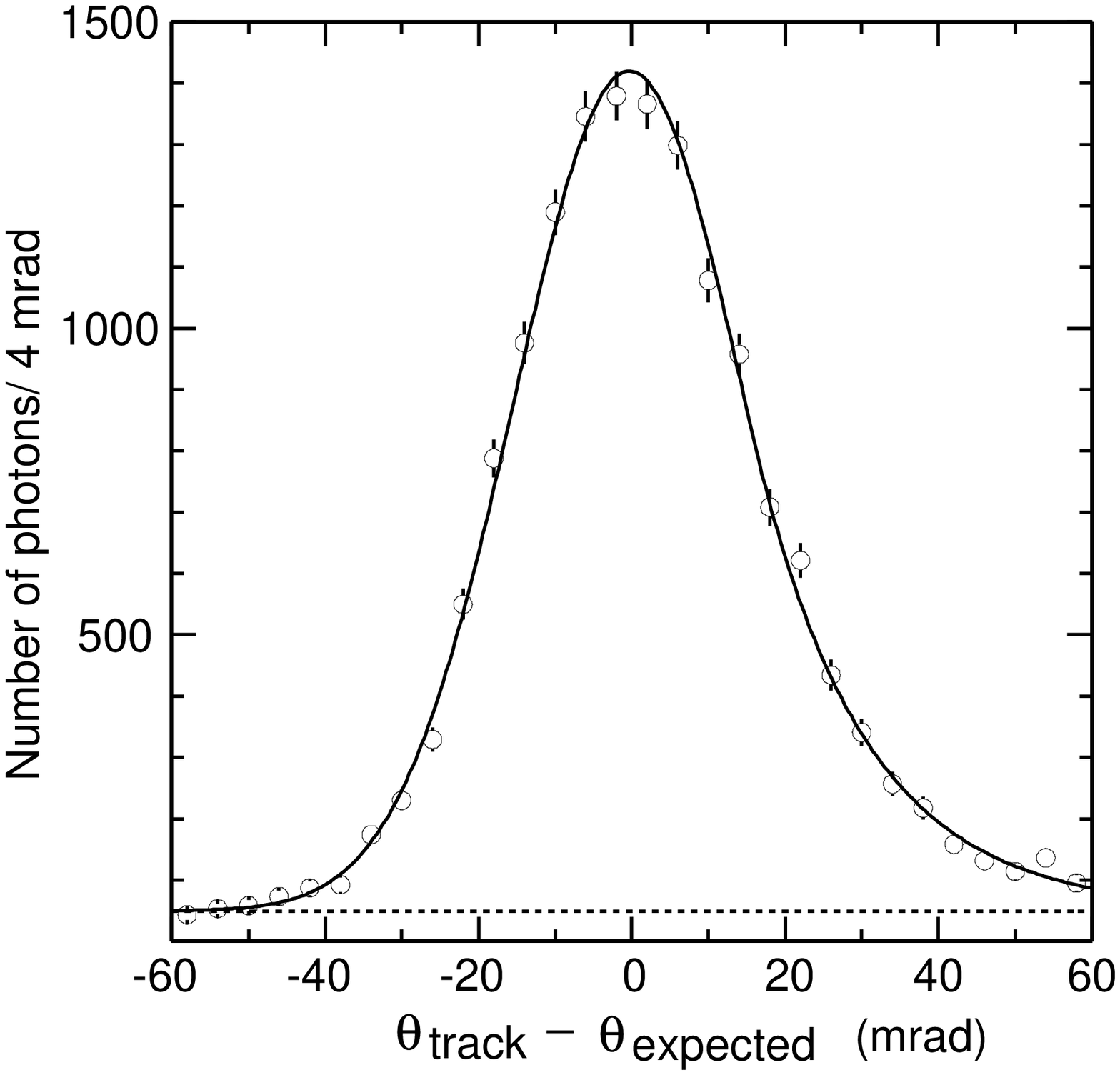}
    }
\vspace{0.2cm} \centerline{\hspace{.4in}
    \epsfxsize1.6in\epsffile{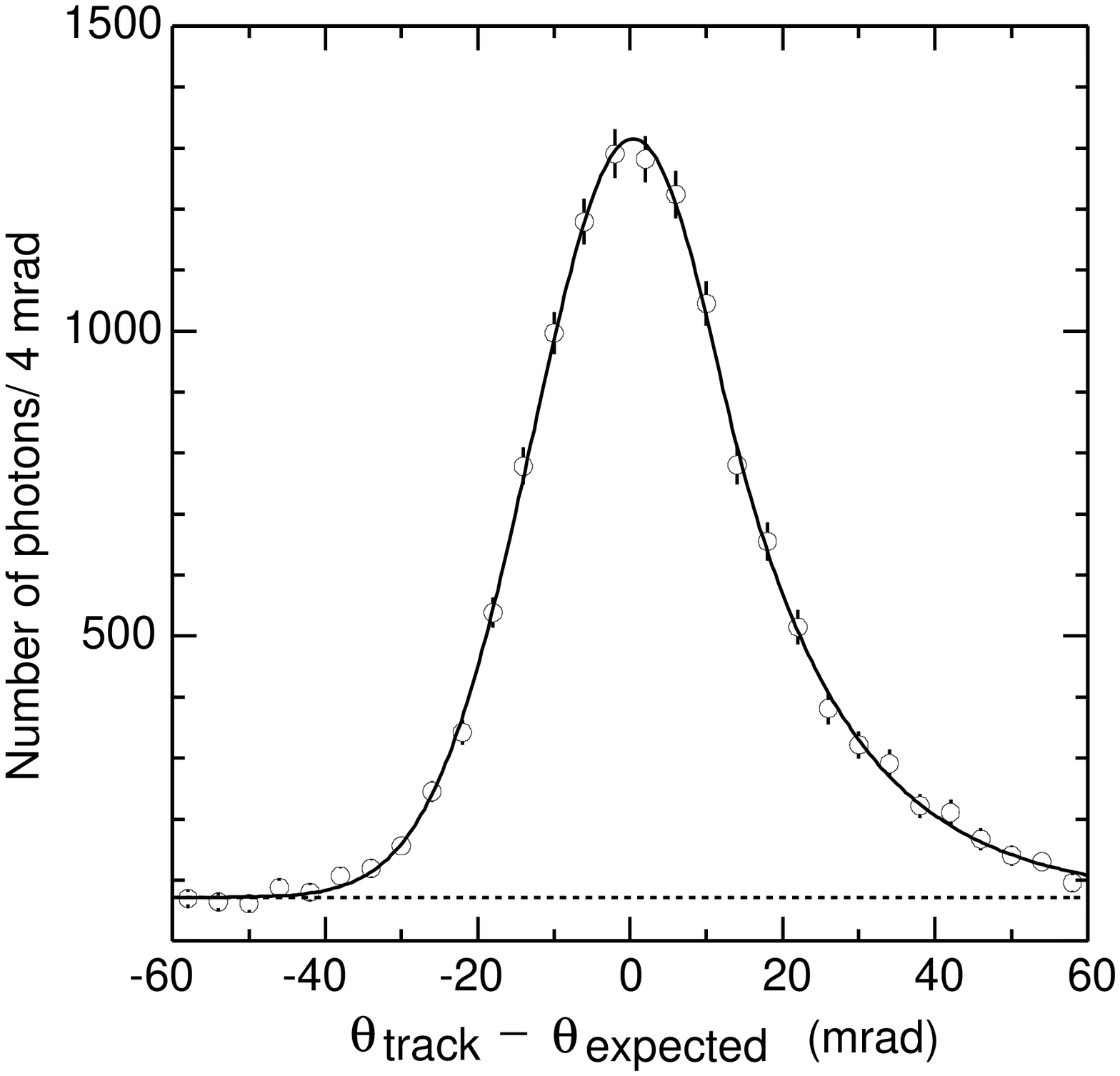}
    }
\vspace{-.007cm} \caption{\label{fig:single_photon_hadrons}
    The measured minus expected
    Cherenkov angle for each photon detected in hadronic events,
    (top) for plane radiators and (bottom) for sawtooth radiators.
    The curves are fits to special line shape function (see text),
    while the lines are fits to a background polynomial.}
\vspace{-0.007cm}
\end{figure}

Single-photon Cherenkov angle distributions obtained on such
identified kaons with the momentum above 0.7 GeV/c are plotted in
Fig.~\ref{fig:single_photon_hadrons}. Averaged over all radiators,
the single-photon resolution is 13.2 mr and 15.1 mr for sawtooth and
flat radiators respectively. The background fraction within
$\pm3\sigma$ of the expected value is 12.8\% and 8.4\%. The
background-subtracted mean photon yield is 11.8 and 9.6. Finally the
per-track Cherenkov angle resolution is 3.7 mr and 4.9 mr.

\subsection{Particle ID Likelihoods}

For parts of the Cherenkov image for the sawtooth radiator, and for
tracks intersecting more than one radiator there are some optical
path ambiguities that impact the Cherenkov angle calculations. In
the previous section we bypassed this problem by selecting the
optical path that produces the closest Cherenkov angle to the
expected one ($\theta^h_{exp}$) for the given particle hypothesis
($h$). There is some loss of information in this procedure,
therefore, we use the likelihood method to perform particle
identification instead of the per-track average angle. The
likelihood method weights each possible optical path by the optical
probability ($P_{opt}$), which includes length of the radiation path
and the refraction probabilities obtained by the inverse ray tracing
method:
\begin{eqnarray*}
L_h = \prod_{j=1}^{No.\,of\,\gamma s} \left\{
P_{{background}} +  \hbox{\qquad\qquad} \phantom{\sum_{opt} P_{opt}^j} \right. \\
   \left. \sum_{opt} P_{opt}^j
    \cdot P_{{signal}}\left( \theta_\gamma^{opt,\,j} |
         \theta_{exp}^h,\,\sigma_\theta^{opt,\,j} \right) \right\}
\end{eqnarray*}
where, $L_h$ is the likelihood for the particle hypothesis $h$ ($e$,
$\mu$, $\pi$, $K$ or $p$), $P_{{background}}$ is the background
probability approximated by a constant and $P_{{signal}}$ is the
signal probability given by the line-shape defined previously. In
principle, the likelihood could include all hits in the detector. In
practice, there is no point in inspecting hits which are far away
from the regions where photons are expected for at least one of the
considered hypotheses (we use $\pm5\sigma$ cut-off).

An arbitrary scale factor in the likelihood definition cancels when
we consider likelihood ratios for two different hypotheses. The
likelihood conveniently folds in information about values of the
Cherenkov angles and the photon yield for each hypothesis. For well
separated hypotheses (typically at lower momenta)  the photon yield
that provides some discrimination.  Since our likelihood definition
does not know about the radiation momentum threshold, the likelihood
ratio method can be only used when both hypotheses are sufficiently
above the thresholds. When one hypothesis is below the radiation
threshold we use the value of the likelihood for the hypothesis
above the threshold to perform the discrimination.

\begin{figure} [tb]
\vspace{-1.3cm} \centerline{
    \epsfxsize3.0in\epsffile{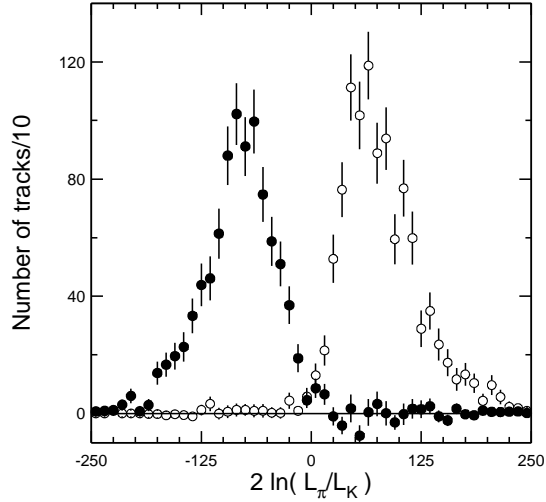}}
\vspace{-.07cm} \caption{\label{fig:likelihood_ratio}
      Distribution of $2\ln\left(L_\pi/L_K\right)$ $\sim\chi^2_K-\chi^2_\pi$
      for 1.0-1.5 GeV/c kaons (filled)
      and pions (open) identified with the $D^*$ method.
       }
\vspace{-0.007cm}
\end{figure}

\begin{figure} [bt]
\vspace{.2cm} \centerline{
    \epsfxsize2.8in\epsffile{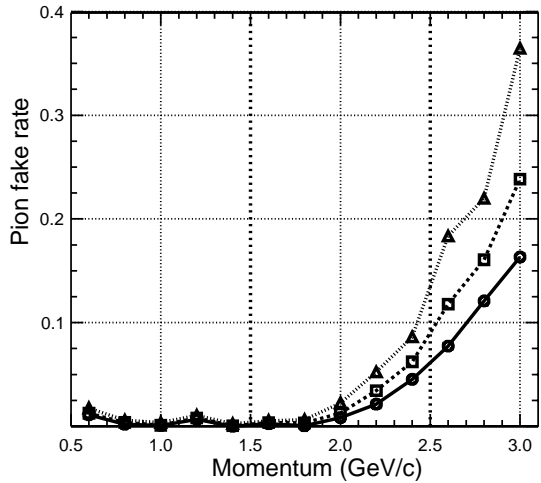}}
\vspace{-.007cm} \caption{\label{fig:id_vs_momentum}
      Pion fake rate as a function of particle momentum for
      kaon efficiency of 80\%\ (circles), 85\%\ (squares) and
      90\%\ (triangles). }
\vspace{-0.007cm}
\end{figure}

The distribution of the $2\ln\left(L_\pi/L_K\right)$, is expected to
behave as the difference $\chi^2_K-\chi^2_\pi$. This $\chi^2$
difference obtained for 1.0-1.5 GeV/c kaons and pions identified
with the $D^*$ method is plotted in Fig.~\ref{fig:likelihood_ratio}.
Cuts at different values of this variable produce identification
with different efficiency and fake rate. Pion fake rates for
different values of kaon identification efficiency are plotted as a
function of particle momentum in Fig.~\ref{fig:id_vs_momentum}. Here
when the fake rates get below a few percent there are other
systematic effects that enter. For example, doubly Cabibbo
suppressed decays where the $D^o$ decays into a $K^+\pi^-$ rather
than a $K^-\pi^+$ have a relative branching fraction of 0.4\%
\cite{PDG}.

\subsection{Efficiency and Fake Rates in CLEO-c}

Here we use 180 pb$^{-1}$ integrated luminosity of CLEO-c data
produced in $e^+e^-$ collisions and recorded at the $\psi''$
resonance (3.770 GeV).  At this energy, the events consist of a
mixture of pure $D^+D^-$, $D^o\overline{D}^o$, three-flavor
continuum event and $\gamma\psi'$ events. There may also be small
amounts of $\tau^+\tau^-$ pairs and two-photon events.

\begin{figure}[htb]
\centerline{ \epsfxsize=2.5in \epsffile{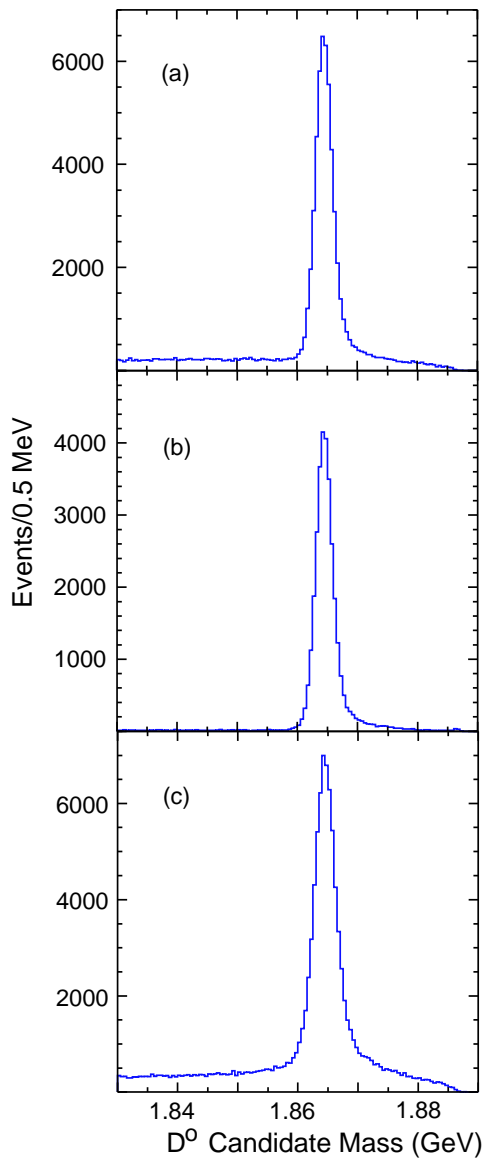} }
\caption{Beam constrained mass distributions for (a)$
K^-\pi^-\pi^+\pi^+$, (b) $K^{-}\pi^{+}$, and (c)
$K^{-}\pi^{+}\pi^{o}$ candidates.} \label{d0-signal}
\end{figure}

\begin{figure}[htb]
\centerline{ \epsfxsize=2.5in \epsffile{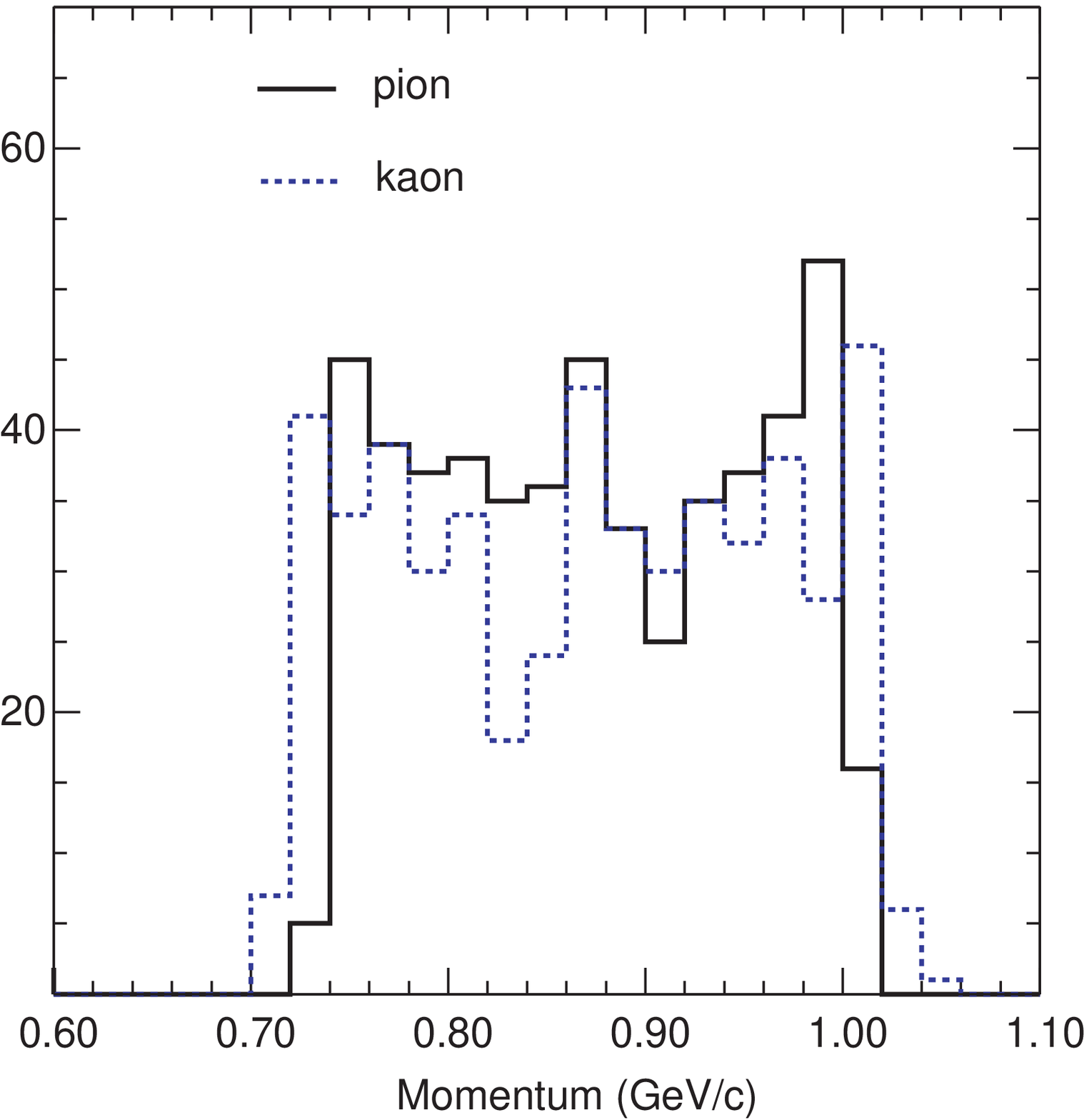} } \caption{The kaon
and pion momentum spectra from $D\to K^{\pm}\pi^{\mp}$ decays.}
\label{mom}
\end{figure}

In this study we select events containing at least one neutral $D$
candidate in the following decays $D^o\to K^-\pi^-\pi^+\pi^+$,
$D^o\to K^{-}\pi^{+}$ and $D^o\to K^{-}\pi^{+}\pi^{o}$. (Charge
conjugate modes are also used.) Event candidates in these modes are
mostly signal with low background fractions. The $D^o$ candidate
invariant mass plots are shown in Fig.~\ref{d0-signal}. These mass
plots are constructed by selecting decays where the sum of the
measured energies is close to the electron beam energy and then
using the measured beam energy to form the mass \cite{CLEOAbs}. We
use this sample to look for two oppositely charged tracks
 present in the other side of the event not containing the tagged
$D$. We then further select events where the $\overline{D}^o$
 decays into
$K^{\pm}\pi^{\mp}$. The momentum spectra of the kaon and the pion
from this decay when the $D$ is produced on the $\psi(3770)$ is
shown in Fig.~\ref{mom}.

We start by describing the analysis of the joint decays
$\overline{D}^o\to K^+\pi^-\pi^+\pi^-$ and $D^o\to
K^{\pm}\pi^{\mp}$. Here we define the decay into $K^-\pi^+$ as
``right" sign and the decay into $K^+\pi^-$ as ``wrong" sign. In
this case the wrong sign decays could result from one of three
sources: background, doubly Cabibbo suppressed decays or $D^o -
\overline{D}^o$ mixing. We note that current measures of mixing
limit it to $\sim <$0.045\% \cite{Bellemix}, while current measures
of doubly Cabibbo suppressed decays are larger. For example, the
modes $K^-\pi^+$, $K^-\pi^-\pi^+\pi^+$ and $K^-\pi^+\pi^o$ have
rates of 0.35\%, 0.42\%, and 0.43\%, respectively.

Fully reconstructed single tags for the $K^+\pi^-\pi^+\pi^-$ mode
(and its charge conjugate, which will not be explicitly mentioned in
what follows) are reconstructed using the beam constrained mass. We
use a 2$\sigma$ cut on $\Delta E$ and require the mass to be between
1.86 and 1.870 GeV. We then form a double tag event using either
right or wrong sign $K\pi$ decays. We use a tight cut on the $K\pi$
of  $|\Delta E|$, the difference between the measured energy and the
beam energy within $2\sigma$, where $\sigma$ is the r.m.s. of the
distribution. Since we need to use the RICH we impose a cut that
both tracks be within $|\cos(\theta)|< 0.81$, where $\theta$ is the
angle of the track with respect to the beam line.

At first we do not use any RICH identification on the $K\pi$. Using
a total of 180 pb$^{-1}$ we have 1158 such events where the kaon and
pion are both in the RICH acceptance that give the right sign and
the same sample yields 642 wrong sign events.

We now make three separate analyses: one where we identify only the
kaon, one where we identify only the pion and one where we identify
both the kaon and the pion. In the latter case we insist that there
is significant discrimination in both cases or we do not accept the
event.

Since the fake rates will be near 2\%, the probability of getting
both the kaon and the pion wrong is $\sim 4\times 10^{-4}$, so that
asking for a double identification is sufficient to ensure that we
are getting the right answer for this level of tags.

The results are shown in Table~\ref{K3pi-Kpi}. We find 1158 right
sign events without using any particle identification, 970 right
sign events with both particles identified, 15 wrong sign events
with both particles identified, 25 events with only the kaon
identified incorrectly and 36 events with only the pion identified
incorrectly.

\begin{table}[htb]
\begin{center}
\begin{tabular}{lrrrrrrrr}
\multicolumn{1}{l}{Mode}&\multicolumn{2}{c}{No ID} &
\multicolumn{2}{c}{Single $K$} & \multicolumn{2}{c}{Single $\pi$} &
\multicolumn{2}{c}{Double ID} \\
&(RS) & (WS)  &(RS) & (WS) &  (RS) & (WS) &  (RS) &   (WS)
\\\hline $K^-\pi^-\pi^+\pi^+$;  $K^{\pm}\pi^{\mp}$
&1158&642&1021&25&105&36
 & 970&  15 \\
$K^-\pi^+$;  $K^{\pm}\pi^{\mp}$ &307&199&273&3&295&6
 & 264&  0\\
 $K^-\pi^+\pi^o$;  $K^{\pm}\pi^{\mp}$ & 1524 &897&1350&14 &1444 &22
 & 1287  &  4 \\\hline
 Sum  & 2989&1738&2644 &  42  & 2842 &64
 & 2521 & 19  \\
\hline\hline
\end{tabular}
\end{center}
\caption{Results of RICH identification on double tag events. RS
indicates right sign and WS indicates wrong sign events.}
\label{K3pi-Kpi}
\end{table}

The 15 doubly identified wrong sign events are the combination of
background, doubly Cabbibo suppressed decays and mixing. They
correspond to a rate of these events of (1.5$\pm$0.4)\%, consistent
with them all being doubly Cabibbo suppressed decays, but somewhat
larger \cite{PDG}. We subtract these events after correcting for the
efficiency for the wrong sign candidates. This gives us a kaon fake
rate of 1.1\%, with an efficiency for events in the RICH of 88.5\%
and a pion fake rate of 3.7\%, with an efficiency of 93.7\%. We note
that background is likely to be absent or small in these double tag
events, but need to make a quantitative assessment.

\begin{figure}[htb]
\centerline{ \epsfxsize=5.0in \epsffile{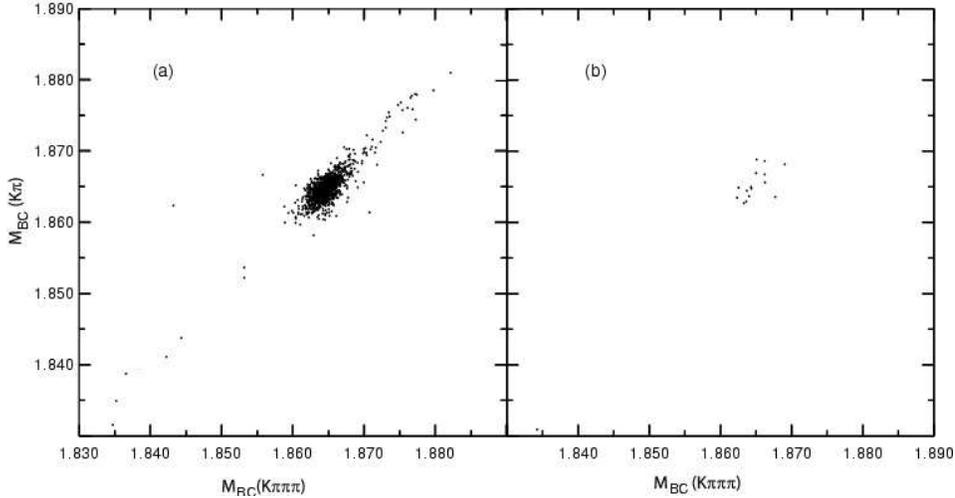} } \caption{
Beam constrained mass for $K^-\pi^-\pi^+\pi^+$  versus
$K^{\pm}\pi^{\mp}$; (a) Right sign events, (b) wrong sign events.}
\label{Kpi-K3pi}
\end{figure}

To ascertain the background level we plot the beam constrained mass
of the $K^-\pi^+\pi^+\pi^-$ tag versus the $K^{\pm}\pi^{\mp}$ tag in
Fig.~\ref{Kpi-K3pi}. Events outside of the region where both masses
are greater than 1.858 GeV are background. There is only one
background event in the wrong sign plot indicating that the
background is much less than one event.

\begin{figure}[htb]
\centerline{ \epsfxsize=4.0in \epsffile{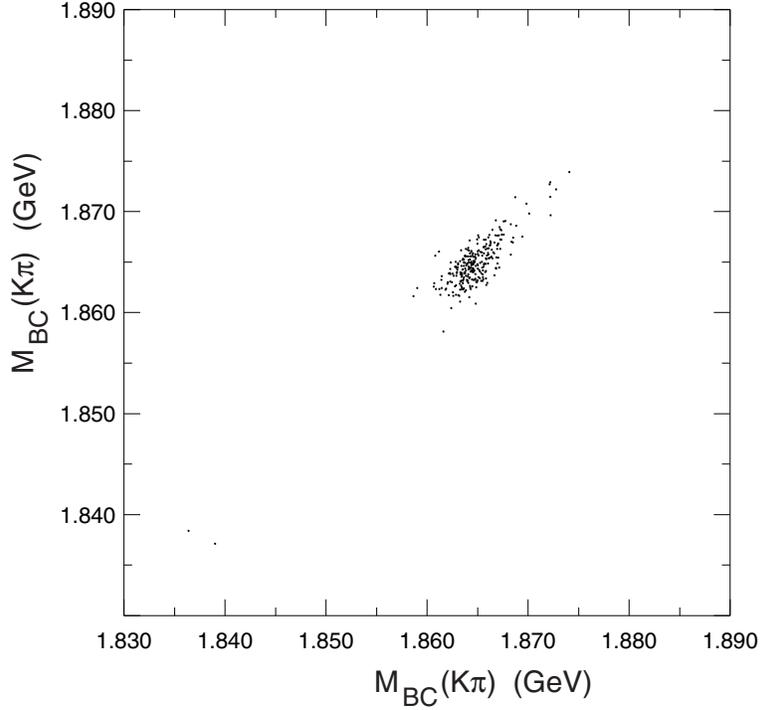} } \caption{
Beam constrained mass for $K^-\pi^+$  versus $K^{\pm}\pi^{\mp}$.}
\label{Kpi-Kpi}
\end{figure}

We now consider the case where both neutral $D$'s decay into
$K^{\pm}\pi^{\mp}$. One difference in this case with other cases is
that doubly Cabbibo decays are forbidden due to Bose-Einstein
statistics \cite{Bigi-Sanda}. The results are presented in
Table~\ref{K3pi-Kpi}.  Here there is wrong sign doubly identified
decay. This could be due to (a) background (b) $D^o$ mixing or (c)
where both particles were incorrectly identified. We find zero
events in the wrong sign doubly identified decay. We plot the beam
constrained mass of the $K^-\pi^+$ tag versus the $K^{\pm}\pi^{\mp}$
tag in Fig.~\ref{Kpi-Kpi}.

\begin{figure}[htb]
\centerline{ \epsfxsize=5.0in \epsffile{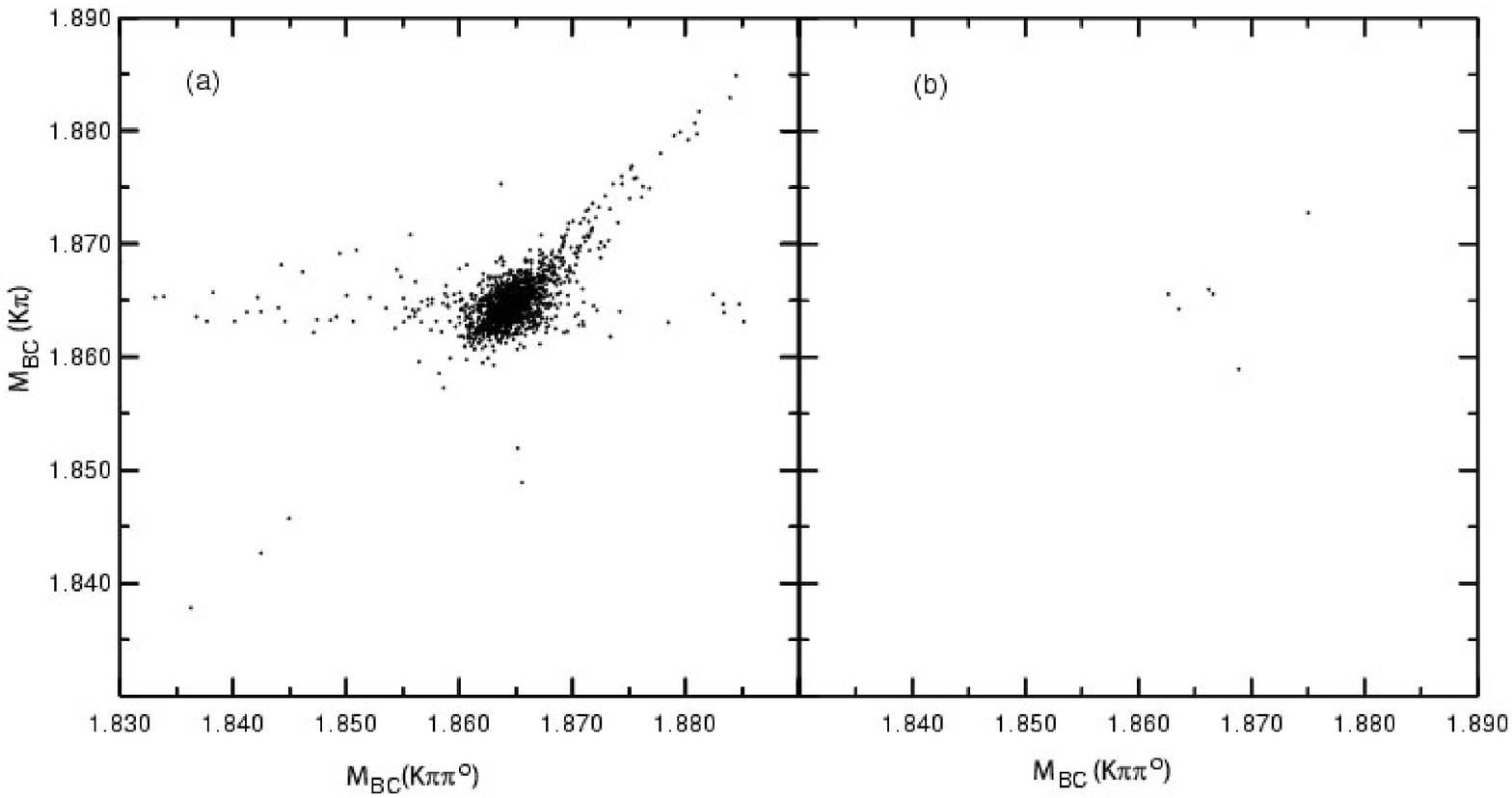} } \caption{
Beam constrained mass for $K^-\pi^+\pi^o$  versus
$K^{\pm}\pi^{\mp}$.} \label{Kpi-Kpipi0}
\end{figure}

Our final mode uses $K^-\pi^+\pi^o$ for the single tag. The results
are also presented in Table~\ref{K3pi-Kpi}. To ascertain the
background level we plot the beam constrained mass of the
$K^-\pi^+\pi^o$ tag versus the $K^{\pm}\pi^{\mp}$ tag in
Fig.~\ref{Kpi-Kpipi0}. Although there appears to be some background
in the right sign plot, the wrong sign shows no evidence of
background.

The 4 doubly identified wrong sign events are the combination of
background, doubly Cabbibo suppressed decays and mixing. They
correspond to a rate of these events of (0.3$\pm$0.2)\%, consistent
with them all being doubly Cabibbo suppressed decays, but somewhat
smaller.

Using all three of these mode combinations we find the rate of pions
faking kaons
 of (1.10$\pm$0.37)\%, with a pion efficiency for events
 in the RICH of (94.5$\pm$0.4)\%. The rate of kaons faking pions is
(2.47$\pm$0.38) \%, with a kaon efficiency for events
 in the RICH of (88.4$\pm$0.6)\%. The lower kaon efficiency arises because a significantly
 larger fraction
 of kaons than pions decay in this momentum region. It should be emphasized that
 these values are obtained for the entire running period between
 October, 2003 and February of 2005, and includes all possible system
 effects.

%
\section{ CONCLUSIONS }

We have successfully constructed and operated a large, complex RICH
detector in a particle physics experiment for over five years. The
oxygen level has been kept below a few ppm in the ``expansion
volume" and the TEA photon conversion gas has been kept out,
allowing for the Cherenkov photon yield to remain almost constant
over the running period. We have lost some photon yield with a small
$\sim$5\% failure of electronics chips. One broken wire has caused
an additional 1.7\% loss and does somewhat effect the track
efficiency. The total ``cylindrical" detector thickness measured
perpendicular to the axis is 13\% of a radiation length.

The RICH is used during the normal course of most physics analyses
using a standard set of criteria based on the minimum number of
observed Cherenkov photons, usually 3, and the relative liklihood
that a track is given type, either pion or kaon, for example.

The particle momenta for $B$ meson decay products seen by CLEO III
are less than $2.65$ GeV/c. The detector provides excellent
separation between pions and kaons at and below this cutoff.
Separation between kaons and protons extends to even higher
momentum, where it is used in charm studies. Thus, the physics
performance has met design criteria. The RICH has provided crucially
important particle separation in a number of important physics
analyses including measurements of charmless hadronic two-body $B$
meson decays and the ratio $\Gamma(B\to D K)/\Gamma(B\to D \pi)$
\cite{Kpi}, and measurement of the form-factors in $D^o\to
\pi^-\ell^+\nu$ and $D^o\to K^-\ell^+\nu$ decays \cite{D0ff}.

CLEO is currently making an extensive study charm mesons and
charmonium decays (called CLEO-c~\cite{cleo-c}). For these
measurements the beam energy is lowered and the maximum particle
momenta is about $1.0-1.5$ GeV/c. At these momentum the particle
identification fake rates are at the 1\% level.

\section{ ACKNOWLEDGEMENTS }

The CLEO RICH project was funded primarily by the U. S. National
Science Foundation which we deeply appreciate. We thank both the
National Science Foundation and Department of Energy for supporting
the University groups. We thank the late Tom Ypsilantis and Jacques
S\'{e}guinot for early work on a similar system and for extensive
discussions. Jeff Cherwinka helped with many engineering aspects of
the system. Lee Greenler of PSL laboratories of Univ. of Wisconsin
did much of the mechanical design. We thank Einar Nygard and Bjorn
Sundal of IDEAS for their work on the front-end hybrid design. Paul
Gelling contributed to the electronics infrastructure. We especially
appreciate the efforts of Charles Brown, Lou Buda and Lester
Schmutzer of the Syracuse Physics Dept. machine shop who made many
of the components. We thank Peter Reed, Heather Lane, Dave Smith,
Don Moulton  at Optovac for their hard work during the four years of
crystal production. Ken Powers helped with the plating of the
windows.
 We
thank the accelerator group at CESR for excellent efforts in
supplying luminosity.


\section*{Appendix A. VUV SPECTROPHOTOMETERS}

The CH$_4$-TEA photosensitive gas inside the RICH chambers has an
appreciable photo-absorption cross-section in a narrow VUV band
$\lambda=150\pm15\,$nm. Satisfactory reconstruction of the Cherenkov
cone geometry requires that the windows and the radiator crystals
(especially their top surface), be sufficiently transparent in this
wavelength band. A crystal surface which is poorly polished or
contaminated would reduce significantly the number of photons
emitted from the radiator crystals and subsequently degrade the
Cherenkov angle measurement resolution.

Three VUV spectrophotometers used to measure the transmission of the
crystals were built and placed at the three locations associated
with crystal production and handling.

\subsection*{A.1.  LiF VUV Transmission Spectrophotometer at SMU}

\begin{figure}[htbp]
  \vspace*{-0.001in}
  \begin{center}
    \includegraphics*[height=2.5in]{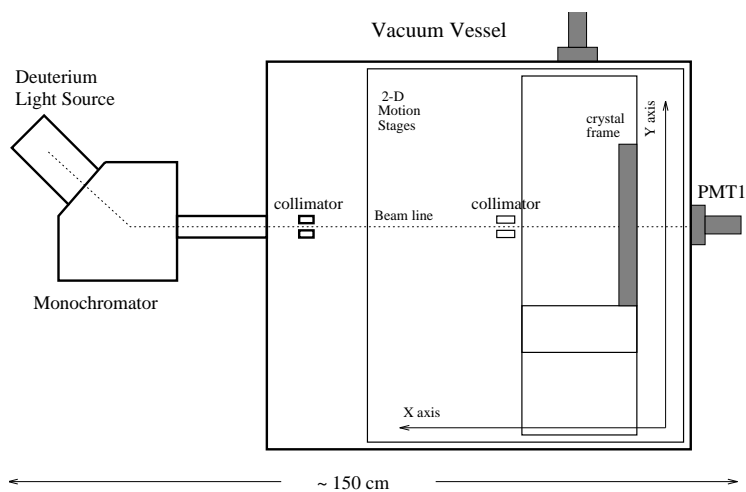}
  \end{center}
  \vspace*{-0.100in}
  \caption{\label{fig:xtal-mono-smu}
        Transmission measurement setup for the LiF planar radiator crystals.
        }
  \vspace*{-0.001in}
\end{figure}

Prior to installation of individual crystals onto the RICH Inner
Cylinder, each crystal's transparency was measured using a specially
constructed VUV spectrophotometer at four wavelengths: $\lambda =
135,\,142.5, \,150\  {\rm and}\ 165\,$nm. The essential components
of this transmission spectrophotometer are shown in
Fig.~\ref{fig:xtal-mono-smu}. A vacuum monochromator with a
deuterium lamp is the VUV light source.  A test box 250~liter vacuum
vessel contains the crystal to be measured, along with a
two-dimensional stage driven by stepper motors. The crystal's top
surface is held perpendicular to the direction of the probe
radiation and collimators produce a beam spot of radius $\sim$1~mm
on the crystal surface. The stage can reliably position a crystal
inside the vessel with a spatial precision of better than $25\,\mu$m
along each of the stage's mutually perpendicular axes. It moves the
crystal perpendicular to the beam direction in a raster style and
periodically stops at discrete points while the transparency
measurement is made. The typical spacing between measurements points
is 1~cm.  For calibration purposes, during each motion along the
width of the crystal, the stage both moves the crystal completely
out of the beam and completely blocks the beam with its frame, so as
to provide measurements along each row of the scan that correspond
to 100\% and 0\% transparency, respectively.

Both, the monochromator and the vacuum vessel, are evacuated by
dedicated turbomolecular vacuum pumps to minimize absorption of the
probe radiation by any residual air or water molecules.  A
photomultiplier tube (PMT), whose front window is coated with an
ultraviolet wave-shifting compound (sodium
salicylate~\cite{Samson}), is attached to the vacuum vessel and
measures the intensity of the transmitted VUV light from the
monochromator after its passage through the crystal.  PMT output
signals are amplified, filtered, digitized and then processed by a
LabVIEW-based program to determine the crystal transparency map in
real time. Data is also written to disk for further off-line
analysis.


The measurement process was necessarily highly automatic and the
control software ran on a PC, which in turn communicated with the
data acquisition system and the stage controllers. The software
controlled the high voltage for the PMT, monitored pressures in the
two vacuum systems (monochromator and test box), and through a
standard GPIB interface adjusted the monochromator grating to the
desired wavelength.  A graphical user interface allowed the user to
set efficiently a wide range of control parameters to measure
radiator crystals in a variety of modes, to measure crystal ingot
test bars, and to perform calibration runs of various types.

Extensive adjustment and calibration of the spectrophotometer system
were performed to minimize systematic errors and to measure overall
system time stability and reproducibility. For example, the
linearity of the electronic readout chain was measured to be much
better than 1\% over the range of readout voltages and transmission
measurements separated in time by more than 1 month of the same LiF
sample show agreement within 1\%. Overall, \mbox{$\simeq 500,000$}
individual LiF transmission measurements were made.

\subsection*{A.2. VUV Spectrophotometer at Syracuse}

A second VUV spectrophotometer similar in design was constructed at
Syracuse University for the purpose of testing the window crystals
as well as the sawtooth radiators.

The window crystals were measured in a manner substantially similar
to that described above for the planar radiators. Transmissions were
measured at three wavelengths (135~nm, 150~nm, and 165~nm), and over
a grid of 30 positions over the surface of the crystal window. For
half-sized windows, there were 15 positions per crystal. The results
are summarized in Section~\ref{sec:xtalwindows}.

In this spectrophotometer, however, systematic uncertainties in the
transmission measurements were reduced using different techniques.
Before the light beam enters the vacuum tank, it is passed through a
chopper which opens and closes at $\sim$40 Hz. The resulting
phototube current appears to be a square wave, with an amplitude
that provides a measure of the light output, while automatically
subtracting stray light and dark current offsets. Also, in general,
the transmission measurement is made by performing
``crystal-in/crystal-out'' measurements: we divide the light output
when the beam passes through a crystal (as registered by the current
in the phototube) by the light output when the crystal is removed
from the beam.  Measuring this ratio attempts to divide out drifts
in the lamp output or gain variations in the PMT over time. Making
this reference measurement frequently means that the measurement
cannot drift appreciably from point-to-point within a scan.
Individual transmission measurements have a ``statistical''
uncertainty of typically 0.5--1.0\% due to averaging the
photocurrent, and a 0.5--1.0\% ``systematic'' uncertainty due to
signal drift over a typical measurement interval.

\begin{figure}[htbp]
  \vspace*{-0.001in}
  \begin{center}
    \includegraphics*[height=3.5in]{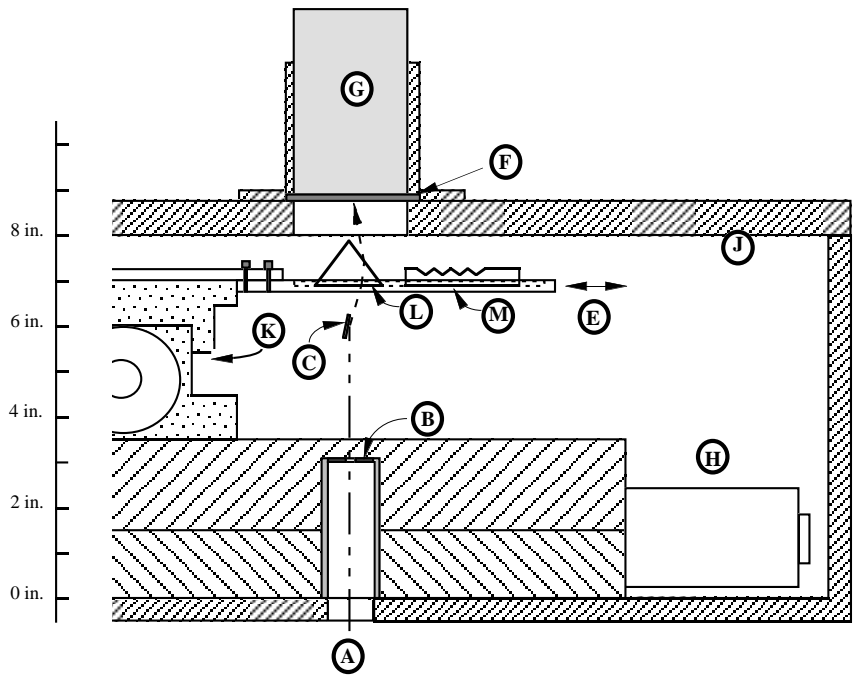}
  \end{center}
  \vspace*{-0.100in}
  \caption{\label{fig:xtal-mono-su}
        Transmission measurement setup for the LiF sawtooth radiator crystals.
Components are: (A) monochromator light beam, (B) collimating slit,
(C) VUV mirror, (E) crystal holder tray mounted on X-Y stage, (F)
sodium salicylate covered glass window, (G) Photomultiplier tube,
(H,K) X,Y-axis linear stage, (J) vacuum tank wall, (L) 42$^\circ$
VUV-polished prism, and (M) sawtooth radiator. (Only four grooves of
the sawtooth radiator are shown, for illustrative purpose.}
  \vspace*{-0.001in}
\end{figure}

The sawtooth radiators required a special technique in order to
measure their transmission. For this, the spectrophotometer is
configured as shown in Fig.~\ref{fig:xtal-mono-su}.
The monochromator light beam enters through the bottom of the vacuum
tank, and passes through a 0.020 inch collimating slit. The pencil
beam has a Gaussian shape with $\sigma=25~\um$. The light reflects
off of an adjustable mirror to produce a 15$^\circ$ deflection of
the beam.  It passes through a 42$^\circ$ VUV-polished prism, which
is mounted on the two-dimensional stage. The light is
wavelength-shifted by the sodium salicylate covered glass window,
and detected in the photomultiplier tube.

The angular deflection of 15$^\circ$ is set so as to probe the
transmission at the smallest possible angle that is not totally
internally reflected and that accrues appreciable signal. Although
the Fresnel coefficients are functions of incident angle at an
interface, the important point is that the prism has the same shape
as the sawtooth radiator grooves.  Hence, by moving the stage such
that the prism and sawtooth crystal passes through the beam, a
relative transmission measurement using the prism as a standard may
be made. Uncertainties were at the 1\% level.

A third spectrophotometer, a duplicate to this system, was
constructed and placed on site at the crystal manufacturing company
in order to make quality assurance measurements of the window and
planar radiator crystals before shipping.

%



\end{document}